\documentclass[sigconf]{acmart} 

\usepackage{cleveref} 
\usepackage[labelfont=bf]{caption}
\usepackage{array}
\usepackage{multirow}
\usepackage{subcaption}    
\usepackage{float}
\usepackage{color}
\usepackage{balance}
\usepackage{enumitem}    
\usepackage{graphicx}
\usepackage{caption}
\usepackage{tabularx}
\usepackage[table]{xcolor}
\usepackage{booktabs}
\usepackage{soul}

\definecolor{labelred}{HTML}{b58484}
\definecolor{labelblue}{HTML}{84a8b5}
\definecolor{labelgreen}{HTML}{7eb0a3}

\definecolor{labelpink}{HTML}{ffd9d7}
\definecolor{labelgrey}{HTML}{e3dcdc}

\AtBeginDocument{%
  \providecommand\BibTeX{{%
    \normalfont B\kern-0.5em{\scshape i\kern-0.25em b}\kern-0.8em\TeX}}}

\copyrightyear{2021}
\acmYear{2021}
\setcopyright{acmlicensed}\acmConference[CHI '21]{CHI Conference on Human Factors in Computing Systems}{May 8--13, 2021}{Yokohama, Japan}
\acmBooktitle{CHI Conference on Human Factors in Computing Systems (CHI '21), May 8--13, 2021, Yokohama, Japan}
\acmPrice{15.00}
\acmDOI{10.1145/3411764.3445250}
\acmISBN{978-1-4503-8096-6/21/05}

\usepackage{blindtext,graphicx}
\usepackage[absolute]{textpos}
\begin{document}
\begin{textblock}{4}(0.5,1)
\noindent\Large\textcolor{red} {To cite: Prerna Juneja and Tanushree Mitra. 2021. Auditing E-Commerce Platforms for Algorithmically Curated Vaccine Misinformation. In Proceedings of the 2021 CHI Conference on Human Factors in Computing Systems (CHI '21). Association for Computing Machinery. DOI: https://doi.org/10.1145/3411764.3445250}
\end{textblock}
\title{Auditing E-Commerce Platforms for Algorithmically Curated Vaccine Misinformation}

\author{Prerna Juneja}
\affiliation{%
  \institution{The Information School\\University of Washington}
  \city{Seattle}
  \country{WA, USA}}
  \email{prerna79@uw.edu}

\author{Tanushree Mitra}
\affiliation{%
  \institution{The Information School\\University of Washington}
  \city{Seattle}
  \country{USA}}
  \email{tmitra@uw.edu}

\begin{abstract}
There is a growing concern that e-commerce platforms are amplify\-ing vaccine-misinformation. To investigate, we conduct two-sets of algorithmic audits for vaccine misinformation on the search and recommendation algorithms of Amazon---world's leading e-retailer. First, we systematically audit search-results belonging to vaccine-related search-queries without logging into the platform---unperson\-alized audits. We find 10.47\% of search-results promote misinformative health products. We also observe ranking-bias, with Amazon ranking misinformative search-results higher than debunking search-results. Next, we analyze the effects of personali\-zation due to account-history, where history is built progressively by performing various real-world user-actions, such as clicking a product. We find evidence of filter-bubble effect in Amazon's recommendations; accounts performing actions on misinformative products are presented with more misinformation compared to accounts performing actions on neutral and debunking products. Interestingly, once user clicks on a misinformative product, home\-page recommendations become more contaminated compared to when user shows an intention to buy that product.
\end{abstract}

\maketitle

\section{Introduction}

The recent onset of coronavirus pandemic has unleashed a barrage of online health misinformation \cite{covidconsp,fakecures} and renewed focus on the anti-vaccine movement, with anti-vax social media accounts witnessing a 19\% increase in their follower base \cite{benefit}.
As scientists work towards creating a vaccine for the disease, health experts worry that  vaccine hesitancy could make it difficult to achieve herd immunity against the new virus \cite{ball2020anti}. Battling health mis\-information, especially anti-vaccine misinformation has never been more important. 

Statistics show that people increasingly rely on the  internet \cite{pewsurvey2000}, and specifically online search engines \cite{pewsurvey2006_2}, for health information including information about medical treatments, immunizations, vaccinations and vaccine-related side effects \cite{pewstat,doi:10.1080/21645515.2017.1264742}. 
Yet, the algor\-ithms powering search engines are not traditionally designed to take into account the credibility and trustworthiness of such inform\-ation. Search platforms being the primary gateway and reportedly the most trusted source \cite{edelman2014digital}, persistent vaccine misinformation on them, can cause serious health ramifications \cite{kata2010postmodern}.
Thus, there has been a growing interest in empirically investigating search engine results for health misinformation. While multiple studies have performed audits on commercial search engines  to investigate problematic behaviour \cite{10.1145/3308558.3313654,robertson2018auditing,hussein2020measuring}, e-commerce platforms have received little to no attention (\cite{10.1145/2872427.2883089,shin2020algorithms} are two  exceptions),  despite critics calling e-commerce platforms, like Amazon, a ``dystopian'' store for hosting anti-vaccine books \cite{wired1}. Amazon specifically has faced  criticism from several technology critics for not regulating health-related products on its platform \cite{wired2,vox1}. Consider the most recent instance. Several medically unverified products for corona\-virus treatment, like prayer healing, herbal treatments and antiviral vitamin supplements proliferated Amazon \cite{quartz1,npr1}, so much so that the company had to remove 1 million fake products after several instances of such treatments were reported by the media \cite{fakecures}. 
The scale of the problematic content suggests that  Amazon could be a great enabler of misinformation, especially health misinformation. It not only hosts problematic health-related content but 
its reco\-mmendation algorithms drive engagement by pushing
potentially dubious health products to
users of the system \cite{amazonrecommendationsuggestingbomb,shin2020algorithms}. Thus, in this paper we investigate Amazon---world's leading e-retailer---for most critical form of health misinformation---vaccine misinformation.

What is the amount of misinformation present in Amazon's search results and recommendations? How does personalization due to user history built progressively by performing real-world user actions, such as clicking or browsing certain products, impact the amount of misinformation returned in subsequent search results and recommendations? 
In this paper, we dabble into these questions. We conduct 2 sets of systematic audit experiments: \emph{Unpersonalized audit} and \emph{Personalized audit}. In the \emph{Unpersonalized audit}, we adopt Information Retrieval metrics from prior work \cite{kulshrestha_bias} to determine the amount of health misinformation  
users are exposed to when searching for vaccine-related queries. In particular, we examine search-results of 48 search queries belonging to 10 popular vaccine-related topics like `hpv vaccine', `immunization', `MMR vaccine and autism', etc. We collect search results   without logging in to Amazon to eliminate the influence of personalization. To gain in-depth insights about the platform's searching and sorting algorithm, our \emph{Unpersonalized audits} ran for 15 consecutive days,  sorting the search results across 5 different Amazon filters each day: ``featured'', ``price low to high'', ``price high to low'', ``average customer review'' and ``newest arrivals''.
The first audit resulted in 36,000 search results and 16,815 product page recommendations which we later annotated for their stance on health misinformation---promoting, neutral or debunking. 

In our second set of audit---\emph{Personalized audit}, we determine the impact of personalization due to user history on the amount of health misinformation returned in search results, recommendations and auto-complete suggestions. User history is built progressively over 7 days by performing several real-world actions, such as  ``search'' \parbox[c]{0.5cm}{
       \includegraphics[width=0.5cm, height=0.4cm]{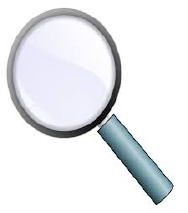}},
       ``search + click''  \parbox[c]{1cm}{
       \includegraphics[width=1cm, height=0.4cm]{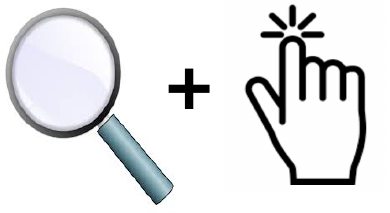}},
       ``search + click + add to cart'' \parbox[c]{1.5cm}{
       \includegraphics[width=1.5cm, height=0.4cm]{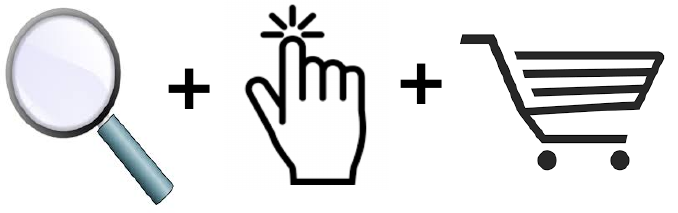}},
       ``search + click + mark top-rated all positive review as helpful''  \parbox[c]{1.5cm}{
       \includegraphics[width=1.5cm, height=0.4cm]{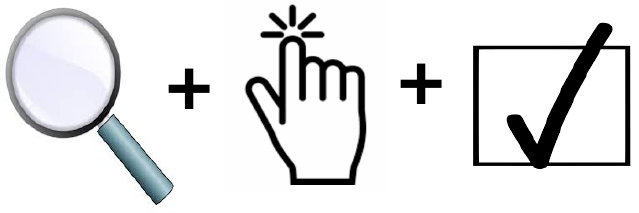}}, 
       ``follow contributor'' \parbox[c]{0.6cm}{
       \includegraphics[width=0.4cm, height=0.4cm]{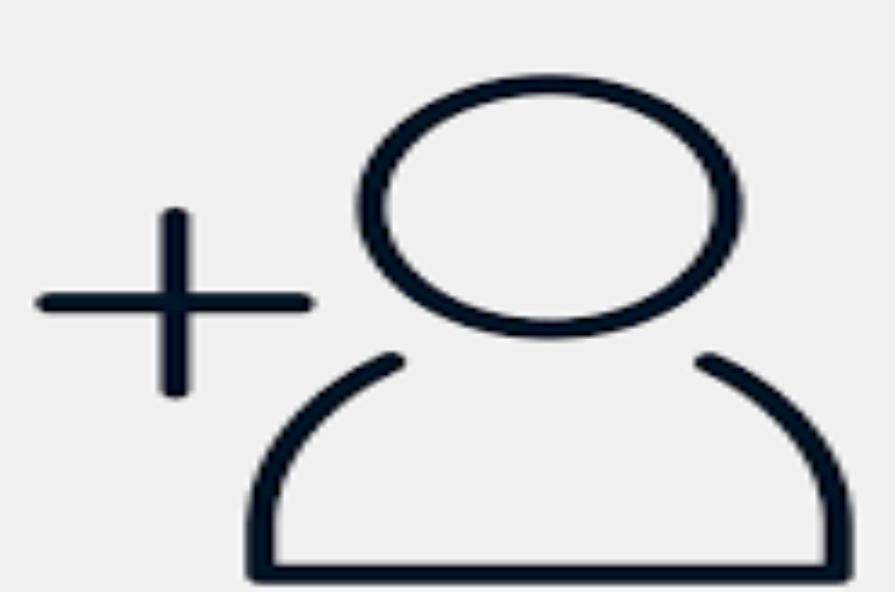}} and ``search on third party website''  ( \parbox[c]{0.5cm}{
       \includegraphics[width=0.5cm, height=0.35cm]{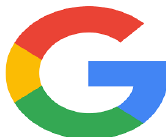}} Google.com in our case) . We collect several Amazon components in 
      our \textit{Personalized audit}, like homepages, product pages, pre-purchase pages, search results, etc. 
    Our audits reveal that Amazon hosts a plethora of health misinformative products belonging to several categories, including Books, Kindle eBooks, Amazon Fashion (e.g. apparel, t-shirt, etc.) and Health \& Personal care items (e.g. dietary supplements). We also establish the presence of a filter-bubble effect in Amazon's recommendations, where recom\-mendations of misinformative health products contain more health misinformation.
    
    Below we present our formal research questions, key findings,  contributions and implication of this study along with ethical consi\-derations taken for conducting platform audits.

\subsection{Research Questions and Findings}
In our first set of audits, we ask,

\noindent\textbf{RQ1 [\emph{Unpersonalized audit}]: What is the amount of health misinformation returned in various Amazon components, given components are not affected by user personalization?}
\begin{itemize}[leftmargin=*]
\item[] \indent RQ1a: How much are the Amazon's search results contaminated with misinformation?
\item[] \indent RQ1b: How much are  recommendations contaminated with misinformation? Is there a filter-bubble effect in recommendations?
\end{itemize}

We find a higher percentage of products promoting health misin\-formation (10.47\%) compared to products that debunk misinforma\-tion (8.99\%) in the unpersonalized search results. We  discover that Amazon returns high number of misinformative search results when users sort their searches by filter ``featured'' and high number of debunking results when they sort results by filter ``newest arrivals''. We also find Amazon ranking misinformative results higher than debunking results especially when results are sorted by filters ``average customer reviews'' and ``price  low to high''. Overall, search results of topics ``vaccination'', ``andrew wakefield'' and ``hpv vaccine'' contain the highest misinformation bias when sorted by default filter ``featured''. Our analysis of product page recommendations suggests that recommendations of products promoting health mis\-information contain more health misinformation when compared to recommendations of neutral and debunking products. 

\noindent\textbf{RQ2 [\emph{Personalized audit}]: What is the effect of personaliza\-tion due to user history on the amount of health misinforma\-tion returned in various Amazon components, where user history is built progressively  by performing certain actions?}
\begin{itemize}[leftmargin=*]

\item[]  \indent RQ2a: How are  \emph{search results} affected by various user actions? 
\item[] \indent  RQ2b: How are  \emph{recommendations} affected by various user actions? Is there a filter-bubble effect in the recommendations?
\item[]  \indent RQ2c: How are the \emph{auto-complete suggestions} affected by  various user actions?
\end{itemize}

Our \emph{Personalized audit} reveals that search results sorted by filters ``average customer review'', ``price low to high'' and ``newest arrivals'' along with auto-complete suggestions are not personalized. Additionally, we find that user actions involving clicking a search product leads to personalized homepages. We find evidence of filter-bubble effect in various recommendations found in homepages, product and pre-purchase pages. Surprisingly, the amount of misinformation present in homepages of accounts building their history by performing actions ``search + click'' and ``mark top-rated all positive review as helpful'' on misinformative products was more than the amount of misinformation present in homepages of accounts that added the same misinformative products in cart. The finding suggests that Amazon nudges users more towards misinformation once a user shows interest in a misinformative product by clicking on it but hasn't shown any intention of purchasing it. 
Overall, our audits suggest that Amazon has a severe vaccine/health misinformation problem exacerbated by its search and recommendation algorithms. Yet, the platform has not taken any steps to address this issue.



\subsection{Contributions and Implications}
In the absence of an online regulatory body monitoring the quality of content created, sold and shared, vaccine misinformation is rampant on online platforms. Through our work, we specifically bring the focus on e-commerce platforms since they have the power to influence browsing as well as buying habits of millions of people. We believe our study is the first large-scale systematic audit of an e-commerce platform that investigates the role of its algorithms in surfacing and amplifying vaccine misinformation.  Our work provides an elaborate understanding of how Amazon's algorithm is introducing misinformation bias in product selection stage and ranking of search results across 5 Amazon filters for 10 impactful vaccine-related topics.  We find that even use of different search filters on Amazon can dictate what kind of content a user can be exposed to. For example,  use of  default filter ``featured''  lead users to more health misinformation while sorting search results by  filter ``newest arrivals'' lead users to products debunking health-related misinformation. Ours is also the first study to empirically establish how certain real-world actions on health misinformative products on Amazon could  drive users into problematic  echo chambers of health misinformation. 
Both our audit experiments resulted in a dataset of 4,997 unique Amazon products distributed across 48 search queries, 5 search filters, 15 recommendation types, and 6 user actions, conducted over 22  (15+7) days \footnote{\url{https://social-comp.github.io/AmazonAudit-data/}}. 
Our findings suggest that traditional  recommendation algorithms should not be blindly applied to all topics equally. There is an urgent need for Amazon to treat vaccine related searches as searches of higher importance and ensure higher quality content for them. Finally, our findings also have several design implications that we discuss in detail in Section \ref{interventions}.

\subsection{Ethical Considerations} We took several steps to minimize the potential harm of our exp\-eriments to retailers. For example, buying and later returning an Amazon product for the purpose of our project can be deemed unethical and thus, we avoid performing this activity. Similarly, writing a fake positive review about an Amazon product containing misinformation could negatively influence the audience. Therefore, in our \emph{Personalized audit} we explored other alternatives that could mimic similar if not the same influence as the aforementioned activities. For example, instead of buying a product, we performed "add to cart" action that shows users' intent to purchase a product. Instead of writing positive reviews for products, we marked top rated positive review as helpful. Since, accounts did not have any purchase history, marking a review helpful did not increase the ``Helpful'' count for that review. Through this activity, the account shows positive reaction towards the product, at the same time avoids manipulation and thus, eliminates impacting potential buyers or users. Lastly, we refrained from performing the experiments on real-world users.  Performing actions on misinformative products could contaminate users' searches and recommendations. It could potentially have long-term consequences in terms of what types of products are pushed at participants. Thus, in our audit experiments, accounts were managed by bots that emulated the actions of actual users.

\section{Related work}

\subsection{Health misinformation in online systems}

The current research on online health misinformation including vaccine misinformation spans three broad themes: 1) quantifying the characteristics of anti-vaccine discourse \cite{mitra2016understanding,monsted2019algorithmic,cossard2020falling}, 2) building machine learning models to identify  users engaging with health misinformation or instances of health misinformation itself \cite{ghenai2018fake,Dai_Sun_Wang_2020,ghenai2017catching} and 3) designing and evaluating effective interventions to ensure that users critically think when presented with health (mis)information \cite{kim2020effects,van2020seeking}. 
Most of these studies are post-hoc inv\-estigations of health misinformation, i.e the misinformation has already propagated. 
Moreover, existing scholarship rarely takes into account how the user encountered health misinformation or what role is played by the source of the misinformation.
With the increasing reliance on online sources for health information, search engines have become the primary avenue of such information, with 55\% of American adults relying on the web to get medical information \cite{pewsurvey2000}. A Pew survey reports that for 5.9M people, web search results influenced their decision to visit a doctor and 14.7M claimed that online information affected their decision on how to treat a disease \cite{pewsurvey2000}. Given how medical information can directly influence one's health and well-being, it is essential  that search engines return quality results in response to health related search queries.
However, currently online health information has been contaminated by several outlets.
These sources could be conspiracy groups or websites spreading misinformation due to vested interests or companies having commercial interests in selling herbal cures or fictitious medical treatments \cite{schwitzer2017pollution}.
Moreover, online curation algorithms themselves are not built to take into account the cred\-ibility of information. Thus, it is of paramount importance that the role of search engines are investigated for harvesting  health misinformation. 
How can we empirically and systematically probe search engines to investigate problematic behaviour like prevalence of health misinformation? In the next section, we briefly describe the emerging research field of ``algorithmic auditing'' that is  focused on investigating search engines to reveal problematic biases. We discuss this field as well as our 
contribution to this growing research space in the next section.

\subsection{Search engine audits} 

Search engines are modern day gatekeepers and curators of infor\-mation. 
Their black-box algorithm  can shape user behaviour, alter beliefs and even affect voting behaviour either by impeding or facilitating the flow of certain kinds of information \cite{epstein2015search,diakopoulos2018vote,knobloch2015science}. Despite their importance and the power they exert, till date,  search engine results and recommendations have mostly been unregulated. Information quality of search engine's output is still measured in terms of relevance and it is up to the user to determine the credibility of information. Thus, researchers have advocated for making algorithms more accountable. One primary method to achieve this is to perform systematic audits to empirically establish the conditions under which problematic behavior surfaces. 
Raji et al provide the following definition of algorithmic audits. \textit{An algorithmic audit involves the collection and analysis of outcomes from a fixed algorithm or defined model within a system.
Through the stimulation of a mock user population, these audits can uncover problematic patterns in models of interest} \cite{raji2019actionable}.

\begin{figure*}[t]
  \centering
  \begin{subfigure}[b]{0.325\textwidth}
      \centering
      \includegraphics[width=0.938\textwidth,keepaspectratio]{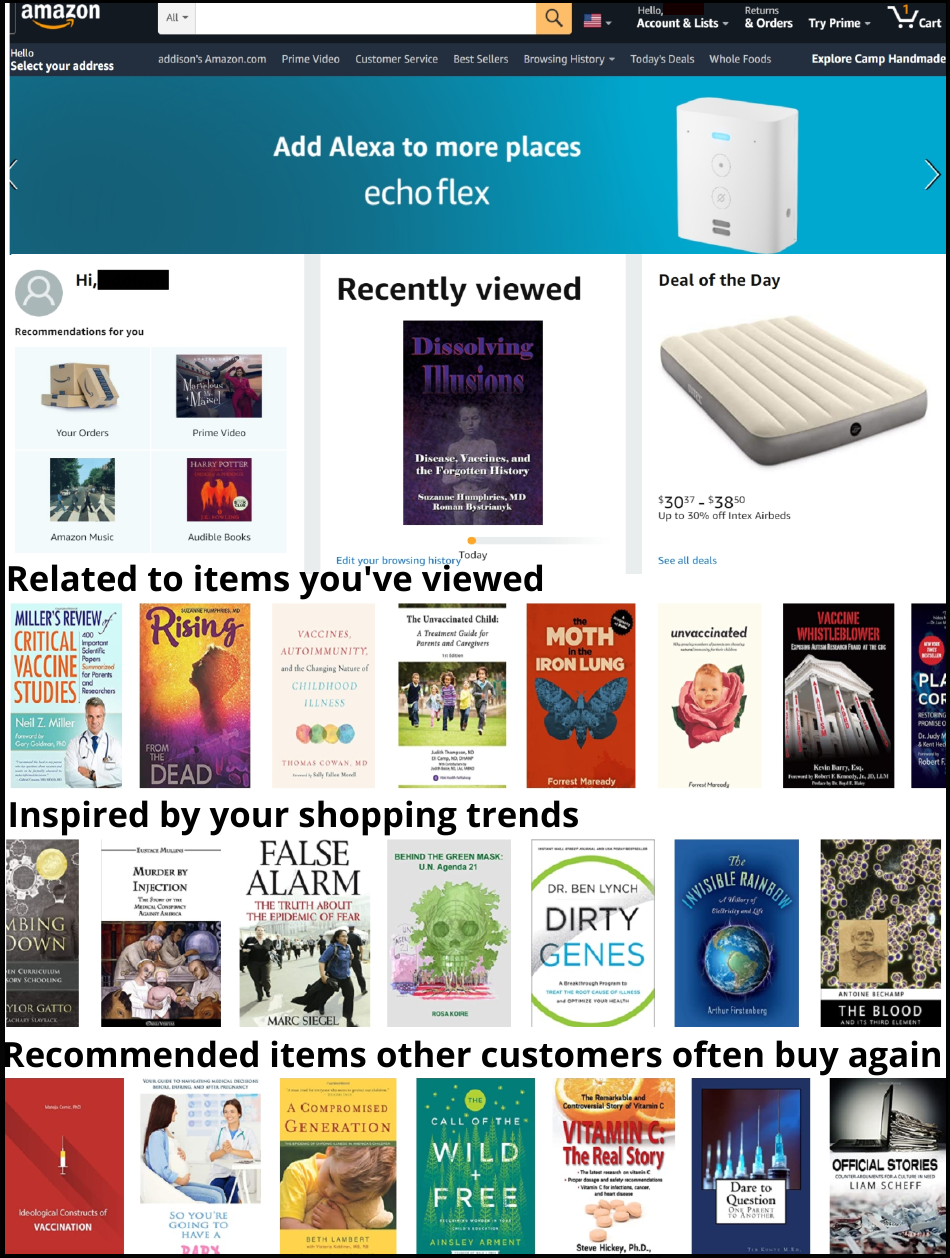}
      \caption{}
      \label{amazon homepage recommendations}
      \Description[Amazon homepage]{Figure illustrates the Amazon homepage containing several books belonging to three different recommendation types specified in Table 1.}
  \end{subfigure}\hspace{-0.16cm}
  \begin{subfigure}[b]{0.38\textwidth}
      \centering
      \includegraphics[width=0.97\textwidth,keepaspectratio]{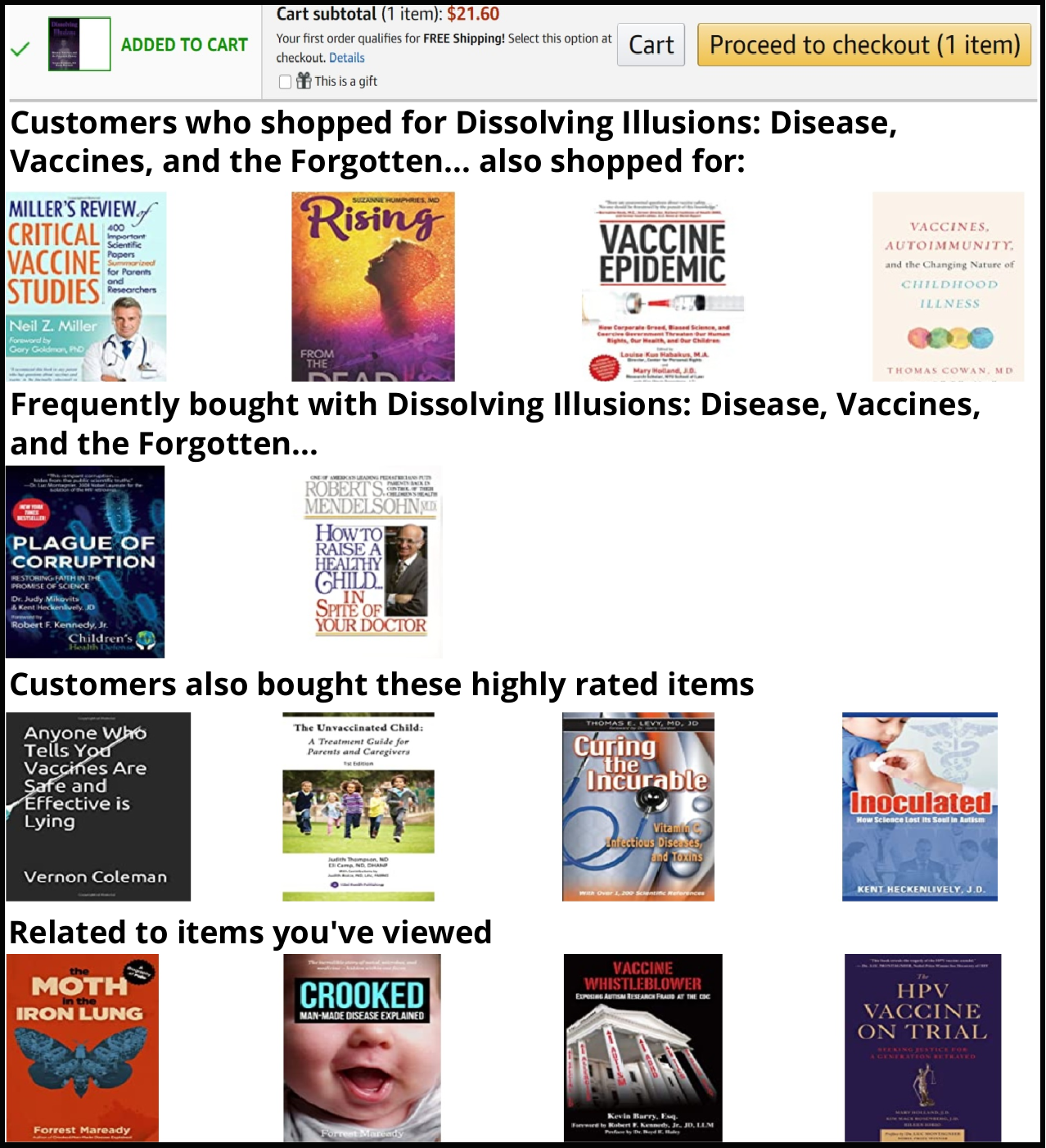}
      \caption{}
      \label{Pre-purchase recommendations}
      \Description[Amazon pre-purchase page]{Figure illustrates the Amazon pre-purchase page with several books belonging to different recommendation types.}
  \end{subfigure}
  \begin{subfigure}[b]{0.2665\textwidth}
   \centering
      \includegraphics[width=1\textwidth,keepaspectratio]{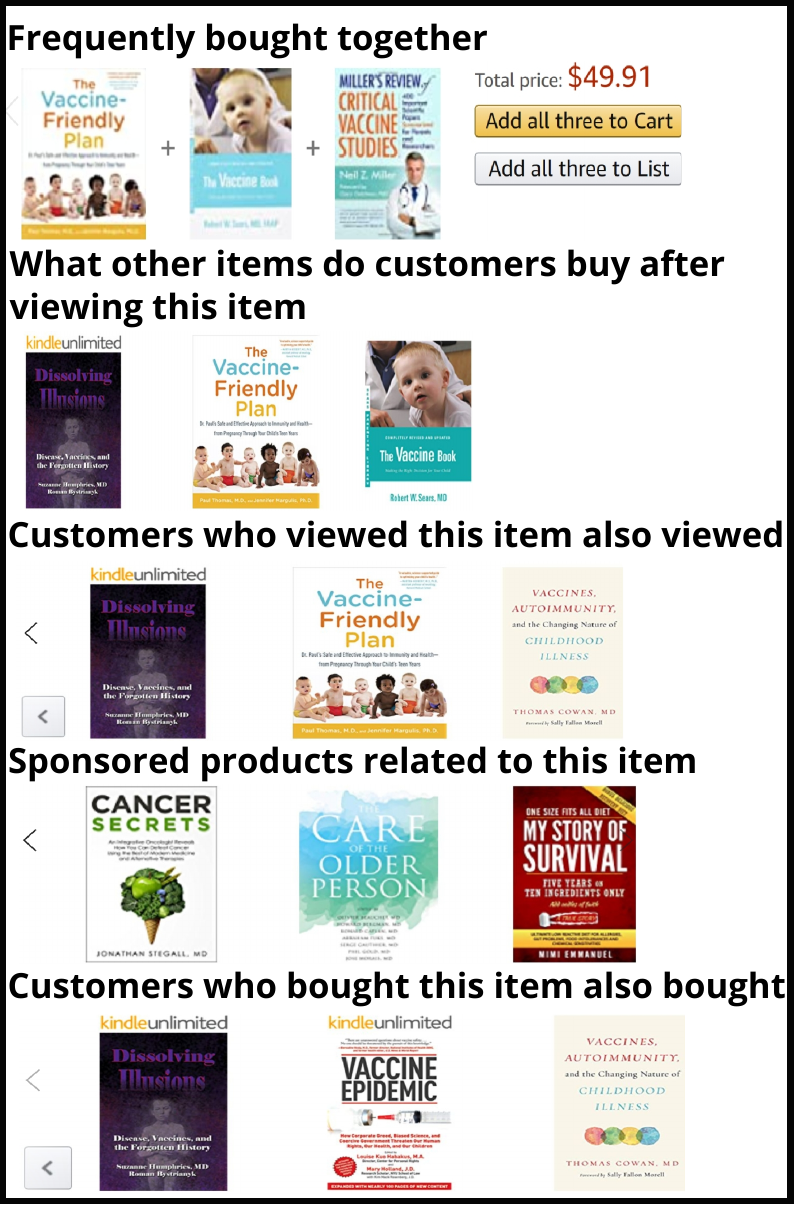}
      \caption{}
       \label{details page recommendations}
       \Description[Amazon product page]{Figure illustrates several books belonging to five different recommendation types  present on the Amazon product page specified in Table 1.}
  \end{subfigure}
%

 \caption{(a) Amazon homepage recommendations. (b) Pre-purchase recommendations displayed to users after adding a product to cart. (c) Product page recommendations.}
 \label{fig:topquery}
\end{figure*}

Previous audit studies have investigated the search engines for partisan bias {\mbox{\cite{robertson2018auditing,10.1145/3351095.3372835}}}, gender bias \cite{chen2018investigating,10.1145/2702123.2702520}, { content diversity \mbox{\cite{10.1145/3290605.3300683,doi:10.1080/1369118X.2020.1776367,doi:10.1080/21670811.2018.1539626}}}, and price discrimination \cite{hannak2014measuring}. However, only a few have systematically investigated search engines' role in surfacing misinformation (\cite{hussein2020measuring} is the only exception). Moreover, there is a dearth of systematic audits focusing specifically on health mis\-information. The past literature, mostly consists of small-scale experiments that probe search engines with a handful of search queries. For example, an analysis of the first 30 pages of search results for query ``vaccines autism'' revealed that Google.com has 10\%  less anti-vaccine search results compared to the other search engines, like Qwant, Swisscows and Bing   \cite{ghezzi2020online}. Whereas, search results present in the first 102 pages for the query ``autism vaccine'' on Google's Turkey version returned 20\% websites with incorrect information \cite{erden2019autism}. One recently published work, closely related to this study, examined Amazon's first 10 pages of search results  in response to the query ``vaccine''. They only collected and annotated books appearing in the searches for misinformation \cite{shin2020algorithms}. The afore\-mentioned studies probed the search engine for one single query and did the analysis on multiple search results pages. We, on the other hand, perform our  \emph{Unpersonalized audit} on a curated  list of 48 search queries belonging to 10 most searched vaccine-related topics, spanning various combinations of search filters and recommenda\-tion types, over multiple days---an aspect missing in prior work.  Additionally, we are the first ones to experimentally quantify the prevalence of misinformation  in various search queries, topics, and filters on an e-commerce platform.  Furthermore, instead of just focusing on books, we analyze the platform for products belonging to different categories, resulting in an extensive all-category inclusive coding scheme for health misinformation. 

Another recent study on YouTube, audited the platform for various misinformative topics including vaccine controversies \cite{hussein2020measuring}. The work established the effect of personalization due to watching  videos on the amount of misinformation present in search results and recommendations on YouTube. However, there are no studies investigating the impact of personalization on misinformation present in the product search engines of e-commerce platforms.  Our work fills this gap by conducting a second set of audit---\emph{Personalized audit} where we shortlist several real-world user actions and investi\-gate their role in amplifying misinformation in Amazon's searches and recommendations.

\begin{table}
\centering
    \begin{footnotesize}
\begin{tabular}{l|l}
\hline
\textbf{\begin{tabular}[c]{@{}l@{}}Recommend-\\ ation page\end{tabular}}      & \textbf{Recommendation types}                                                                    \\ \hline
\multirow{3}{*}{Homepage}                                                     & Related to items you’ve viewed                                                                   \\ \cline{2-2} 
                                                                              & Inspired by your shopping trends”                                                                \\ \cline{2-2} 
                                                                              & \begin{tabular}[c]{@{}l@{}}Recommended items other customers often buy again\end{tabular}      \\ \hline
\multirow{7}{*}{\begin{tabular}[c]{@{}l@{}}Pre-purchase \\ page\end{tabular}} & \begin{tabular}[c]{@{}l@{}}Customers also bought these highly  rated items\end{tabular}        \\ \cline{2-2} 
                                                                              & Customers also shopped these items                                                               \\ \cline{2-2} 
                                                                              & Related to items you’ve viewed                                                                   \\ \cline{2-2} 
                                                                              & Frequently bought together                                                                       \\ \cline{2-2} 
                                                                              & Related to items                                                                                 \\ \cline{2-2} 
                                                                              & Sponsored products related                                                                       \\ \cline{2-2} 
                                                                              & Top picks for                                                                                    \\ \hline
\multirow{5}{*}{Product page}                                                 & Frequently bought together                                                                       \\ \cline{2-2} 
                                                                              & \begin{tabular}[c]{@{}l@{}}Customers who bought this item also bought\end{tabular}             \\ \cline{2-2} 
                                                                              & \begin{tabular}[c]{@{}l@{}}Customers who viewed this item also viewed\end{tabular}             \\ \cline{2-2} 
                                                                              & \begin{tabular}[c]{@{}l@{}}Sponsored products related to this item\end{tabular}                \\ \cline{2-2} 
                                                                              & \begin{tabular}[c]{@{}l@{}}What other items customers buy after viewing this item\end{tabular}
\end{tabular}
\end{footnotesize}
\caption{Table showing 15 recommendation types spread across 3 recommendation pages.}
\label{reco_types}
\vspace{-0.8cm}
\end{table}

\begin{figure*}
  \centering
  \begin{subfigure}{0.33\textwidth}
      \centering
      \includegraphics[width=1\textwidth,keepaspectratio]{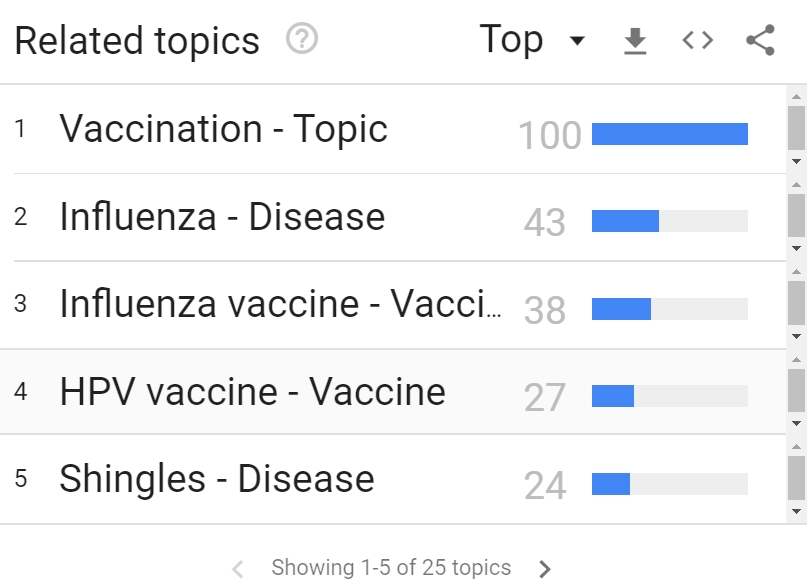}
      \caption{}
      \label{searchtopicfig}
      \Description[Google Trends' Related topics]{Figure illustrates the Related topics section of the Google Trends page for topic vaccine. The section displays topics related to vaccine such as vaccination, influenza, HPV vaccine etc.}
  \end{subfigure}\hspace{5mm}
  \begin{subfigure}{0.33\textwidth}
      \centering
      \includegraphics[width=1\textwidth,keepaspectratio]{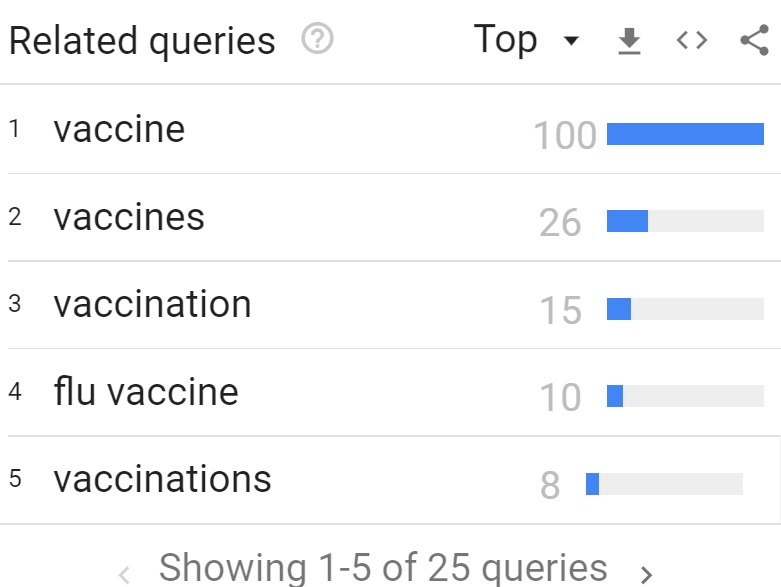}
      \caption{}
      \label{searchqueryfig}
      \Description[Google Trends' Related queries]{Figure illustrates the Related queries section of the Google Trends page for topic vaccine. The section contains search queries related to topic vaccine such as vaccine, vaccines, vaccination, flu vaccine etc.}
  \end{subfigure}\hspace{5mm}
 \begin{subfigure}{0.25\textwidth}
      \centering
      \includegraphics[width=1\textwidth,keepaspectratio]{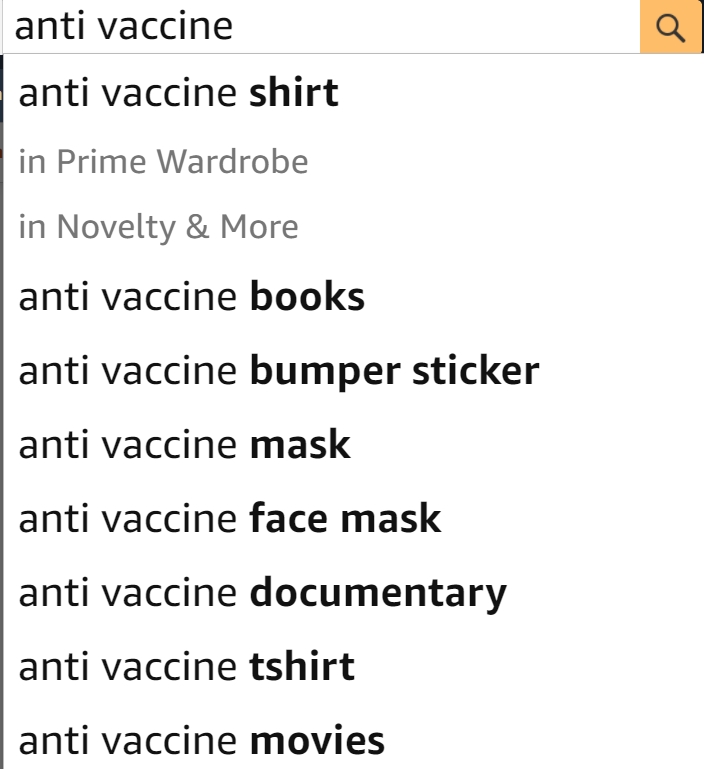}
      \caption{}
      \label{autocompletefig}
      \Description[Amazon's auto-complete suggestions]{Figure illustrates Amazon's auto-complete suggestions for query anti vaccine. Some of the search query suggestions displayed are anti vaccine shirt, anti vaccine books,anti vaccine mask etc. }
  \end{subfigure}
  \caption{(a) Google Trends' Related Topics list for \textit{topic} vaccine. People who searched for vaccine topic also searched for these topics. (b) Google Trends' Related queries list for \textit{topic} vaccine. These are the top search queries searched by people related to vaccine topic. (c) Amazon's auto-complete suggestions displaying popular and trending search queries.}
  \label{fig:topic_query_selection}
\end{figure*}

\section{Amazon components and terminology}\label{compo}
For the audits, we collected 3 major Amazon components and numerous sub-components. We list them below.

\begin{enumerate}
	\item \textbf{Search results: }These are products present on Amazon's  Search Engine Results Page (SERP) returned in response to a search query. SERP results can be sorted using five filters: ``featured'', ``price low to high,'' ``price high to low,'' ``average customer review'' and ``newest arrivals.''
	\item \textbf{Auto-complete suggestions:} These are the popular and trending search  queries suggested by Amazon when a query is typed into the search box (see Figure \ref{autocompletefig}). 
	\item \textbf{Recommendations:} Amazon presents several recommen\-dations as users navigate through the platform. For the purpose of this project, we collect recommendations present on three different Amazon pages: homepage, pre-purchase page and product pages. Each page hosts several types of recommendations. Table \ref{reco_types} shows the 15 recommendation types collected across 3 recommendation pages. We describe all three recommendations below.
	
	\begin{enumerate}
	\item \textbf{Homepage recommendations:} These recommendations are present on the homepage of a user's Amazon account. They could be of three types namely, ``Related to items you've viewed'', ``Inspired by your shopping trends'' and ``Recommended items other customers often buy again'' (see Figure \ref{amazon homepage recommendations}). Any of the three types together or separately could be present on the homepage depending on the actions performed by the user. For example, ``Inspired by your shopping trends'' recommendation type appears when a user performs one of two actions: either makes a purchase or adds a product to cart.
	
	\item \textbf{Pre-purchase recommendations:} These recommenda\-tions consist of product suggestions that are presented to users after they add product(s) to cart. These recommenda\-tions could be considered as a  nudge  to purchase other similar products. Figure \ref{Pre-purchase recommendations} displays pre-purchase page. The page has several recommenda\-tions like ``Frequently bought together'', ``Customers also bought these highly rated items'', etc. We collectively call these recommendations as pre-purchase recommendations.
	
	\item \textbf{Product recommendations:} These are the recommen\-dations present on the product page, also known as details page\footnote{https://sellercentral.amazon.com/gp/help/external/51}. 
	The page contains details of an Amazon product, like product title, category (e.g., Amazon Fashion, Books, Health \& Personal care, etc.),  description, price, star rating, number of reviews, and other metadata. 
	The details page is home to several different types of recommendations. We extracted five: ``Frequently bought together'', ``What other items customers buy after viewing this item'', ``Customers who viewed this item also viewed'', ``Sponsored products related to this item'' and ``Customers who bought this item also bought''. Figure \ref{details page recommendations} presents an example of product page recommendations.
\end{enumerate}
\end{enumerate}

  \begin{figure*}
  \centering
      \includegraphics[scale=0.7]{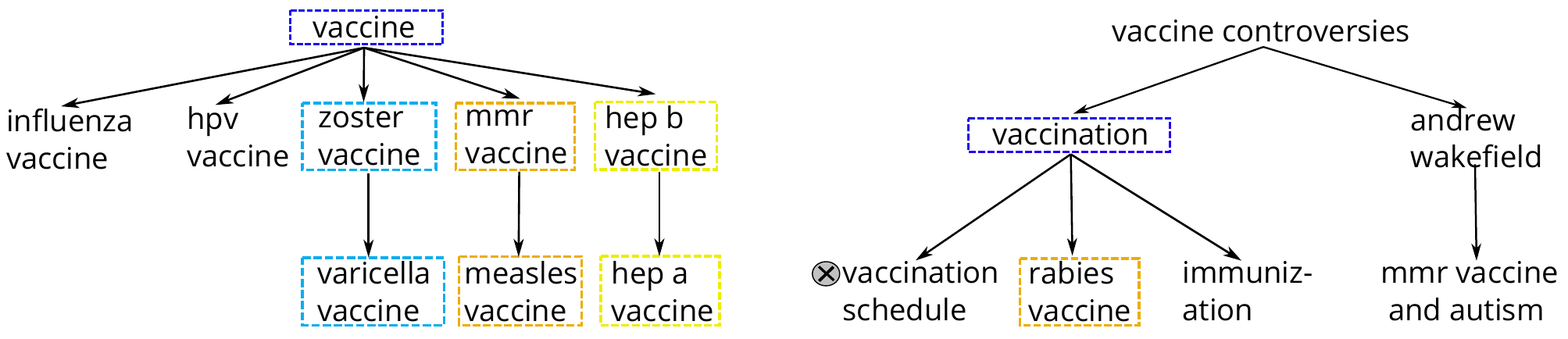}
  \caption{Figure illustrating the breadth-wise topic discovery approach used to collect vaccine-related topics from Google Trends starting from two seed topics: vaccine and vaccine controversies. Each node in the tree denotes a vaccine-related topic. An edge A$\rightarrow$ B indicates that topic B was discovered from the Trends' Related Topic list of topic A. For example, topics ``vaccination'' and ``andrew wakefield'' were obtained from the Trends' Related Topic list of ``vaccine controversies'' topic. Then, topic ``mmr vaccine and autism'' was obtained from topic ``andrew wakefield'' and so on. \protect\includegraphics[height=0.3cm]{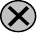} indicates the topic was discarded during filtering. Similar colored square brackets indicate similar topics that were merged together. }
  \label{topic_tree_pruned}
  \Description[Topic discovery approach]{The figure contains two trees with roots as our seed topics namely, vaccine and vaccine controversies. The children of each node are the topics discovered from the Related Topic list of the parent. For example, topics vaccination and andrew wakefield are children of root node vaccine controversies. Furthermore, topics vaccination schedule, rabies vaccine and immunization are child nodes of topic vaccination.}
\end{figure*}

\begin{table*}
\resizebox{\textwidth}{!}{%
\begin{tabular}{l|lllll|lll}
\textbf{\#}        & \textbf{Search topic}                                                             & \textbf{Seed query}                                                                          & \textbf{\begin{tabular}[c]{@{}l@{}}Sample search\\ queries\end{tabular}} &  & \#                 & \textbf{Search topic}                                                              & \textbf{Seed query}                                                                   & \textbf{\begin{tabular}[c]{@{}l@{}}Sample search\\ queries\end{tabular}} \\ \cline{1-4} \cline{6-9} 
\multirow{2}{*}{1} & \multirow{2}{*}{\begin{tabular}[c]{@{}l@{}}vaccine \\ controversies\end{tabular}} & \multirow{2}{*}{\begin{tabular}[c]{@{}l@{}}vaccine controversy/\\ anti vaccine\end{tabular}} & anti vaccination                                                         &  & \multirow{2}{*}{6} & \multirow{2}{*}{\begin{tabular}[c]{@{}l@{}}mmr vaccine \\ and autism\end{tabular}} & \multirow{2}{*}{\begin{tabular}[c]{@{}l@{}}mmr autism/\\ vaccine autism\end{tabular}} & autism                                                                   \\
                   &                                                                                   &                                                                                              & anti vaccine shirt                                                       &  &                    &                                                                                    &                                                                                       & autism vaccine                                                           \\ \cline{1-4} \cline{6-9} 
\multirow{2}{*}{2} & \multirow{2}{*}{vaccination}                                                      & \multirow{2}{*}{\begin{tabular}[c]{@{}l@{}}vaccine/\\ vaccination\end{tabular}}              & vaccine                                                                  &  & \multirow{2}{*}{7} & \multirow{2}{*}{\begin{tabular}[c]{@{}l@{}}influenza \\ vaccine\end{tabular}}      & \multirow{2}{*}{varicella vaccine}                                                    & flu shot                                                                 \\
                   &                                                                                   &                                                                                              & vaccine friendly me                                                      &  &                    &                                                                                    &                                                                                       & influenza vaccine                                                        \\ \cline{1-4} \cline{6-9} 
\multirow{2}{*}{3} & \multirow{2}{*}{\begin{tabular}[c]{@{}l@{}}andrew \\ wakefield\end{tabular}}      & \multirow{2}{*}{andrew wakefield}                                                            & andrew wakefield                                                         &  & \multirow{2}{*}{8} & \multirow{2}{*}{\begin{tabular}[c]{@{}l@{}}hepatitis\\ vaccine\end{tabular}}       & \multirow{2}{*}{\begin{tabular}[c]{@{}l@{}}hepatitis\\ vaccine\end{tabular}}          & hepatitis b vaccine                                                      \\
                   &                                                                                   &                                                                                              & wakefield autism                                                         &  &                    &                                                                                    &                                                                                       & hepatitis a vaccine                                                      \\ \cline{1-4} \cline{6-9} 
\multirow{2}{*}{4} & \multirow{2}{*}{hpv vaccine}                                                      & \multirow{2}{*}{hpv vaccine}                                                                 & vaccine hpv                                                              &  & 9                  & \multirow{2}{*}{\begin{tabular}[c]{@{}l@{}}varicella \\ vaccine\end{tabular}}      & \multirow{2}{*}{\begin{tabular}[c]{@{}l@{}}varicella \\ vaccine\end{tabular}}         & chicken pox                                                              \\
                   &                                                                                   &                                                                                              & hpv vaccine on trial                                                     &  &                    &                                                                                    &                                                                                       & varicella vaccine                                                        \\ \hline
\multirow{2}{*}{5} & \multirow{2}{*}{immunization}                                                     & \multirow{2}{*}{immunization}                                                                & immunization                                                             &  & 10                 & \multirow{2}{*}{mmr vaccine}                                                       & \multirow{2}{*}{mmr vaccine}                                                          & mmr vaccine                                                              \\
                   &                                                                                   &                                                                                              & immunization book                                                        &  &                    &                                                                                    &                                                                                       & measles vaccination                                                     
\end{tabular}}
\caption{Sample search queries for each of the ten vaccine-related search topics.}
\label{search_topics}
\end{table*}

\section{Methodology}
Here we present our audit  methodology in detail. This section is organized as follows. We start by describing our approach to compile high impact vaccine related topics and associated search queries (\cref{compiletopics}). Then, we present overview of each audit experiment followed by the details of numerous methodological decisions we took while designing our audits (\cref{rq1method} and \cref{rq2method}). Next, we describe our qualitative coding scheme for annotating Amazon products for health misinformation (\cref{anno_section}). Finally, we discuss our approach to calculate misinformation bias in search results (\cref{misinfo_bias_metric}).

\subsection{Compiling high impact vaccine-related topics and search queries} \label{compiletopics}

Here, we present our methodology to curate high impact vaccine-related topics and search queries.

\subsubsection{Selecting high impact search topics:}
The first step of any audit is to determine input---a viable set of topics and associated search queries that will be used to query the platform under investigation. We leveraged Google Trends (\emph{Trends} henceforth) to select and expand vaccine-related search topics. \emph{Trends} is an optimal choice since it shares past search trends and popular queries searched by people across the world. Since it is not practical to audit all topics present on \emph{Trends}, we designed a method to curate a reasonable number of high impact topics and associated search queries, i.e., topics that were searched by a large number of people for the longest period of time. 
We started with 2 seed topics and employed a breadth-wise search to expand our topic list.

\emph{Trends} allows to search for any subject matter either as a \textit{topic} or a \textit{term}. Intuitively, \textit{topic} can be considered as a collection of terms that share a common concept. Searching as a \textit{term} returns results that include terms present in the search query while searching as a \textit{topic} returns all search terms having same meaning as the topic\footnote{\url{https://support.google.com/trends/answer/4359550?hl=en}}. 
We began our search with two seed words namely ``vaccine'' and ``vaccine controversies'' and decided to search them as \textit{topics}. Starting our topic search by the aforementioned seed words ensured that the related topics will cover general vaccine-related topics and topics related to controversies surrounding the vaccines, offering us a holistic view of search interests. We set location to United States, date range to 2004-Present (this step was performed in Feb, 2020), categories to ``All'' and search service to ``Web search''. The date range ensured  that the topics are perennial, and have been popular for a long time (note that \emph{Trends} data is available from 1/1/2004 onwards). 
We selected the category setting as ``All” so as to get a holistic view of the search trends encompassing all the categories together. Search service filter has options like `web search', `YouTube search', `Google Shopping', etc. Although Google shopping is an e-commerce platform like Amazon,  its selection returned handful to no results. Thus, we opted for `web search' service filter. 

We employed \emph{Trends'} Related Topics feature for breadth-wise expansion of search topics (see Figure \ref{searchtopicfig}).  We viewed the Related Topics using ``Top'' filter which presents popular  search topics in the selected time range that are related to the topic searched. We manually went through the top 15 Related Topics and  retained relevant topics using the  following guidelines. All generic topics like Infant, Travel, Side-Effects, Pregnancy CVS,  etc. were discarded. Our focus was to only pick topics representing vaccine information.  
Thus, we discarded topics that were names of diseases but kept their corresponding vaccines. For example, we discarded topic Influenza but kept the topic Influenza vaccine.
We  kept track of duplicates and discarded them from the search. To further expand the topics list, we again went through the Related Topics list of the shortlisted topics and used the aforementioned filtering strategy to shortlist relevant topics. This step allowed us to expand our topic list to a reasonable number.
After two levels of breadth-wise search, we obtained a list of 16 vaccine-related search topics (see Figure \ref{topic_tree_pruned}). 

 \begin{figure*}
  \centering
      \includegraphics[scale=0.65]{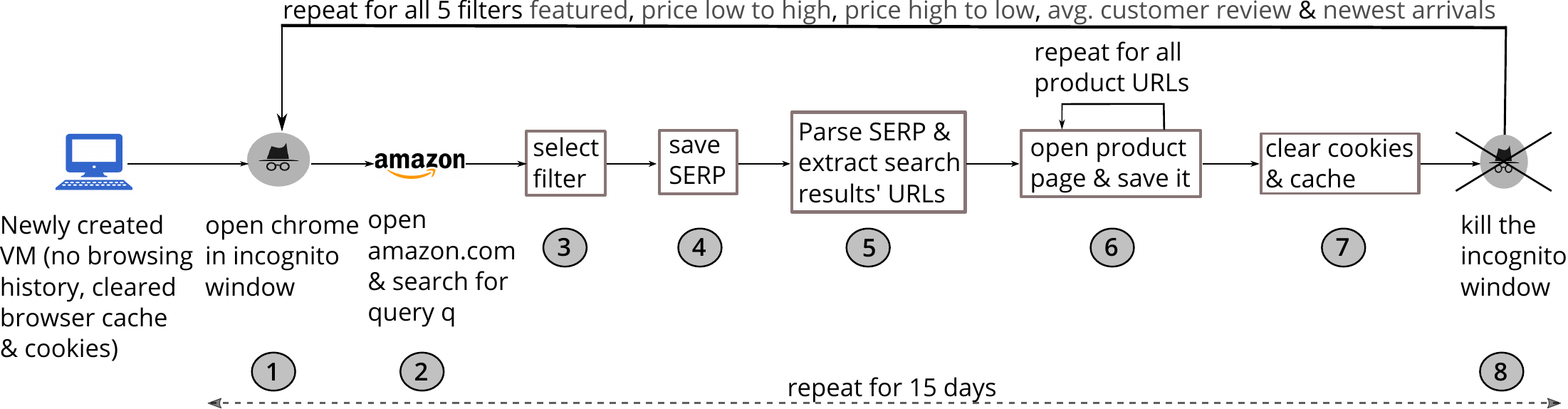}
  \caption{Eight steps performed in \textit{Unpersonalized audit}. The steps are described in detail in Section \ref{unperonalized_implemention}}
  \label{unpers_diag}
  \Description[Experimental steps of Unpersonalized audit]{Figure illustrates the eight steps performed in our Unpersonalized audit. These steps are explained in the method section in detail.}
\end{figure*}
Next, we combined multiple similar topics into a single topic. The idea is to collect search queries for both topics separately and then combine them under one single topic. For example,  topics zoster vaccine and varicella vaccine were combined since both the vaccines are used to prevent chickenpox. Thus, later search queries of both topics were  combined under topic varicella vaccine. All topics enclosed with similar colored boxes in Figure \ref{topic_tree_pruned} were merged together.  11 topics remained after merging.

\subsubsection{Selecting high impact search queries:} \label{query_curation} 
After shortlisting a reasonable number of topics, next we determined the associated search queries per topic, to be later used for querying Amazon's search engine. To compile search queries, we relied on both \emph{Trends} and Amazon's auto-complete suggestions; \emph{Trends}, because it gives a list of popular  queries that people searched on Google---the most popular search service,  and Amazon, because it is the platform under investigation and it will provide popular trending queries specific to the platform. 

Searching for a topic on \emph{Trends} displays popular search queries related to the topic (see Figure \ref{searchqueryfig}). We obtained top 3 queries per topic. 
Next, we collected Top 3 auto-complete suggestions obtained by typing seed query of each topic into Amazon's search box (see Figure \ref{autocompletefig}). 
We removed all animal or pet related search queries (e.g ``rabies vaccine for dogs''), overly specific queries (e.g. ``callous disregard by andrew wakefield'') and replaced redundant and similar queries with a single search query selected at random. 
For example  search queries ``flu shots'' and ``flu shot'' were replaced with a single search query “flu shot”. 
After these filtering steps, only one query remained in the query list of topic vaccination schedule, and thus, it was removed from the topic list. Finally, we had 48 search queries corresponding to 10 vaccine-related search topics. Table \ref{search_topics} presents sample search queries for all 10 search topics.

\subsection{RQ1: Unpersonalized Audit} \label{rq1method}

\subsubsection{Overview}
The aim of the \textit{Unpersonalized audit} is to determine the amount of misinformation present in Amazon's search results and recommendations without the influence of  personalization. 
We measure the amount of misinformation by determining the misinformation bias of the returned results. We explain the misinf\-ormation bias calculation in detail in Section \ref{misinfo_bias_metric}. Intuitively, more the number of higher ranked misinformative results, higher the overall bias. We ran the \textit{Unpersonalized audit} for 15 days, from 2 May, 2020 to 16 May, 2020. We took two important methodological decisions regarding which components to audit and what sources of noise to control for. We present these decisions as well as implementa\-tion details of the audit experiment below.

\subsubsection{What components should we collect for our Unpersonalized audits?}
We collected SERPs sorted by all 5 Amazon filters: ``featured'', ``price low to high'', ``price high to low'', ``average customer review'' and ``newest arrivals''. For analysis, we extracted the top 10 search results from each SERP. {Since 70\% of Amazon users never click on search results beyond the first page \mbox{\cite{amazon_search_results}}, count 10 is a reasonable approximation of the number of search results users are likely to engage with.} Recent statistics have {also} shown that the first three search results receive 75\% of all clicks \cite{top3ctr}. Thus, we extracted the recommendations present on the product pages of the first three search results.  We collected  following 5 types of product page recommendations: ``Frequently bought together'', ``What other items customers buy after viewing this item'', ``Customers who viewed this item also viewed'', ``Sponsored products related to this item'' and ``Customers who bought this item also bought''. Refer Figure \ref{details page recommendations} for an example. We extracted the first product present in each recommendation type for analysis. Next, we annotated all collected components as promoting, neutral or debunking health misinformation.  We describe our annotation scheme shortly in Section \ref{anno_section}.

\subsubsection{How can we control for noise?}
We controlled for potential confounding factors that may add noise to our audit measurements. To eliminate the effect of personalization, we ran the experiment on newly created virtual machines (VM) and freshly installed browser with empty browsing history, cookies and cache. Additionally, we ran search queries from the same version of Google Chrome in incognito mode to ensure that no history is built during our audit runs. To avoid cookie tracking, we erased cookies and cache before and after opening the incognito window and destroyed the window after each search. In sum, we performed searches on newly created incognito windows everyday. All VMs operated from same geolocation so that any effects due to location would affect all machines equally. To prevent machine speeds from affecting the experiment, all VMs had the same architecture and configuration. To control for temporal effect, we searched every single query  at one particular time everyday for consecutive 15 days. Prior studies have established the presence of carry-over effect in search engines, where  previously executed queries affect the results of the current query when both queries are issued subsequently within a small time interval \cite{hannak2013measuring}. Since, we destroyed browser windows and cleared session cookies and cache after every single search, carry over effect did not influence our experiment. 

\begin{table*}
\resizebox{\textwidth}{!}{%
{ \footnotesize
\begin{tabular}{l|m{4cm}|m{1.5cm}|m{3cm}|m{3.4cm}}
\hline
\# & \multicolumn{2}{m{3.5cm}|}{\textbf{User action}}                                & \textbf{Type of history}                      & \textbf{Tested values}                                                                                                                                                                                                                                                                              \\ \hline
1  & Search product                                                        & \parbox[c]{0.5cm}{
       \includegraphics[width=0.5cm, height=0.4cm]{symbol/s.pdf}}  & Product search history                        & \multirow{6}{*}{\begin{tabular}[c]{@{}l@{}}Product debunks vaccine or other \\health related misinformation \\ (annotation value -1)  \&\\ \\ Neutral health information\\ (annotation value 0)  \&\\ \\ Product promotes vaccine or other \\health related misinformation \\(annotation value 1)\end{tabular}}                               \\ \cline{1-4}
2  & Search + click product                                              &  \parbox[c]{1cm}{
      \includegraphics[width=1cm, height=0.4cm]{symbol/s+c.pdf}}   & Product search and click history              &                                                                                                                                                                                                                                                                                                                                  \\ \cline{1-4}  
3  & Search + click + add to cart                                          & \parbox[c]{1.8cm}{
      \includegraphics[width=1.5cm, height=0.4cm]{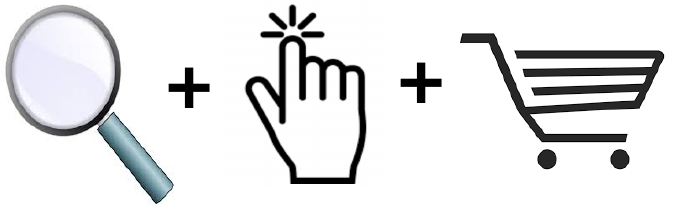}} & Intent to purchase history                    &                                                                                                                                                                                                                                                                                                                                \\ \cline{1-4}
4  & Search + click + mark “Top rated, All positive review”  helpful       & \parbox[c]{1.5cm}{
      \includegraphics[width=1.5cm, height=0.4cm]{symbol/mr.pdf}} & Searching, clicking and marking reviews helpful history &                                                                                                                                                                                                                                                                                                                                 \\ \cline{1-4}  
5  & Following contributor by clicking follow button on contributor’s page & \parbox[c]{0.6cm}{
      \includegraphics[width=0.4cm, height=0.4cm]{symbol/f.pdf}} & Following history                             &                                                                                                                                                                                                                                                                                                                               \\ \cline{1-4}  
6  & Search product on Google (third party application)                    & \parbox[c]{0.5cm}{
      \includegraphics[width=0.5cm, height=0.35cm]{symbol/g.pdf}} & Third party search history                    &                                                                                                                                                                                                                                                                                                                               \\ \hline
\end{tabular}}}
\caption{List of user actions employed to build account history. Every action and product type (misinformative, neutral or debunking) combination was performed on two accounts.  One account sorted search results by filters ``featured'' and ``average customer review''. The other account built history in the same way but sorted the search results by filters ``price low to high'' and ``newest arrivals''.  Overall, we created 40 Amazon accounts (6 actions X 3 tested values X 2 replicates for filters + 2 control accounts + 2 twin accounts).}
\label{tab:per_features}
\end{table*}
\subsubsection{Implementation details}  \label{unperonalized_implemention}
Figure \ref{unpers_diag} illustrates the eight steps for the \emph{Unperonalized audit}. We used  Amazon Web Services (AWS) infrastructure to create all the VMs. We created selenium bots to automate web browser actions. 
As a first step, each day at a particular time, the bot opened {\tt amazon.com} in incognito window. 
Next, the bot searched for a single query, sorted the results by an Amazon filter and saved the SERPs.
The bot then extracted the top 10 URLs of the products present in the results. The sixth step is an iterative step where the bot iteratively opened the product URLs and saved the product pages. In the last two steps, the bot cleared the browser cache and killed the browser window. We repeated steps 1 to 8 to collect search results sorted by all 5 Amazon filters. We added appropriate wait times  after each step to prevent Amazon from detecting the account as a bot and blocking our experiment. We repeated these steps for 15 consecutive days for each of the 48 search queries. After completion of the experiment, we parsed the saved product pages to extract product metadata, like product category, contributors' names (author, editor, etc.), star rating and number of ratings. We extracted product page recommendations for the top 3 search results only.

\subsection{RQ2: Personalized Audit} \label{rq2method}
\subsubsection{Overview}
The goal of our  Personalization Experiments is twofold. First, we assess whether user actions, such as clicking on a product, adding to cart would trigger personalization on Amazon. Second, and more importantly, we determine the impact of a user's account history on the amount of misinformation presented to them in the search results page, recommendations, and auto-complete suggestions; account history is built progressively by performing a particular action for seven consecutive days.  We ran our \emph{Personalized audit} from 12th to 18th August, 2020. We took several methodological decisions while 
designing this experimental setup. 
We discuss each of these decisions below.


\subsubsection{What real-world user actions should we select to build account history?}\label{actions}
Users' click history and purchase history trigger personal\-ization and influence the price of commodities on e-commerce websites \cite{hannak2014measuring}. Account history also affects the amount of misinformation present in the personalized results \cite{hussein2020measuring}.  Informed by the results of these studies, 
we selected six real-world user actions  that could trigger personalization and thus, could potentially  impact the amount of misinformation in search results and recommendations. The actions are (1) ``search'' \parbox[c]{0.5cm}{
       \includegraphics[width=0.5cm, height=0.4cm]{symbol/s.pdf}}
       (2) ``search + click'' \parbox[c]{1cm}{
       \includegraphics[width=1cm, height=0.4cm]{symbol/s+c.pdf}}  
       (3) ``search + click + add to cart'' 
       \parbox[c]{1.5cm}{
       \includegraphics[width=1.5cm, height=0.4cm]{symbol/scc.pdf}}
       (4) ``search + click + mark top-rated all positive review as helpful''
        \parbox[c]{1.5cm}{
       \includegraphics[width=1.5cm, height=0.4cm]{symbol/mr.pdf}}
       (5) ``follow contributor'' 
       \parbox[c]{0.6cm}{
       \includegraphics[width=0.4cm, height=0.4cm]{symbol/f.pdf}}
       and (6) ``search on third party website'' (Google.com in our case) 
       \parbox[c]{0.5cm}{
       \includegraphics[width=0.5cm, height=0.35cm]{symbol/g.pdf}}. 
Table \ref{tab:per_features} provides an overview. First two actions involve searching for a product and/or clicking on it. Through the third and fourth action, a user shows positive reaction towards a product by adding it to cart and marking its top rated critical review as helpful respectively. Fifth action investigates the impact of following a contributor.  For example, for a product in the Books category, the associated list of contributors include the author and editor of the book. The contributors have dedicated profile pages that a user can follow. 
Sixth action investigates the effect of searching for an Amazon product on Google.com. The user logs into Google using the email id used to register the Amazon account. The hypothesis is that Amazon search results could be affected by third party browsing history. 
After selecting the actions, we determined the products on which the actions needed to be performed.

\subsubsection{What products and contributors should we select for building account history?}  

To build user history, all user actions except ``follow contributor'' need to be performed on products. First, we annotated all products collected in the \textit{Unpersonalized audit} run as debunking (-1), neutral (0) or promoting (1) health misinformation. \textbf{}We present the annotation details in Section \ref{anno_section}. 
For each annotation value (-1, 0, 1), we selected top-rated products that had received maximum engagement and belonged to the most occurring category---`Books'
. We started by filtering Books belonging to each annotation value and eliminated the ones that did not have an ``Add to cart'' button on their product page at the time of product selection. {Since users make navigation and engagement decisions based on information cues on the web \mbox{\cite{sensemaking}}, we considered cues present on Amazon such as customer ratings as a criteria to further shortlist Books.} First, we sorted Books based on the accumulated engagement---number of customer ratings received
. Next, we sorted the top 10 Books obtained from the previous sorting based on  star ratings received by the Books to end up with highly rated, high-impact and high-engagement products. We selected top 7 books from the second sorting for the experiment (see Appendix, Table \ref{tab:books} for the shortlisted books).

Action ``follow contributor'' is the only action that is performed on contributors' Amazon profile pages \footnote{The contributors could be authors,  editors, people writing foreward of a book, publisher, etc.}. We selected contributors who contributed to the most number of debunking (-1), neutral (0) and  promoting (1) books. We retained only those who had a profile page on Amazon. 
Table \ref{tab:contributors} lists the selected contributors. 

\begin{table*}
\resizebox{\textwidth}{!}{%
{\scriptsize
\begin{tabular}{l|ll|ll|ll}
\hline
\multirow{2}{*}{\textbf{\#}} & \multicolumn{2}{l|}{\textbf{\begin{tabular}[c]{@{}l@{}}Contributors to debunking\\ health products\end{tabular}}} & \multicolumn{2}{l|}{\textbf{\begin{tabular}[c]{@{}l@{}}Contributors to neutral \\ health products\end{tabular}}} & \multicolumn{2}{l}{\textbf{\begin{tabular}[c]{@{}l@{}}Contributors to misinformative \\ health products\end{tabular}}} \\ \cline{2-7} 
                             & \textbf{name}                                            & \textbf{url code}                                      & \textbf{name}                                            & \textbf{url code}                                     & \textbf{name}                                               & \textbf{url code}                                        \\ \hline
1                            & Paul-A-Offit                                            & B001ILIGP6                                             & Jason-Soft                                               & B078HP6TBD                                            & Andrew-J-Wakefield                                         & B003JS8YQC                                               \\
2                            & Seth-Mnookin                                             & B001H6NG7A                                             & Joy-For-All-Art                                          & B07LDMJ1P4                                            & Mary-Holland                                                & B004MZW7HS                                               \\
3                            & Michael-Fitzpatrick                                      & B001H6L348                                             & Peter-Pauper-Press                                       & B00P7QR4RO                                            & Kent-Heckenlively                                           & B00J08DNE8                                               \\
4                            & Ziegler-Prize                                            & B00J8VZKBQ                                             & Geraldine-Dawson                                         & B00QIZY0MA                                            & Jenny-McCarthy                                              & B001IGJOUC                                               \\
5                            & Ben-Goldacre                                             & B002C1VRBQ                                             & Tina-Payne-Bryson                                        & B005O0PL3W                                            & Forrest-Maready                                             & B0741C9TKH                                               \\
6                            & Jennifer-A-Reich                                        & B001KDUUHY                                             & Vassil-St-Georgiev                                      & B001K8I8XC                                            & Wendy-Lydall                                                & B001K8LNVQ                                               \\
7                            & Peter-J-Hotez                                           & B001HPIC48                                             & Bryan-Anderson                                           & B087RL79G8                                            & Neil-Z-Miller                                              & B001JP7UW6                                               
                                              \\ \hline
\end{tabular}}}
\caption{List of contributors who have contributed to the most number of books that are either debunking, neutral or promote health misinformation, selected for building account history for action ``Follow contributors''. For example, Andrew J Wakefield, Mary Holland (both prominent vaccine deniers) have contributed to the most number of books that promote health misinformation.\protect\footnotemark}
\label{tab:contributors}
\end{table*}

\begin{figure*}
  \centering
      \includegraphics[scale=0.72]{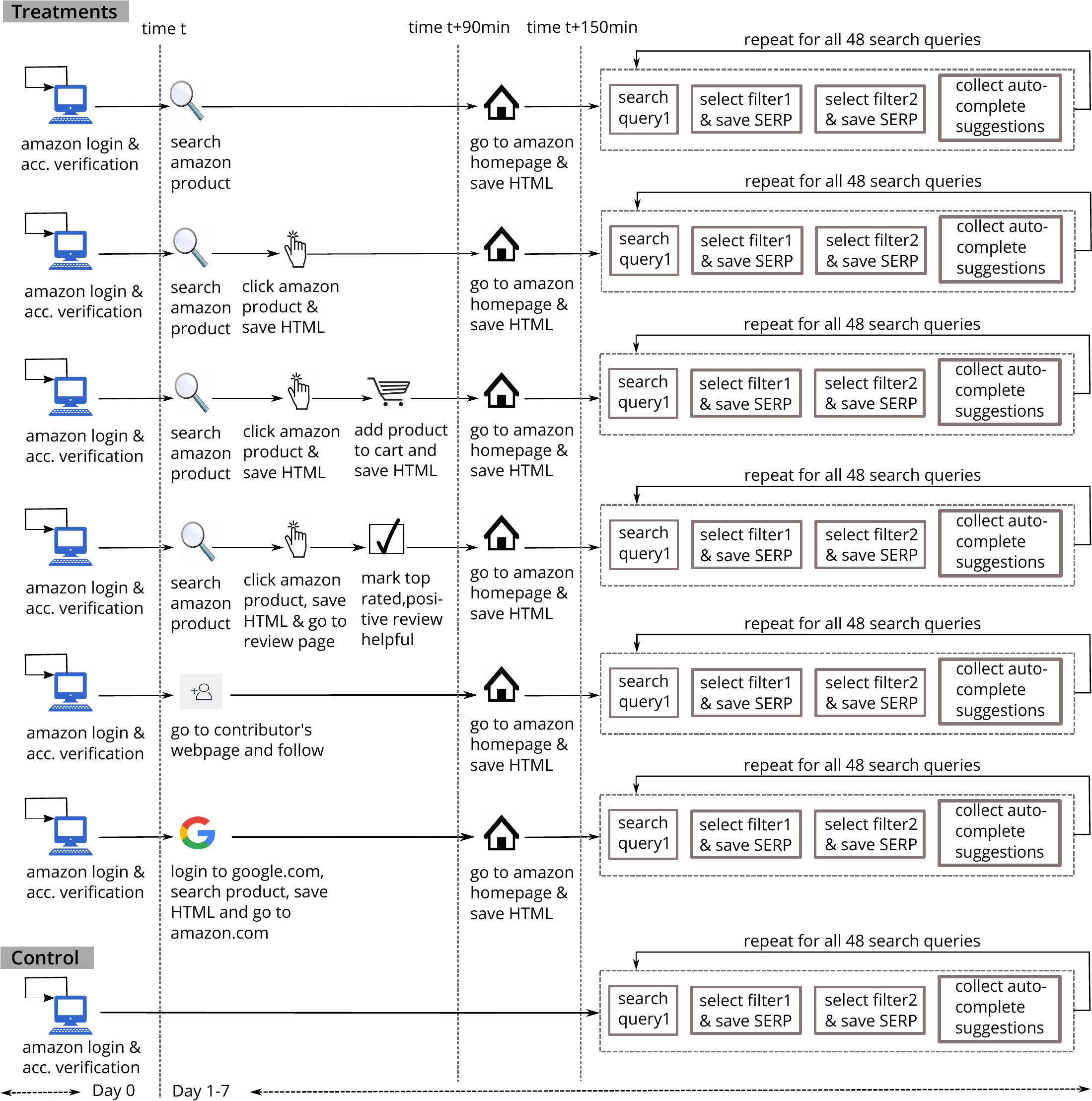}
  \caption{Steps performed by treatment and control accounts in \textit{Personalized audit} corresponding to the 6 different features.}
  \label{p}
  \Description[Experimental steps of Personalized audit]{Figure illustrates how treatment accounts built histories by performing various actions and later collected homepage recommendations, SERPs for all 48 search queries followed by auto-complete suggestions. The control accounts did not build account history but collected SERPs and auto-complete suggestions for the 48 search queries at the same time as the treatment accounts.}
\end{figure*}

\subsubsection{How do we design the experimental setup?} 
We performed all six actions explained in Section \ref{actions} and Table \ref{tab:per_features} on Books (or contributors of the books in case of action ``follow contributor'') that are either all debunking, neutral or promoting health misinformation.
\footnotetext{The contributor's Amazon web page can be accessed by forming the url ``www.amazon.com/ + name + /e/ + url\_code''.}
Each action and product type combination was acted upon by two treatment accounts. One account built its search history by first performing searches on Amazon and then viewing search results sorted by filters ``featured'' and ``average customer review'' while the other did the same but sorted results by ``price low to high'' and ``newest arrivals''\footnote{Every account created for this experiment was run by a bot. It was not possible for a bot to complete the following order of tasks in 24 hours because of wait times added after every action-- building history using a particular action, searching for 48 search queries sorted by 4 filters and collecting auto-complete suggestions for those queries etc. Thus,
every action-product type combination was performed on two accounts. First account, sorted the search results by two filters and second account sorted results using remaining two filters. We call these two accounts replicates since they built their history in the same way.}.  We did not use the filter ``price high to low'' since intuitively it is less likely to be used during searches.

We also created 2 control accounts corresponding to 2 treatments that emulated the same actions as the treatments except that they did not build account histories by performing one of the 6 user actions. 
Like 2 treatment accounts, the first control account searched for 48 queries curated in Section \ref{query_curation} and sorted them by filters ``Featured'' and ``Average customer Review'' while the other control sorted them by the remaining two filters.
Figure \ref{p} outlines the experimental steps performed by treatment and control accounts. 
We also created twins for each of the control accounts. The twins performed the exact same tasks as the corresponding control. 
Any inconsistencies between a control account and its twin can be attributed to noise, and not personalization. Remember, Amazon's algorithms are a black box. Even after controlling for all known possible sources of noise, there could be some sources that we are not aware of or the algorithm itself could be injecting some noise in the results. If the difference between search results of control and treatment is greater than the baseline noise, only then it can be attributed to personalization. Prior audit work  have also adopted the strategy of creating a control and its twin to differentiate between the effect due to noise versus personalization \cite{hannak2014measuring}.  Overall, we created 40 Amazon accounts (6 actions X 3 tested values X 2 replicates for filters + 2 control accounts + 2 twin accounts). Next, we discuss the components collected from each account.


\subsubsection{What components should we collect for the personalized audit?}
We collected search results and auto-complete suggestions for treatment and control accounts to measure the extent of personaliza\-tion. We collected  recommendations only for the treatment accounts since they built history by clicking on product pages, pre-purchase pages, etc. Search results were sorted by filters `featured'',  ``average customer review'', ``price low to high'' and ``newest arrivals''.
Once users start building their account history, Amazon displays several recommendations to drive engagement on the platform. We coll\-ected various types of recommendations spread across three reco\-mmendation pages: homepage, product page and pre-purchase page. Pre-purchase pages  were only collected for the accounts that performed ``add to cart'' action. Additionally, product pages were collected for accounts that clicked on search results while creating their respective account history. Each of the aforementioned pages consist of several reco\-mmendation types, such as ``Customers who bought this item also bought'', etc. We collected the first product present in each of these recommendation types from both product pages and pre-purchase pages and two products from each type from the homepages for further analysis.\textbf{} Refer to Table \ref{reco_types} and  Figures \ref{amazon homepage recommendations}, \ref{Pre-purchase recommendations} and \ref{details page recommendations} for examples of these recommendation types. 

\subsubsection{How do we control for noise?}
Just like our \emph{Unpersonalized audit}, we first controlled for VM configuration and geolocation.
Next, we controlled for demographics by setting the same gender and age for newly creating Google accounts. Recall, that these Google accounts were used to sign-up for the Amazon accounts. Since, the VMs were newly created, the browser had no search history that could otherwise hint towards users' demographics. All accounts created their histories at the same time. They also performed the searches at the same time each day, thus, controlling for temporal effects. 
Lastly, we did not account for carry over effects since it affected all the treatment and control accounts equally.

\subsubsection{Implementation details}
Figure \ref{p} illustrates the experimental steps. We ran 40 selenium bots on 40 VMs. Each selenium bot operated on a single Amazon account. On day 0, we manually logged in to each of the accounts by entering login  credentials and performing account verification. 
Next day, experiment began at time t.  All bots controlling treatment accounts started performing various actions to build history. Note, everyday bots built history by  performing actions on a single Book/contributor. We gave bots sufficient time to build history (90 min) after which they collected and saved Amazon homepage.  Later,  all  40 accounts (control + treatment) searched for 48 queries with different search filters and saved the SERPs. Next, the bots collected and saved auto-complete suggestions for all 48 queries. We included appropriate wait times between every step to prevent accounts from being recognized as bots and getting banned in the process.
We repeated these steps for a week. At the end of the week, for each treatment account we had collected personalized search results, recommendations  and auto-complete suggestions. Next, we annotated the collected search results and recommendations to determine their stance on misinformation so that later we could analyze them to study the effect of user actions on the amount of misinformation presented to users in each component.

\begin{table*}[t]
\centering
{ \scriptsize
\begin{tabular}{c|m{2cm}|m{9.5cm}|p{2.1cm}}
\hline
\multicolumn{1}{l|}{\multirow{2}{*}{\textbf{\begin{tabular}[c]{@{}l@{}}A. Scale\\ Value\end{tabular}}}} & \multirow{2}{*}{\textbf{\begin{tabular}[c]{@{}l@{}}Annotation \\ Description\end{tabular}}} & \multirow{2}{*}{\textbf{Annotation Heuristics}}                                                                                                                                                                                                                                                                                                                                                                                                                                                                                                                                                        & \multicolumn{1}{c}{\multirow{2}{*}{\textbf{Sample Amazon Products}}} \\
\multicolumn{1}{l|}{}                                                                                   &                                                                                             &                                                                                                                                                                                                                                                                                                                                                                                                                                                                                                                                                                                                        & 
\\ \hline
 -1                                                                                                      & debunks vaccine misinformation                                                              & Product debunks, derides OR provides evidence against the myths/controversies surrounding vaccines OR helps understand anti-vaccination attitude OR promotes use of vaccination OR describes history of a disease and details how its vaccine was developed OR describes scientific facts about vaccines that help users to understand how they work OR debunks other health-related misinformation                                                                                                                                                                                                   &  

\parbox[c]{2.5cm}{
      \includegraphics[width=2.5cm,keepaspectratio]{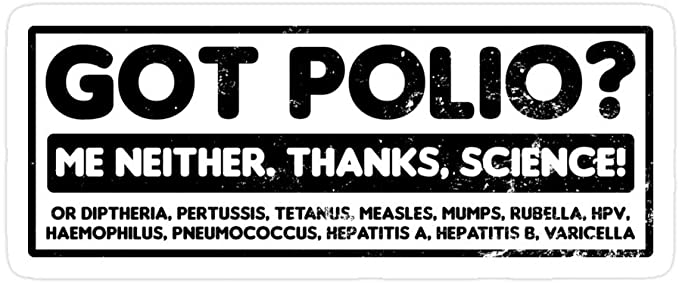}}                                                                     
\\ \hline
0                                                                                                       & neutral health related information                                                          & All medicines and antibodies OR medical equipment (thermometer, syringes, record-books, etc.) OR dietary supplements that do not violate Amazon’s policy OR products about animal vaccination and diseases OR health-related products not promoting any conspiratorial views about health and vaccines                                                                                                                                                                                                                                                                                                  &     \parbox[c]{2.5cm}{\centering
       \includegraphics[height=2.1cm,keepaspectratio]{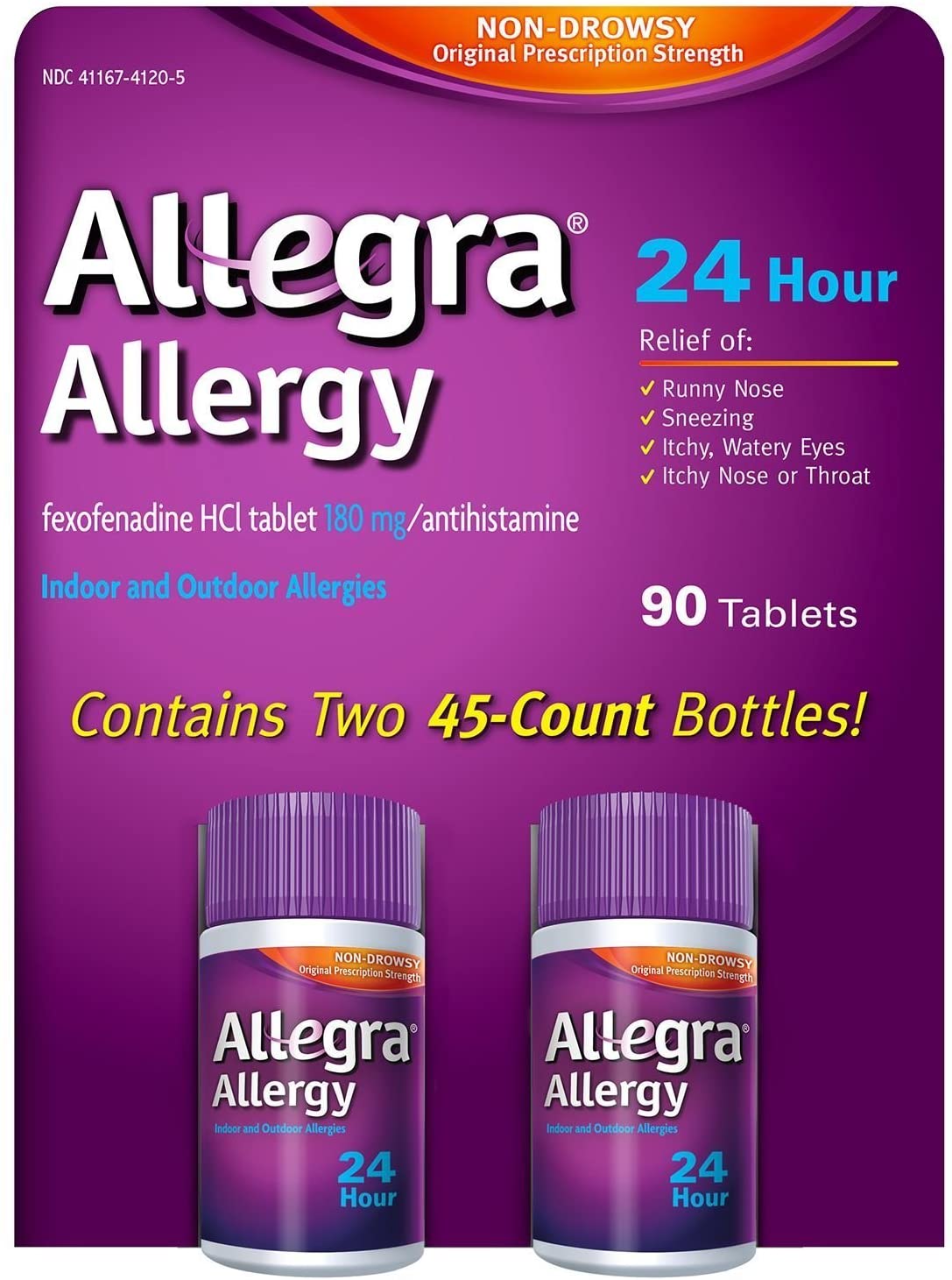}}                                                             \\ \hline
1                                                                                                       & promotes vaccine and other health related misinformation                                    & Product promotes disuse of vaccines OR promotes anti-vaccine myths, controversies or conspiracy theories surrounding the vaccines OR advocates alternatives to vaccines and/or western medicine (diets, pseudoscience methods like homeopathy, hypnosis, etc.) OR product is a misleading dietary supplement that violates Amazon’s policy on dietary supplements- the supplement states that it can cure, mitigate, treat, or prevent a disease in humans, but the claim is not approved by the FDA OR it promotes other health-related misinformation  &  

\parbox[c]{2.5cm}{\centering
       \includegraphics[height=2.1cm, keepaspectratio]{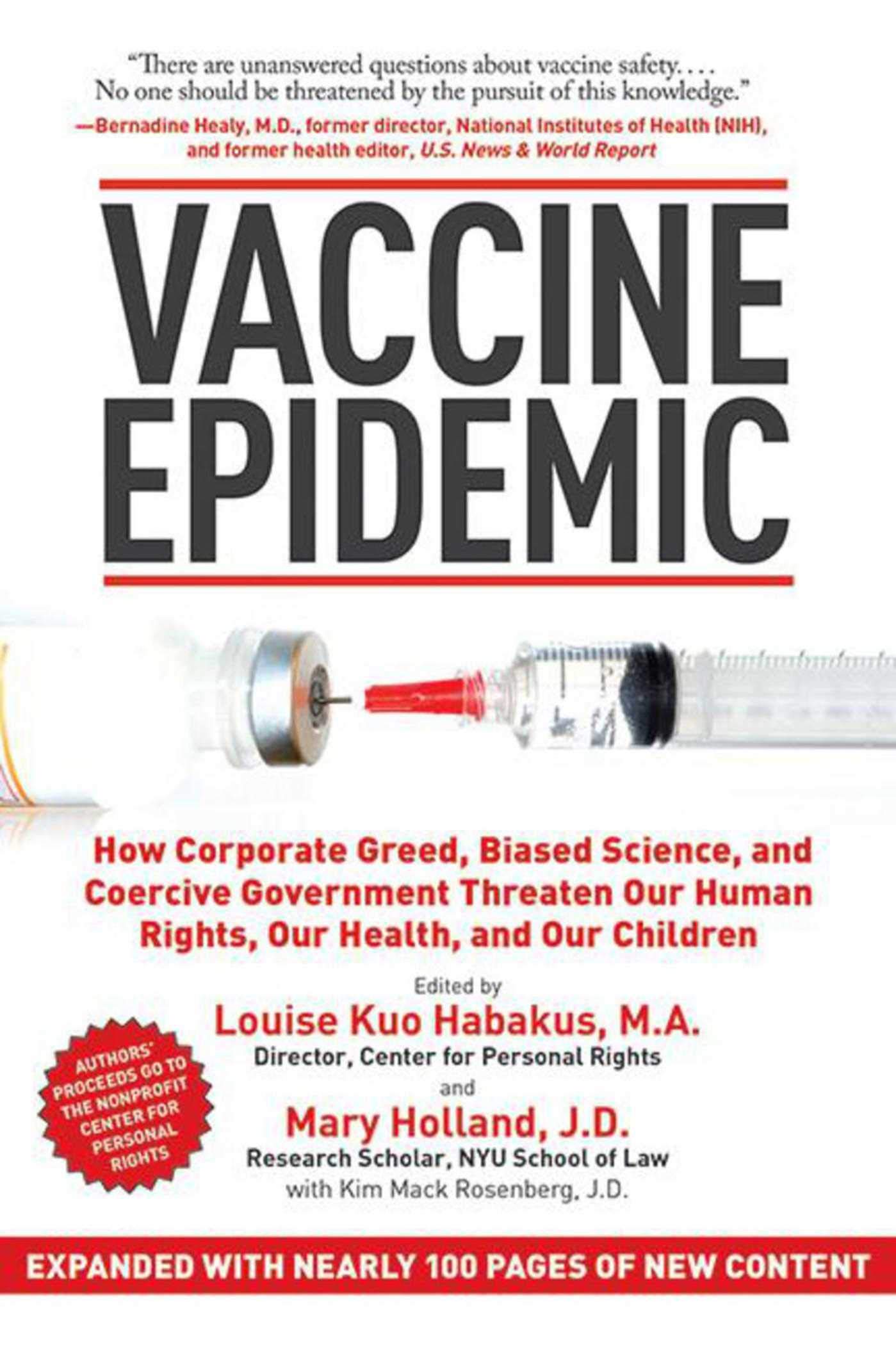}}                                                                  \\ \hline
2                                                                                                       & unknown                                                                                     & Product's description and metadata is not sufficient to annotate it as promoting, debunking or neutral information                                                                                                                                                                                                                                                                                                                                                                                                                                                                                     &      \parbox[c]{2.5cm}{\centering
       \includegraphics[height=2.1cm, keepaspectratio]{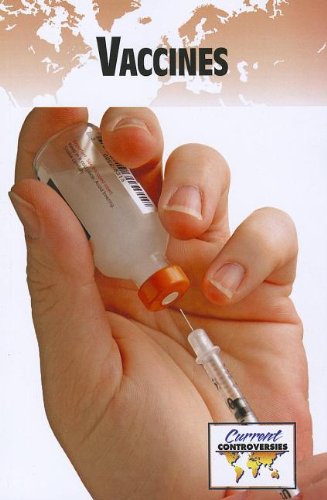}}                                                                \\ \hline
3                                                                                                       & removed                                                                                     & Product's URL is not accessible at the time of annotation                                                                                                                                                                                                                                                                                                                                                                                                                                                                                                                                              &    -                                                                \\ \hline
4                                                                                                       & Other language                                                                              & Product’s title and description is in language other than english                                                                                                                                                                                                                                                                                                                                                                                                                                                                                                                                      &     \parbox[c]{2.5cm}{\centering
       \includegraphics[height=2.1cm, keepaspectratio]{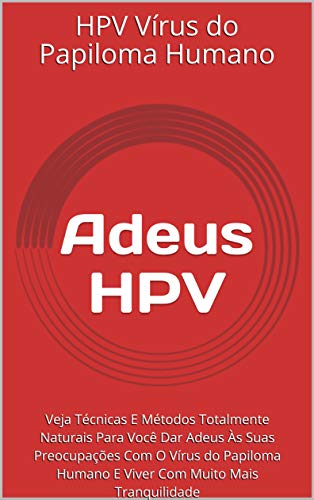}}                                                                 \\ \hline
5                                                                                                       & Unrelated                                                                                   & Non-health related products                                                                                                                                                                                                                                                                                                                                                                                                                                                                                                                                                                            &     \parbox[c]{2.5cm}{\centering
       \includegraphics[height=2.1cm, keepaspectratio]{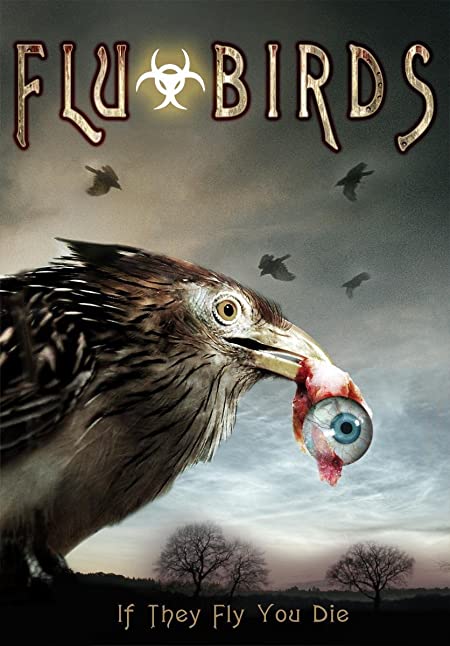}}  \\ \hline                                                              
\end{tabular}}
\caption{Description of annotation scale, heuristics along with sample products corresponding to each annotation value.}
\label{tab:annotation}
\end{table*}

 \subsection{Annotating Amazon data for health misinformation} \label{anno_section}
 Unlike partisan bias where bias could be determined by using features such as news source bias \cite{robertson2018auditing}, labelling a product for misinformation is hard and time-consuming. There are no pre-determined sources of misinformation such as list of sellers or authors of misinformative products on Amazon. Additionally, we found that the annotation process for some categories of products, like Books, Kindle ebooks, etc. required us to consider the product image, read the book's preview, if available, and even perform external search about the authors. Therefore, we opted to manually annotate our data collection. We developed a qualitative coding scheme to label our Amazon data collection through an iterative process that required several rounds of discussions to reach an agreement on the annotation scale. 

In the first round, first author randomly sampled 200 Amazon products  across different topics and categories. After multiple itera\-tions of analyzing and interpreting each product, the author came up with an initial 7-point annotation scale. Then,  six researchers with extensive work experience on online misinformation indepen\-dently annotated 32 products,  randomly selected from the 200 products. We discussed every product's annotation value and the researchers' annotation process. We refined the scale as well as the scheme based on the  feedback.
This process was repeated thrice after which all six annotators reached a consensus on the annotation scheme and process. In the fourth round, we gathered additional feedback from an external researcher from the Credibility Coalition group\footnote{https://credibilitycoalition.org/}---an international organization of interdisciplinary researchers and practitioners dedicated to developing standards for news credibility and tackling the problem of online misinformation. 
The final result of the multi-stage iterative process (see Appendix, Figure \ref{q_c}) is a 7-point annotation scale comprising of values ranging from -1 to 5 (see Table \ref{tab:annotation}).  The scale measures the scientific quality of products that users are exposed to when they make vaccine-related searches on Amazon. 
 
\subsubsection{Annotation Guidelines}
In order to annotate an Amazon pro\-duct, the annotators 
were required to go through several fields present on the product's detail page in the following order: title, description, top critical and top positive reviews about the product, other metadata present on the detail page, such as editorial reviews, legal disclaimers, etc. If the product was a book, the annotators were also recommended to do the following three steps: (1) go through the first few pages in the book preview \footnote{Amazon has introduced a Look Inside feature that allows users to preview  few pages from the book.},  
(2) see other books published by the authors, (3) perform a google search on the book and go through the first few links to discover more information about the book. Annotators were asked to see contextual information about the product from multiple sources to  gain more context and perspective. This technique is grounded in lateral reading that has proven to be a good approach for credibility assessment \cite{lateralreading}.

\subsubsection{Annotation scale and heuristics:} 
Below we describe each value in our annotation scale. Table \ref{tab:annotation} presents examples.

\noindent\textbf{Debunking (-1): } Annotation value `-1' indicates that the product debunks vaccine misinformation 
or derides any vaccine-related myth or conspiracy theory or promotes the use of vaccination. As an example, consider the poster titled \textit{Immunization Poster 1979 Vintage Star Wars C-3PO R2-D2 Original} (B00TFTS194)\footnote{Every title of the Amazon product is followed by a URL id. This URL id can be converted into a url using the format: http://www.amazon.com/dp/url\_id} that encourages parents to vaccinate their children. 
Products helping users understand anti-vaccination attitude or those that describe the history about the development of vaccines or the science behind how vaccines work were also included in this category.\\

\noindent\textbf{Promoting (1): } 
This category includes all products that support or substantiate any vaccine related myth or controversy and encourages parents to raise a vaccine-free child. For example, consider the following books that promote anti-vaccination agenda. In \textit{A Summa\-ry of the Proofs that Vaccination Does Not Prevent Small-pox but Really Increases It} (B01G5QWIFM), the author talks about dangers of large scale vaccination and in \textit{Vaccine Epidemic: How Corporate Greed, Biased Science, and Coercive Government Threaten Our Human Rights, Our Health, and Our Children} (B00CWSONCE), the authors question vaccine safety and present several narratives of vaccine injuries.   We included several Amazon Fashion (B07R6PB2KP) and Amazon Home (B01HXAB7TM) merchandise in this category too since they contained anti-vaccine slogans like ``Educate before you Vaccinate'', ``Jesus wasn't vaccinated'', etc. 


We also included all products advocating any alternatives to vaccines, products that promote other health-related misinformation, dietary supplements that claim to cure diseases in their description but are not approved by Food and Drug Administration (FDA) \footnote{Note that for dietary supplements category, Amazon asks sellers not to state that the products cure, mitigate, treat, or prevent a disease in humans in their details page, unless that statement is approved by the FDA \cite{amazonsellercentral}} in this category.\\


\noindent\textbf{Neutral (-0): } We annotated all medical equipment and medicines as neutral (annotation value `0'). Note that it is beyond the scope of this project to determine the safety and veracity of the claims of each medicine sold on the Amazon platform. This means that the number of products that we have determined to be promoting (1) serve as the lower bound of the amount of misinformation present on the platform. This category also includes dietary supplements that do not violate Amazon's policy, pet-related products and health-related products not advocating a conspiratorial view.  \\

\noindent\textbf{Other annotations: } We annotated a product as `2' if the product's description and metadata were not sufficient to determine its stance. 
We assigned values `3' and `4' to all products whose URL was not accessible at the time of the annotation and whose title and description was in a language other than English, respectively.  We annotated all non-health related products (e.g. diary, carpet, electronic products, etc.) with value `5'. 

Both our audits resulted in a dataset of 4,997 Amazon products that were annotated by the first author and Amazon Mechanical Turk  workers (MTurks). The first author being the expert  annotated majority of products (3,367) to determine what would be a good task representation to obtain high quality annotations for the remaining 1,630 products from novice MTurks. We obtained three Turker ratings for each remaining product and used the majority response to assign the annotation value. Our task design worked. For 97.9\% of the products, annotation values converged. Only 34 products had diverging responses. The first author then annotated these 34 products to obtain the final set of annotation values. We describe the AMT job in detail in Appendix \ref{amt_job}.




\begin{table}[b]
\centering
{ \scriptsize
\begin{tabular}{lllll}
\hline
\multicolumn{1}{|l|}{\textbf{\begin{tabular}[c]{@{}l@{}}Rank\\r\end{tabular}}} &
\multicolumn{1}{|l|}{\textbf{Product}} &
\multicolumn{1}{l|}{\textbf{\begin{tabular}[c]{@{}l@{}}Bias of each\\ product\end{tabular}}} & 
\multicolumn{1}{l|}{\textbf{\begin{tabular}[c]{@{}l@{}}Bias till \\ rank r\end{tabular}}} & \multicolumn{1}{l|}{\textbf{Bias value}}             \\ \hline
\multicolumn{1}{|l|}{1}       & \multicolumn{1}{l|}{$p_1$}  
& \multicolumn{1}{l|}{$s_1$}                                                                    & \multicolumn{1}{l|}{B(1)}                      & \multicolumn{1}{l|}{$s_1$}                            \\ \hline
\multicolumn{1}{|l|}{2}   & \multicolumn{1}{l|}{$p_2$}
& \multicolumn{1}{l|}{$s_2$}                                                                    & \multicolumn{1}{l|}{B(2)}                      & \multicolumn{1}{l|}{$\frac{1}{2}$ ($s_1$ + $s_2$)}       \\ \hline
\multicolumn{1}{|l|}{3}    & \multicolumn{1}{l|}{$p_3$}
& \multicolumn{1}{l|}{$s_3$}                                                                    & \multicolumn{1}{l|}{B(3)}                      & \multicolumn{1}{l|}{$\frac{1}{3}$($s_1$ + $s_2$ + $s_3$)} \\ \hline
\multicolumn{4}{|l|}{Input Bias (ib)}                                                                                                                                                 & \multicolumn{1}{l|}{$\frac{1}{3}$($s_1$ + $s_2$ + $s_3$)}              \\ \hline
\multicolumn{4}{|l|}{Output Bias (ob)}                                                                                                                                                & \multicolumn{1}{l|}{$\frac{1}{3}$ [$s_1$(1 + $\frac{1}{2}$ + $\frac{1}{3}$) +  $s_2$($\frac{1}{2}$  + $\frac{1}{3}$) +  $s_3$($\frac{1}{3}$)]       }               \\ \hline
\multicolumn{4}{|l|}{Rank Bias (rb)}                                                                                                                                                  & \multicolumn{1}{l|}{ob-ib}                           \\ \hline
\end{tabular}}
\caption{Example illustrating the bias calculations. For a given query, Amazon's search engine presents users with the following products in the search results $p_1$, $p_2$ and $p_3$. The misinformation bias scores of the products are $s_1$, $s_2$ and $s_3$ respectively. The table has been adopted from previous work \cite{kulshrestha_bias}. A bias score larger than 0 indicates a lean towards  misinformation.}
\label{bias_cal}
\vspace{-0.9cm}
\end{table}

\begin{figure*}[t]
  \centering
  \hspace{-0.5cm}
  \begin{subfigure}{0.5\textwidth}
      \centering
      \includegraphics[scale=0.35]{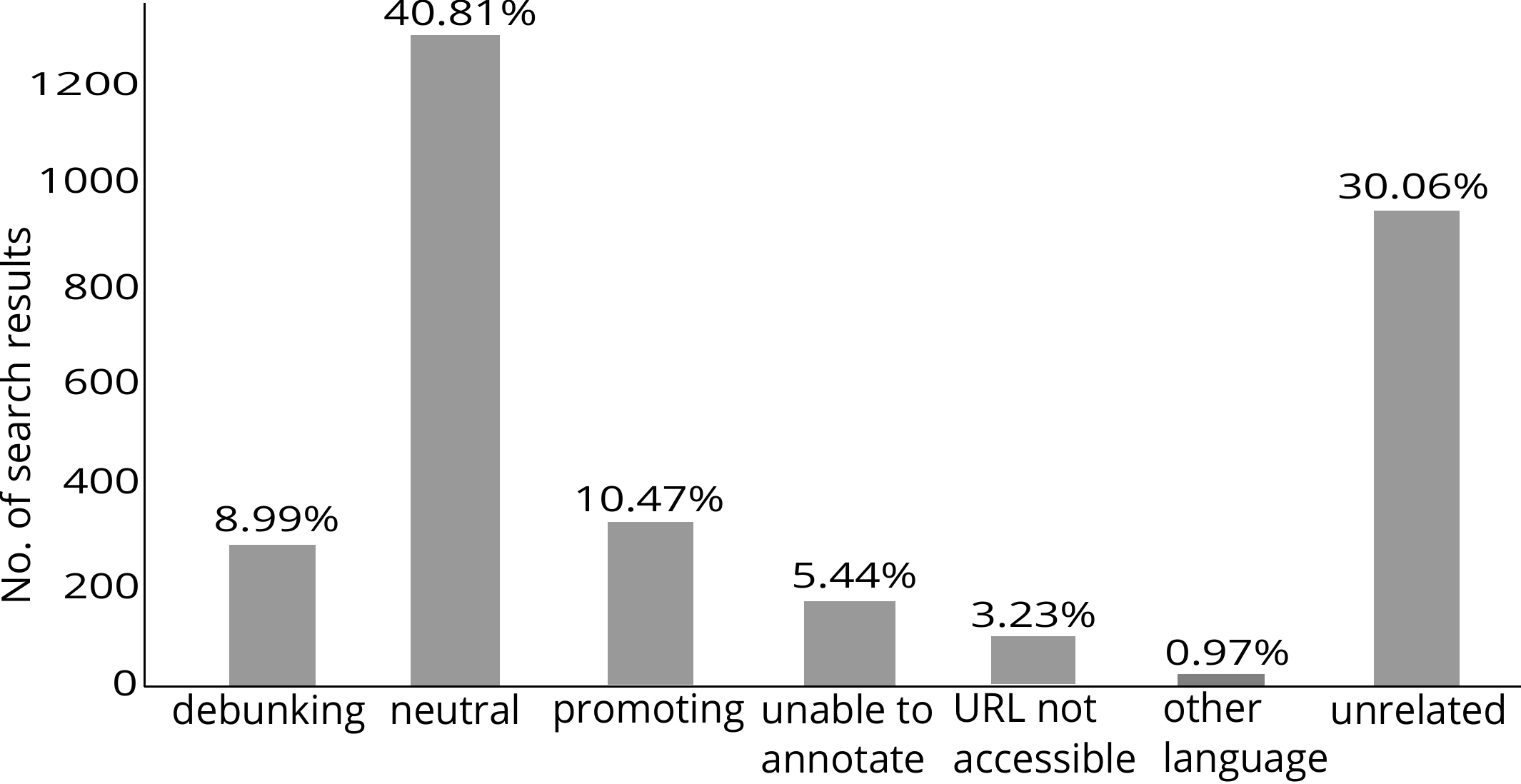}
      \caption{Search results}
      \label{searchcount}
      \Description[Bar graph showing the percentage of search results belonging to different annotation values ]{ debunking (8.99\%), neutral (40.81\%), promoting (10.47\%), unable to annotate (5.44\%), URL not accessible (3.23\%), other language (0.97\%) and unrelated (30.06\%).}
  \end{subfigure}
  \begin{subfigure}{0.5\textwidth}
      \centering
      \includegraphics[scale=0.35]{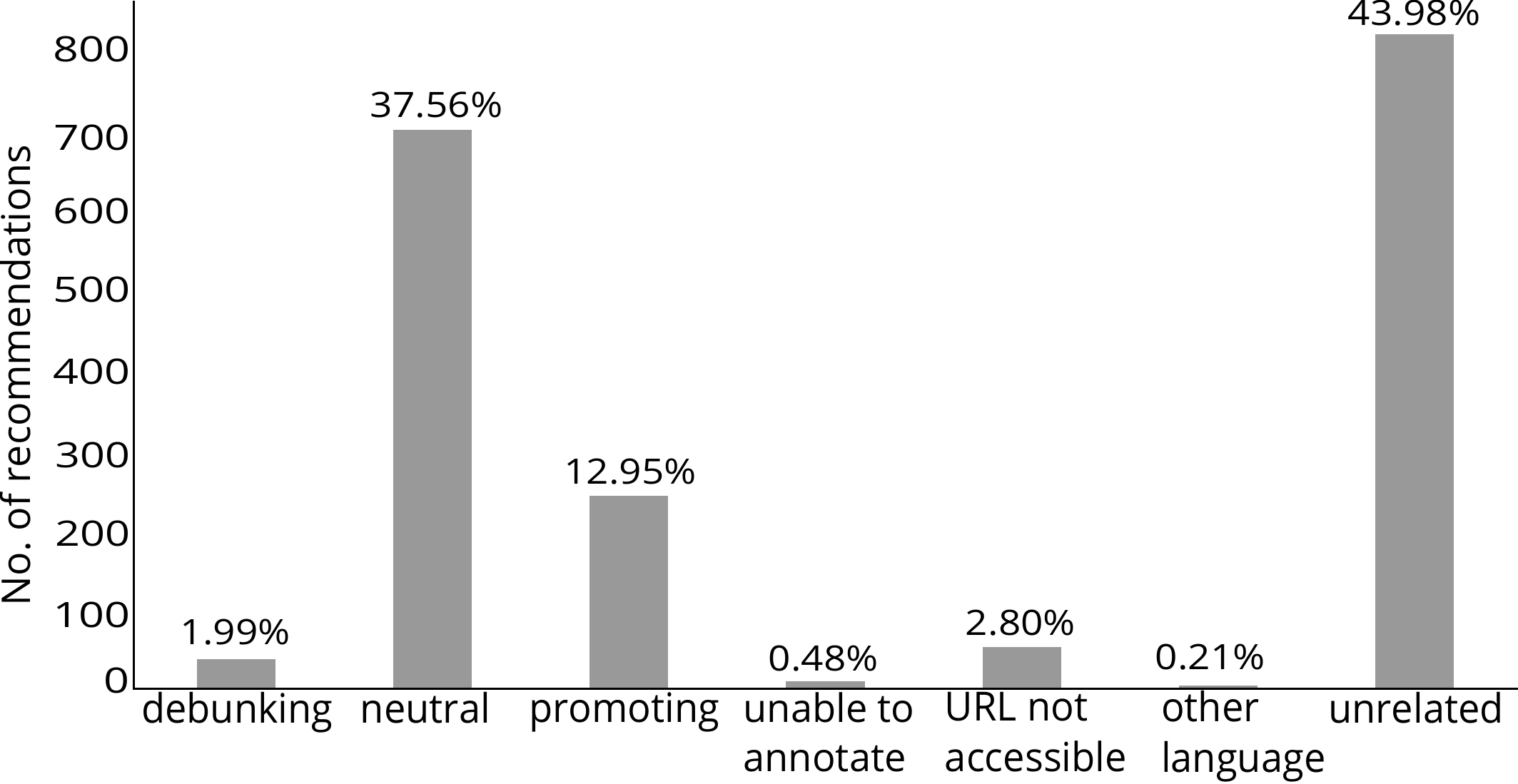}
      \caption{Recommendations}
      \label{reco_count}
      \Description[Bar graph showing the percentage of recommendations belonging to different annotation values ]{debunking (1.99\%), neutral (37.56\%), promoting (12.95\%), unable to annotate (0.48\%), URL not accessible (2.80\%), other language (0.21\%) and unrelated (43.98\%).}
  \end{subfigure}

  \caption{\textbf{RQ1a:} (a) Number (percentage) of search results belonging to each annotation value. While majority of products have a neutral stance (40.81\%), products promoting health misinformation (10.47\%) are greater than products debunking health misinformation (8.99\%). 
  (b)  Number (percentage) of recommendations belonging to each annotation value. A high percentage of product recommendations promote misinformation (12.95\%) while percentage of recommendations debunking health misinformation is very low (1.99\%).}
  \label{basic_un}
\end{figure*}

\begin{figure*}[!t]
  \centering

      \includegraphics[scale=0.5]{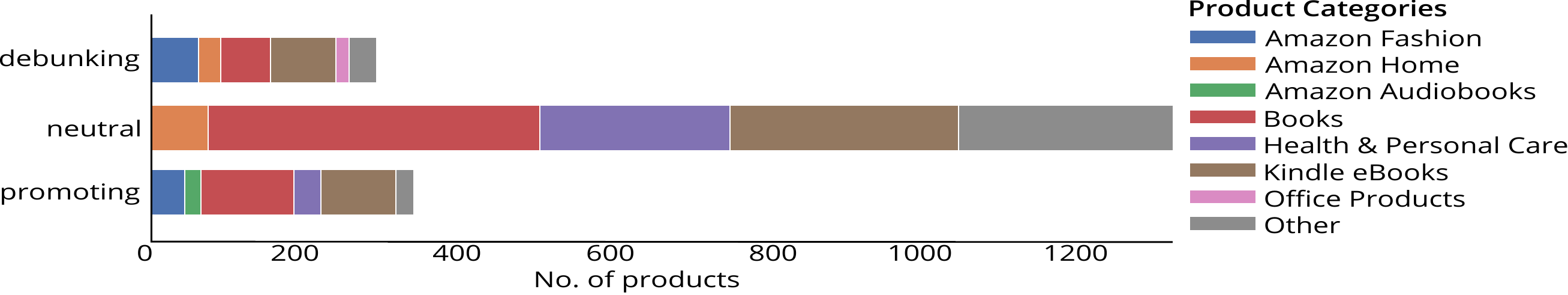}

  \caption{\textbf{RQ1a:} Figure showing categories of promoting, neutral and debunking Amazon products (search results). All categories occurring less than 5\% were combined and are presented as \textit{other} category. Note that misinformation exists in various forms on Amazon. Products promoting health misinformation include books (Books, Kindle eBooks, Audible Audiobooks), apparel (Amazon Fashion) and dietary supplements (Health \& Personal Care). Additionally, proportion of books promoting health misinformation is much greater than proportion of books debunking misinformation.}
  \label{cat}
  \Description[Categories of debunking, neutral and promoting Amazon products]{ Debunking products mostly belong to categories, Kindle eBooks, Books, Amazon fashion and Amazon home. Neutral products mostly belong to categories Books, Kindle eBooks, Health \& Personal care and Amazon home. Promoting products belong to categories Books, Kindle eBooks, Health \& Personal care and Amazon fashion.}
\end{figure*}

\subsection{Quantifying misinformation bias in SERPs:}
\label{misinfo_bias_metric} 
In this section, we describe our method to determine the amount of misinformation present in search results.
How do we estimate the misinformation bias present in Amazon's SERPs? First, we used our annotation scheme to assign misinformation bias scores ($s_i$)  to individual products present in SERPs. 
We converted our 7 point (-1 to 5)  scale to misinformation bias scores with values -1, 0 and 1. We mapped annotation values 2, 3, 4, and 5 to bias score 0. 
Merging ``unknown'' annotations to neutral will result in a conservative estimate of misinformation bias present in the search results.  Now, a product can be assigned one of the three bias scores: -1 suggests that product debunks misinformation, 0 indicates a neutral stance and 1 implies that the product promotes misinformation. 
Next, to quantify misinformation bias in Amazon's SERPs, we adopt the framework and metrics proposed in prior work to quantify partisan bias in Twitter search results  \cite{kulshrestha_bias}. 
Below we discuss three kinds of bias proposed by the framework and delineate how we estimate each bias with respect to misinformation. Table \ref{bias_cal} illustrates how we calculated the bias values.

\begin{enumerate}[label=(\roman*)]

\item The \textit{input bias} (\textit{ib})  of a list of Amazon products is the mean of misinformation bias scores of the constituting products  \cite{kulshrestha_bias}. Therefore, \textit{ib} = ${\sum_{i=1}^{n} {s_i}}$, where \textit{n} is the length of the  list \& ${s_i}$ is the misinformation bias score of \textit{ith} product in the list. Input bias is an unweighted bias, i.e it is not affected by the rank/ordering of the items.  

\item The \textit{output bias} (\textit{ob}) of a ranked list is the overall bias present in the SERPs and is the sum of biases introduced due to input and ranks of the input. 
We first calculate weighted bias score B(r) of every rank r, which is the average misinformation bias of products ranked from 1 to r. Thus, B(r) = $\frac{\sum_{i=1}^{r} {s_i}}{r}$, where ${s_i}$ is the misinformation bias score of \textit{ith} product. Output bias (ob) is the average of weighted bias score B(r) for all ranks. Thus, by definition ob = $\frac{\sum_{i=1}^{r} {B(i)}}{r}$.  

\item The \textit{ranking bias} (\textit{rb}) is  introduced by the ranking algorithm of search engine \cite{kulshrestha_bias}. It is calculated by subtracting input bias from output bias. Thus, \textit{rb} = \textit{ob}-\textit{ib}. In our case, high ranking bias indicates that search algorithm ranks misinformative products higher than neutral or debunking products.
\end{enumerate}

Why do we need three bias scores? Amazon's search algorithm is not only selecting the products to be shown in the search results but it is also ranking them according to their internal algorithm. Therefore, the overall bias (ob) could be introduced either at the product selection stage (ib), or ranking stage (rb) or both. Studying all three biases gives us an elaborate understanding of how biases are introduced by the search algorithm. All three bias values (\textit{ib}, \textit{ob} and \textit{rb}) lie between -1 and 1. A bias score larger than 0 indicates a lean towards  misinformation. Conversely, a bias score less than 0 indicates a propensity towards debunking information. We only consider top 10 search results in each SERP. Thus, in the bias calculations, rank always varies from 1 to 10.

\section{RQ1 Results [Unpersonalized audit]: Quantify misinformation bias}
The aim of the \textit{Unpersonalized audit} is to determine the amount of misinformation bias in search results. Below we present the input, rank, and output bias detected by our audit in search results of all 10 vaccine-related topics with respect to 5 search filters. 

\begin{figure*}
  \centering
      \centering
      \includegraphics[scale=0.28]{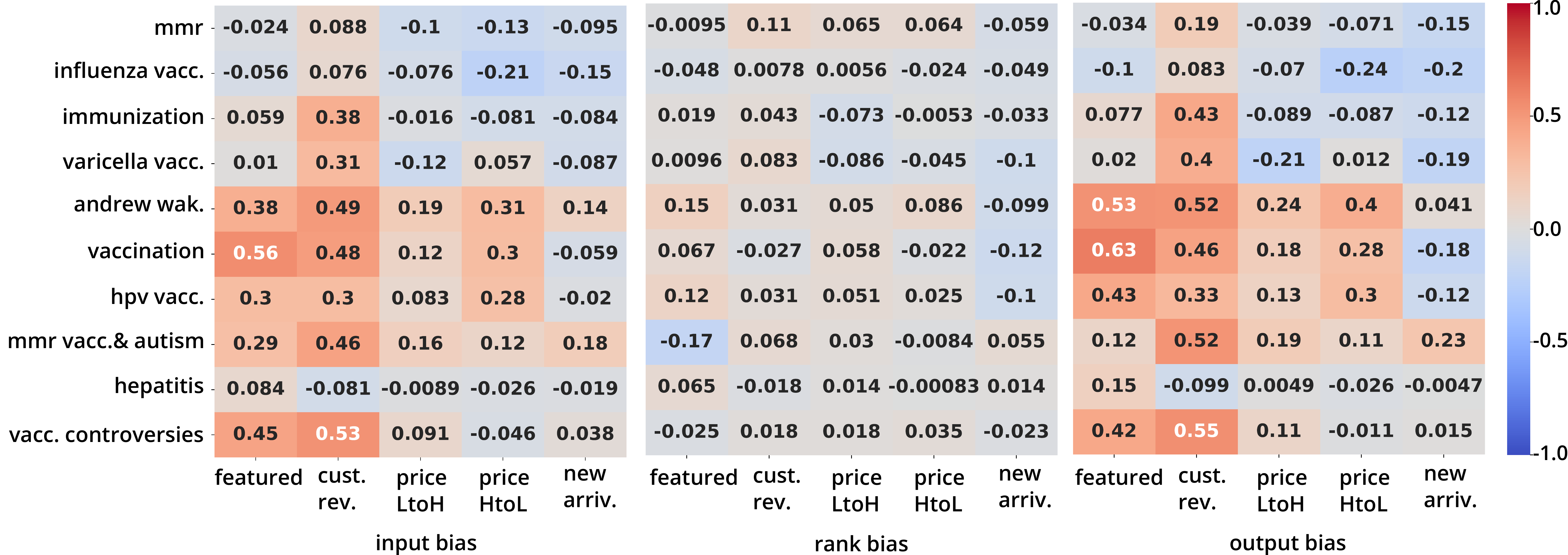}
  \caption{\textbf{RQ1a:} Input, rank and output bias for all 10 vaccine-related topics across five search filters. The bias scores are average of scores obtained for each of the 15 days. Input and rank bias is positive (>0) in the search results of majority of topics for filters ``featured'' and ``average customer review''. A bias value greater than 0 indicates a lean towards misinformation. Topics ``andrew wakefield'' and ``mmr vaccine \& autism'' have a positive input bias across all five filters indicating that search results of these topics contain large number of products promoting health misinformation irrespective of the filter used to sort the search results.  Topic ``vaccination'' has the highest overall bias (output bias) of 0.63 followed by topic ``andrew wakefield'' that has output bias of 0.53 for filter ``featured''.}
  \label{heatmaps}
  \Description[Input, rank and output bias values]{All topics except hepatitis have positive input and output bias for filter customer reviews. Additionally, all topics except mmr and influenza vaccine have positive input and output bias values for filter featured. Furthermore, all topics except andrew wakefield, mmr vaccine and autism, and vaccine controversies have a negative input and output bias for filter newest arrivals. On the other hand, ranking bias of all topics except vaccination and hepatitis for filter customer review and immunization and varicella vaccine for filter price low to high is positive.}
\end{figure*}

\subsection{RQ1a: Search results}\label{RQ1asection}
We collected 36,000  search results  from our \textit{Unpersonalized audit} run, out of which 3,180 were unique. Recall, we collected these products by searching for 48 search queries belonging to vaccine-related topics and sorting results by each of the 5 Amazon filters. We later extracted and annotated top 10 search results from all the collected SERPs resulting in 3,180 annotations.
Figure \ref{searchcount} shows the number (and percentage) of products corresponding to each annotation value. Through our audits, we find a high percentage (10.47\%) of misinformative products in the search results. Moreover, the number of misinformative products outnumbered the debunking products. Figure \ref{cat} illustrates the distribution of categories of Amazon products annotated as debunking (-1), neutral (0) and promoting (1). Note that the products promoting health misinformation primarily belong to categories Books (35.43\%), Kindle eBooks (28.52\%), Amazon Fashion (12.61\%)---a category that includes t-shirts, apparel, etc. and Health \& Personal Care (10.21\%)---a category consisting of dietary supplements.
Below we discuss the misinformation bias observed across all the vaccine-related topics, the Amazon search filters and  search queries. 

\subsubsection{Misinformation bias in vaccine related  topics}

We calculate the input, rank and output bias for each  of the 10 search topics. All the bias scores presented are average of scores obtained across the 15 days of audit. The bias score for a topic is also the average across each of the constituting search queries.  Figure \ref{heatmaps} shows the bias scores for all the topics, search filters and bias combinations.\\ 

\noindent\textbf{Input bias: }We observe a high input bias (>0) for all topics except ``hepatitis'' for ``average customer review'' filter indicating presence of a large number of misinformative products in the SERPs  when search results are sorted by this filter. 
Similarly, input biases for most topics is also positive for ``featured'' filter. Note, ``featured'' is the default Amazon filter. Thus, by default Amazon is presenting more misinformative search results to users searching for vaccine related queries. Topics ``andrew wakefield'', ``vaccination'' and ``vaccine controversies'' have highest input biases for the both ``featured'' and ``average customer review'' filters.  Another noteworthy trend is the negative input bias for 7 out of 10 topics with respect to filter ``newest arrivals'' indicating that there are more debunking products present in the SERP when users look for newly appearing products on Amazon. ``Andrew wakefield'' and ``mmr vaccine \& autism'' are the only two topics that have positive input bias (>0) across all the five filters. Interestingly, there is no topic that has negative input bias across all filters. Recall, a negative (<0) bias indicates a debunking lean. Topics ``mmr'', ``influenza vaccine'' and ``hepatitis'' have negative bias scores in four out of five filters.\\
\begin{figure}[b]
  \centering
      \centering
      \includegraphics[scale=0.31]{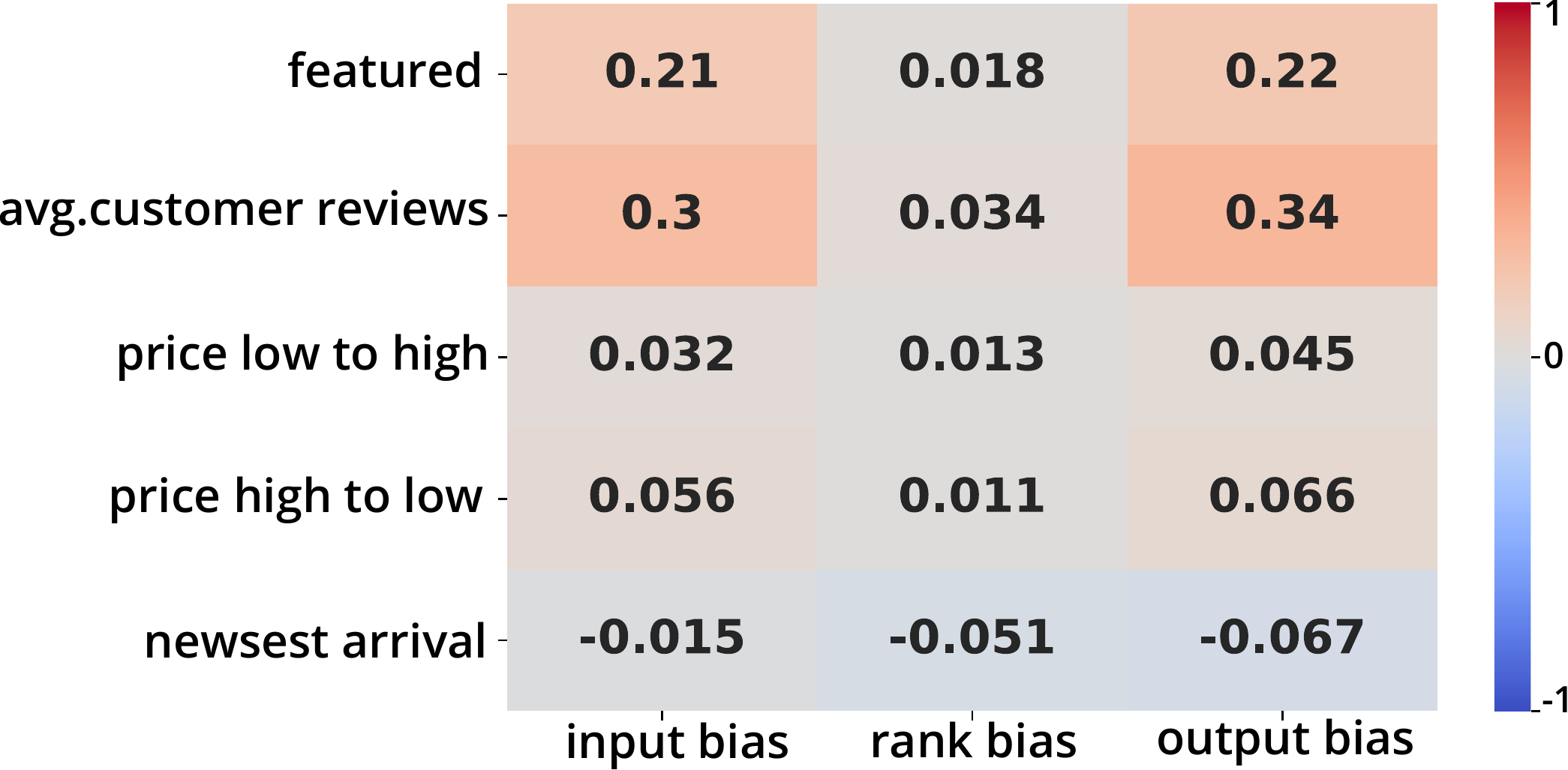}
\caption{Input, rank and output bias for all filter types.}
\label{filters}
\Description[Bias values for all filter types]{Input bias values: featured (0.21), avg customer reviews (0.3), price low to high (0.032), price high to low (0.056), newest arrival (-0.015). Rank bias: featured (0.018), avg customer reviews (0.034), price low to high (0.013), price high to low (0.011), newest arrival (-0.051). Output bias: featured (0.22), avg customer reviews (0.34), price low to high (0.045), price high to low (0.066), newest arrival (-0.067).}
\end{figure}

\begin{figure*}
  \centering
  \begin{subfigure}{\textwidth}
      \centering
      \includegraphics[scale=0.35]{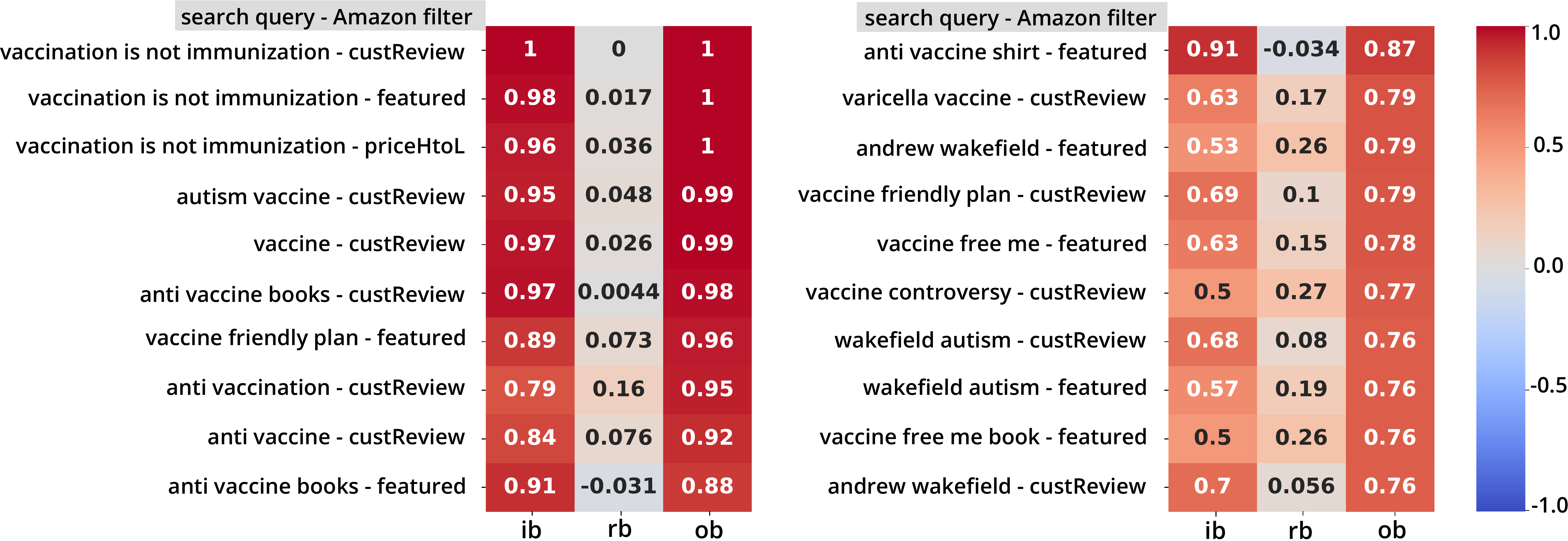}
  \end{subfigure}
  \caption{Top 20 search query-filter combinations when sorted by output bias (ob). In other words, these query-filter combinations are the most problematic ones containing highest amount of misinformation (highest ob).}
  \label{query-filter}
  \Description[Search query - Amazon filter combinations containing highest amount of misinformation]{Top five combinations: vaccination is not immunization - custReview (ob = 1), vaccination is not immunization - featured (ob = 1), vaccination is not immunization - priceHtoL (ob = 1), autism vaccine - custReview (ob = 0.99), vaccine - custReview (0.99)  }
\end{figure*}

\noindent\textbf{Rank bias: } 8 out of 10 topics have positive (>0) rank bias for  filters ``price low to high'' and ``average customer reviews'' and 6 out of 10 topics have positive rank bias for filter ``featured''. These results  suggest that Amazon's ranking algorithm favors misinformative products and ranks them higher when customers filter their search results by the aforementioned filters. Some topics have negative input bias but positive rank bias. Consider topic ``mmr'' with respect to filter ``price low to high'' whose input bias is -0.1 but the rank bias is 0.065. This observation suggests that although the SERPs obtained had more debunking products, a few misinformative products were still ranked higher. Rank bias for 8 out of 10 topics with respect to filter ``newest arrivals'' was negative, similar to what we observed for input bias. 
\\

\noindent\textbf{Output bias:} Output bias is positive (>0) for most topics with respect to filters ``featured'' and ``average customer reviews''. Recall, a bias value greater than 0 indicates a lean towards misinformation. 
Topic ``vaccination'' has the highest output bias (0.63) for filter ``featured''. On the other hand, topic ``influenza vaccine'' has least output bias (-0.24) for filter ``price high to low''. 
\subsubsection{Misinformation bias in search  filters}
Figure \ref{filters} shows the results for all 5 filters. Bias scores are averaged across all search queries. All filters except ``newest arrivals'' have positive input, rank, and output misinformation bias. Filter ``average customer review'' has the highest positive output bias indicating that misinformative products  belonging to vaccine related topics receive higher ratings. We present the implications of these results in our discussion (Section \ref{discussion}).
\subsubsection{Misinformation bias in search queries}
Figure \ref{query-filter} shows the top 20 search queries and filter combinations with highest output bias. Predictably, filter ``newest arrivals'' does not appear in any instance. Surprisingly, 9 search query-filter combinations have very high output biases (ob > 0.9). Search query ``vaccination is not immunization'' has output bias of 1 for three filter types. Most of the search queries in Figure  \ref{query-filter} have a negative connotation, i.e the queries themselves have a bias (e.g search queries anti vaccine books, vaccination is not immunization indicates an intent to search for misinformation). This observation reveals that if you search for anti vaccine stuff, you will get high amount of vaccine and health misinformation. This indicates how Information Retrieval systems currently work; they curate by relevance with no notion of veracity. The most troublesome observation is the presence of high output bias for generic and neutral search queries, ``vaccine'' (ob = 0.99) and ``varicella vaccine'' (ob = 0.79). These results indicate that, unlike companies like Pinterest, who have altered their search engines in response to vaccine related queries \cite{pinterest}, Amazon has not made any modification to its search algorithm to push less anti vaccine products to users.

\begin{figure*}
  \centering
 \begin{subfigure}{0.45\textwidth}
      \centering
      \includegraphics[width=6cm,height=4.6cm]{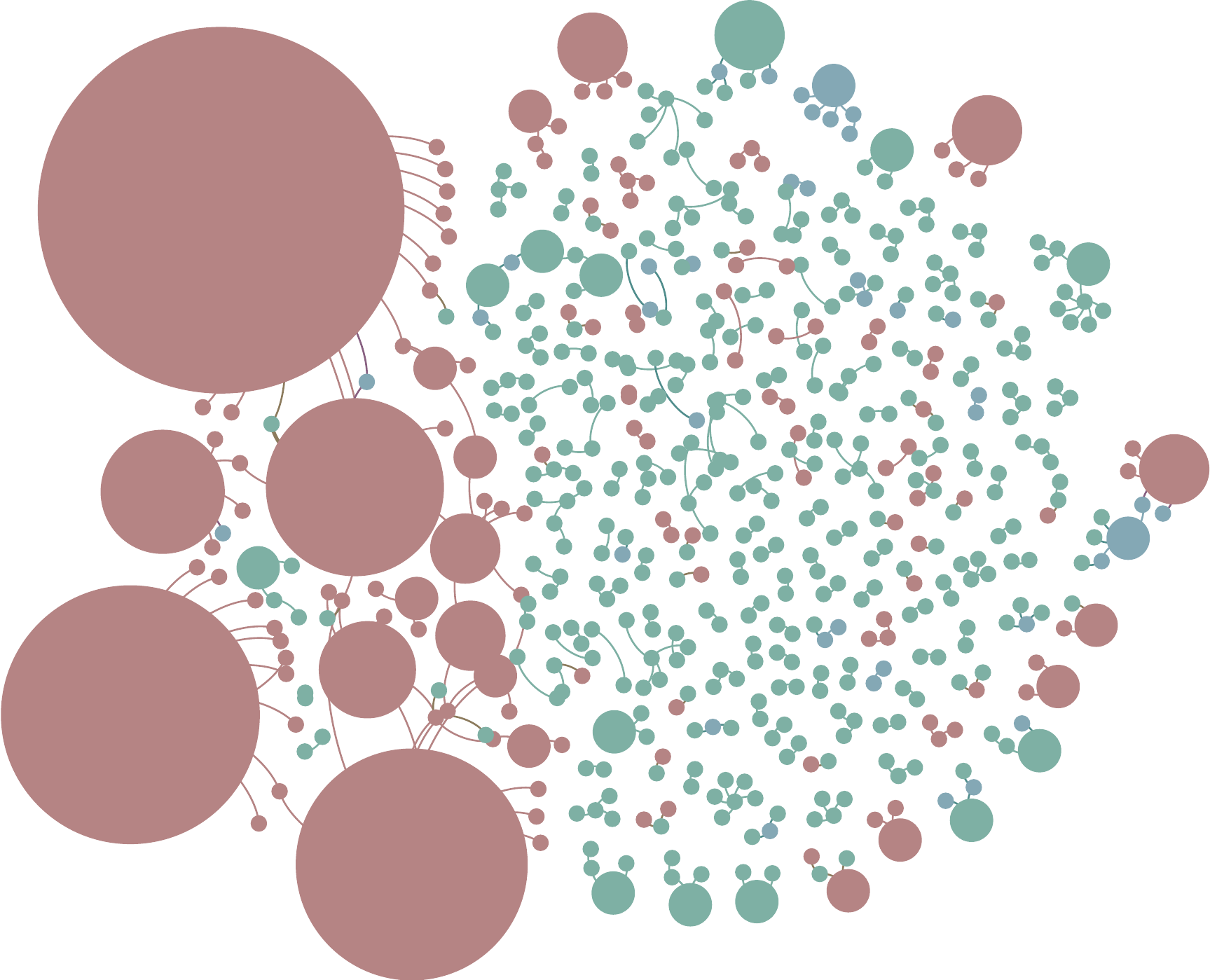}
      \caption{Customers who bought this item also bought (CBB)}
      \label{buy_buy}
      \Description[CBB]{There are several instances of red nodes connected to each other and green nodes connected to each other. Few of the green nodes are attached to red nodes too.}
  \end{subfigure}
   \begin{subfigure}{0.45\textwidth}
      \centering
      \includegraphics[width=7cm,height=4.6cm]{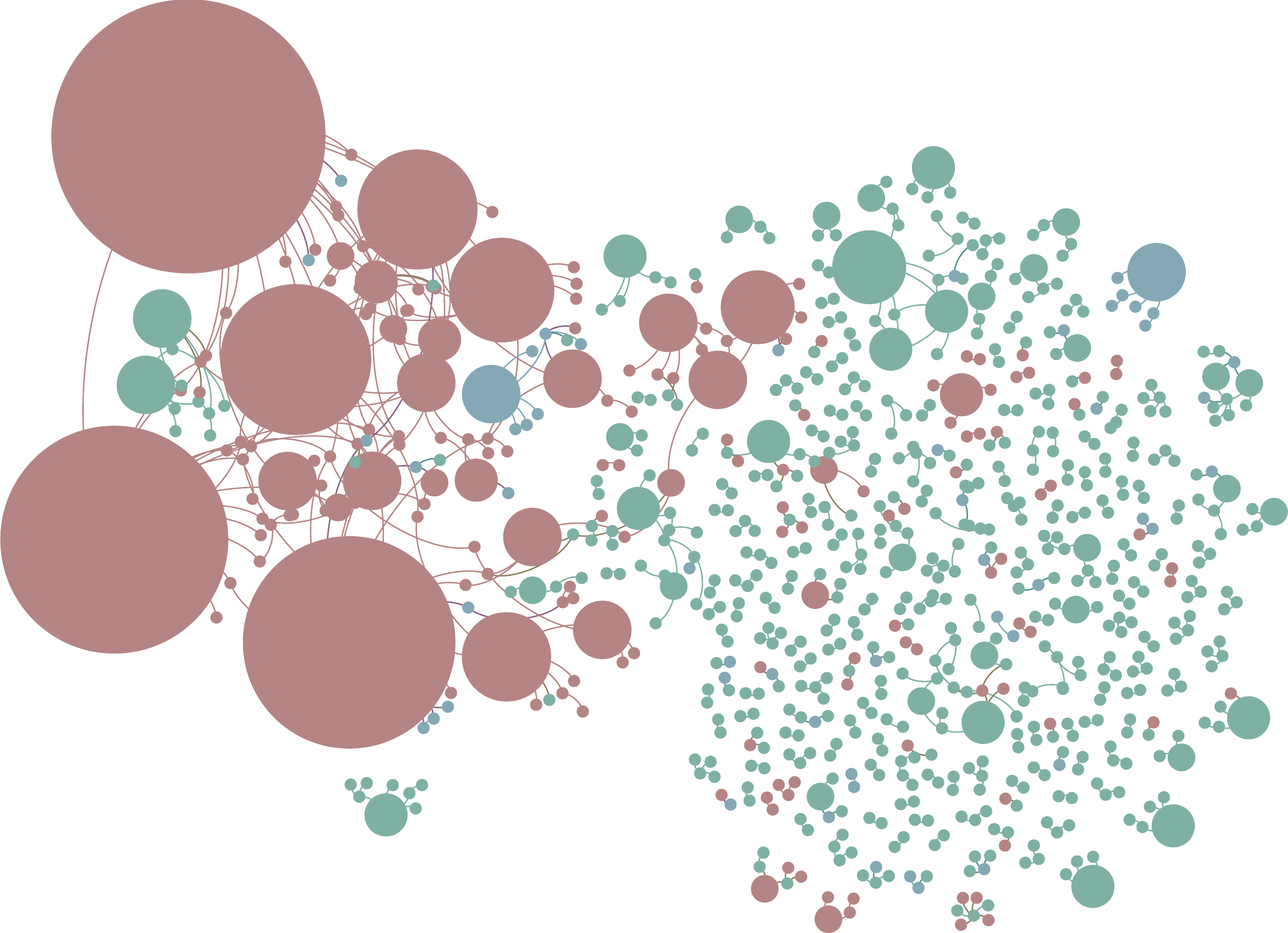}
      \caption{Customers who viewed this item also viewed (CVV)}
      \label{view_view}
      \Description[CVV]{There are several instances of red nodes connected to each other and green nodes connected to each other. Few of the green nodes are attached to red nodes too.}
  \end{subfigure}\\ 

   \begin{subfigure}{0.45\textwidth}
      \centering
      \includegraphics[width=6cm,height=4.6cm]{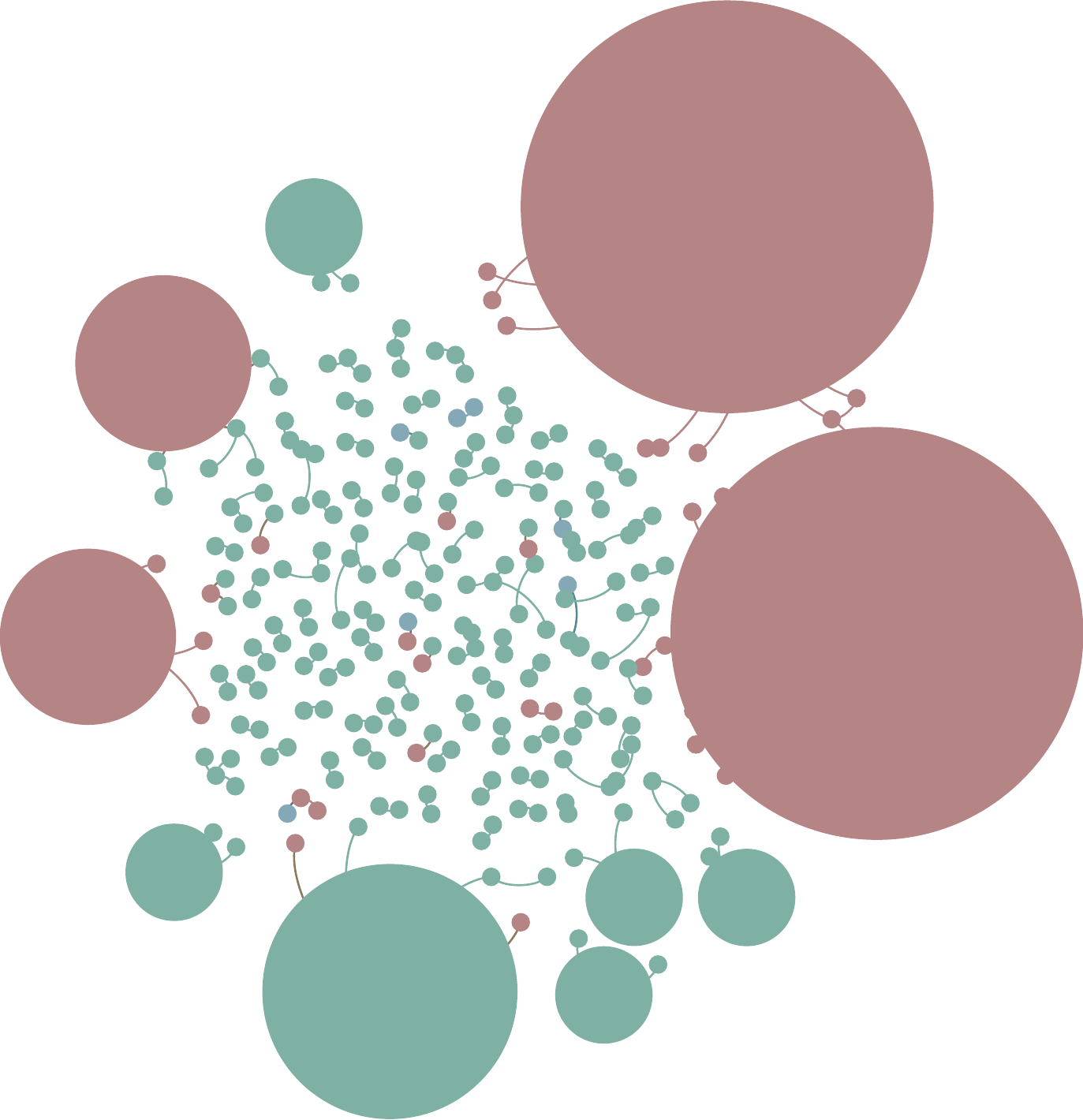}
      \caption{Frequently bought together (FBT)}
      \label{fig:search_results_freq}
      \Description[FBT]{There are large sized red nodes. Red nodes are attached to other red nodes and several green nodes are also attached together.}
  \end{subfigure}
   \begin{subfigure}{0.45\textwidth}
      \centering
      \includegraphics[width=7cm,height=4.6cm]{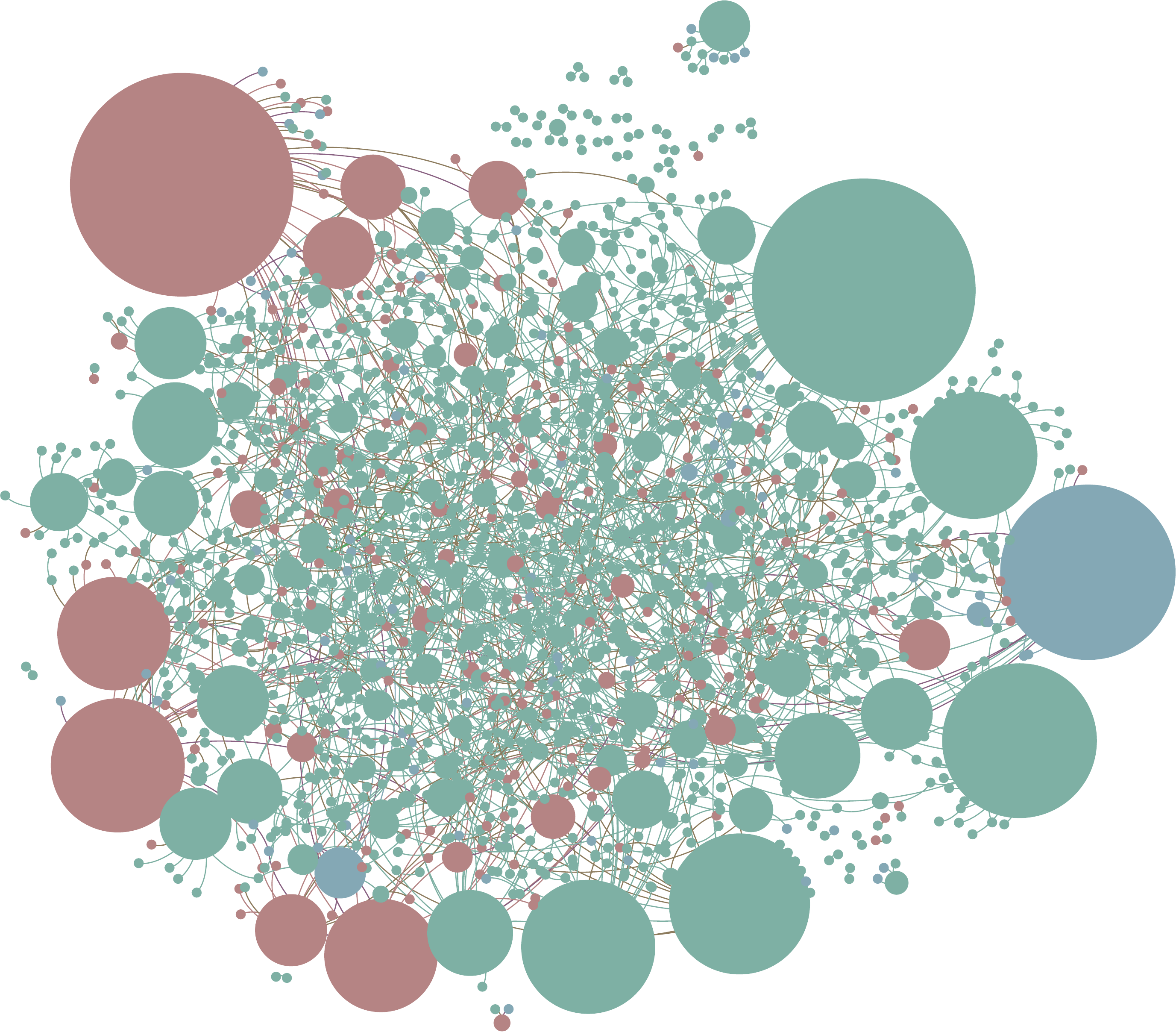}
      \caption{Sponsored products related to this item}
      \label{Sponsored}
      \Description[Sponsored]{The graph contains many large green nodes attached to other green nodes. Few large red nodes are also present that are attached to other red and green nodes.}
  \end{subfigure}\\

  
   \begin{subfigure}{1\textwidth}
      \centering
      \includegraphics[width=12cm,height=4.6cm]{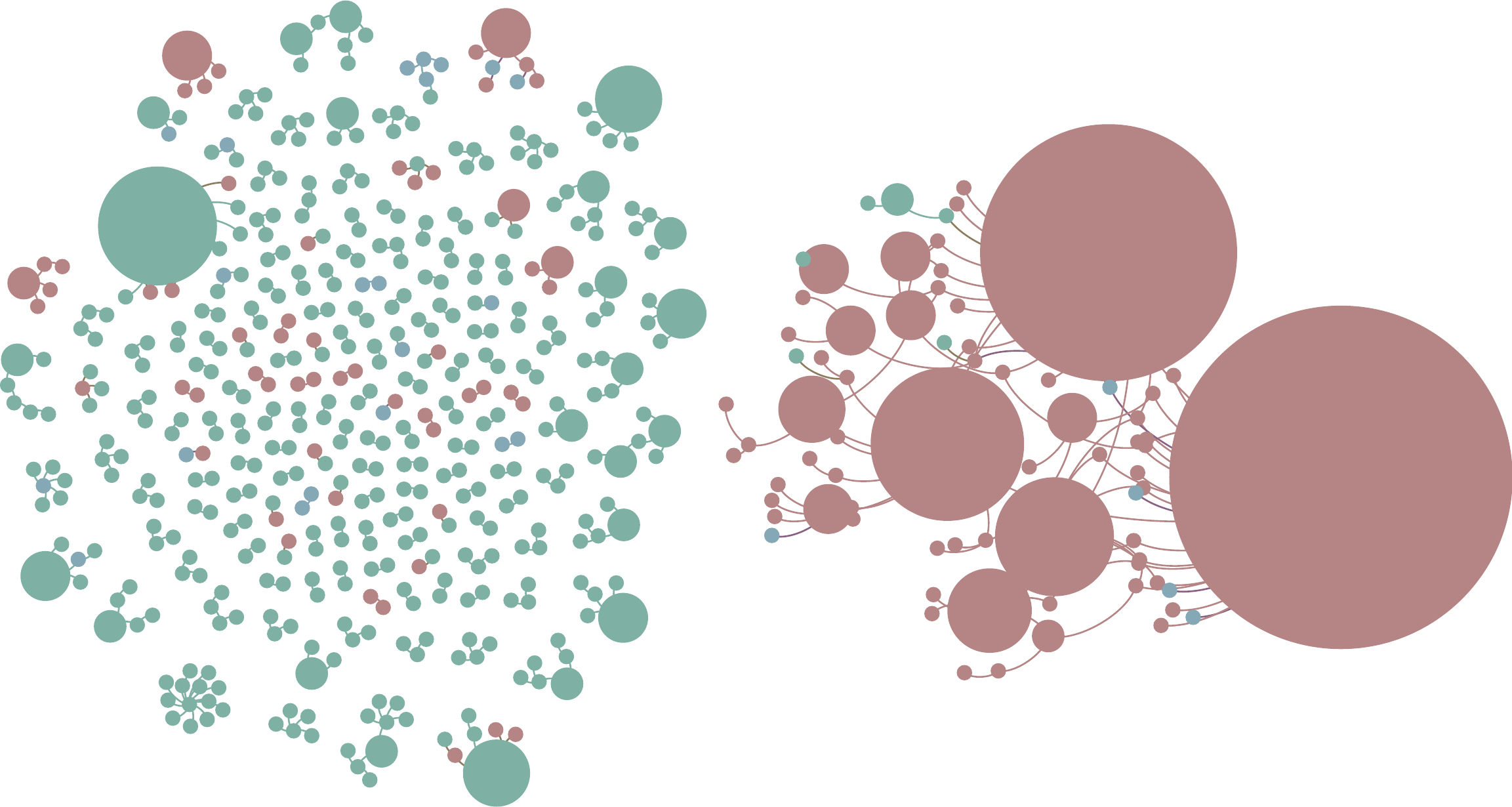}
      \caption{What other items customers buy after viewing this item (CBV). Note that the recommendation graph for CBV recommendation type is indeed one figure. It consists of two disconnected components, indicating strong filter bubble effect.}
      \label{buy_view}
      \Description[CBV]{The graph has two disconnected components. One component mostly consists of red nodes attached to one another. Other component comprises of green nodes and a few red nodes.}
  \end{subfigure}

  \caption{Recommendation graphs for 5 different types of recommendations collected from the  product pages of top three search-results obtained in response to 48 search queries, sorted by 5 filters over a duration of 15 days during \textit{Unpersonalized audit} run. \fcolorbox{labelred}{labelred}{\rule{0pt}{4pt}\rule{4pt}{0pt}} denotes products annotated as misinformative, \fcolorbox{labelgreen}{labelgreen}{\rule{0pt}{4pt}\rule{4pt}{0pt}} as neutral and \fcolorbox{labelblue}{labelblue}{\rule{0pt}{4pt}\rule{4pt}{0pt}} as debunking.  Node size is proportional to the times the product was recommended in that recommendation type. Large sized red nodes coupled with several interconnections between red nodes indicate a strong filter-bubble effect where recommendations of misinformative products returned more misinformation.}
  \label{un_reco_graphs}
\end{figure*}

\subsection{RQ1b: Product page recommendations}

We extracted the product page recommendations of top 3 search results present in the SERPs. The product page constitutes of various types of recommendations. For analysis, we considered the first prod\-uct present in 5 types of recommendations  ``Customers who bought this item also bought'' (CBB),  ``Customers who viewed this item also viewed'' (CVV), ``Frequently bought together'' (FBT), ``Sponsored products related to this item'' and ``What other items customers buy after viewing this item'' (CBV). The process resulted in 16,815 recommendations out of which 1,853 were unique. Figure \ref{reco_count} shows the number and percentage of recommendations belonging to different annotation values. The percentage of misinformative reco\-mmendations (12.95\%) is much higher than the debunking reco\-mmendations (1.95\%). 
The total input bias in all 16,815 reco\-mm\-endations is 0.417 while in all 1,853 unique recommendations is 0.109, indicating a  lean towards misinformation. 

Does filter-bubble effect occur in product page recommendations? 
To answer, we compared the misinformation bias scores of all types of recommendations considered together (refer Table \ref{RQ1b}). Kruskal Wallis Anova test revealed the difference to be significant (KW H(2, N=16815) = 6,927.6, p=0.0). Post-hoc Tukey HSD test showed that the product page recommendations of misinformative products contain more misinformation when compared to recommendations of neutral and debunking products. Even more concerning is that the recommendations of debunking products have more misinforma\-tion than neutral products. 
To investigate further, we qualitatively studied the recommendation graphs of each of the five recommenda\-tion types (Figure \ref{un_reco_graphs}). Each node in the graph represents an Amazon product.  An edge A$\rightarrow$ B indicates that B was recommended in the product page of A. Node size is proportional to the number of times the product was recommended.

\begin{table*}
\resizebox{\textwidth}{!}{%
\begin{tabular}{l|l|l|l|l|l}
\hline
\textbf{Type of product page recommendations}                                                       & \textbf{Kruskal Wallis Anova Test}   & \textbf{Post hoc Tukey HSD}                              & \textbf{d} & \textbf{n} & \textbf{m} \\ \hline
All                                                                                                 & KW H(2, N=16815) = 6,927.6, p=0.0    & M\textgreater{}D \& M\textgreater{}N \& D\textgreater{}N & 37         & 1576       & 240        \\ \hline
Cust. who bought this item also bought (CBB)                                                        & KW H(2, N=3133) = 2136.03, p=0.0     & M \textgreater D \& M\textgreater{}N \& N\textgreater{}D & 11         & 225        & 66         \\ \hline
Cust. who viewed this item also viewed (CVV)                                                        & KW H(2, N=4485) = 2673.95, p=0.0 & M\textgreater{}D \& M\textgreater{}N \& D\textgreater{}N & 18         & 331        & 100        \\ \hline
Frequently bought together (FBT)                                                                    & KW H(2, N=388) = 277.08, p=6.8e-61     & M\textgreater{}D \& M\textgreater{}N \& D\textgreater{}N & 1          & 111        & 16         \\ \hline
Sponsored products related to this item                                                             & KW H(2, N=6575) = 628.52, p=3.2e-137   & M\textgreater{}D \& M\textgreater{}N   \& D\textgreater{}N                  & 7          & 953        & 98         \\ \hline
\begin{tabular}[c]{@{}l@{}}What other items cust. buy after viewing\\  this item (CBV)\end{tabular} & KW H(2, N=2234) = 1611.34, p=0.0     & M\textgreater{}D \& M\textgreater{}N \& D\textgreater{}N & 9          & 230        & 57         \\ \hline
\end{tabular}%
}
\caption{\textbf{RQ1b:} Analyzing echo chamber effect in product page recommendations. M, N and D are the means of misinformation bias scores of products recommended in the product pages of misinformative, neutral and debunking Amazon products respectively. Higher means indicate that recommendations contain more misinformative products. For example, M\textgreater D indicates that recommendations of misinformative products  have more misinformation than recommendations of debunking products. d, n and m are number of unique products annotated as debunking, neutral and promoting for each recommendation type.}
\label{RQ1b}
\end{table*}

\subsubsection{Recommendation type- Customers who bought this item also bought (CBB)}
Misinformation bias scores of CBB are significantly different for debunking, neutral, and promoting products (KW H(2, N=3133) = 2136.03, p=0.0). Post hoc tests reveal that CBB recommendations of misinformative products have more misinfor\-mation when compared to CBB recommendations of neutral and debunking products. Additionally CBB recommendations of neutral products have more misinformation than CBB recommendations of debunking products. The findings are evident from Figure \ref{buy_buy} too. For example, there are several instances of red nodes connected to each other. In other words, if you click on a misinformative search result, you will get misinformative products in CBB recommenda\-tions. Few of the green nodes are attached to red ones indicating that CBB recommendation of a neutral product sometimes contain a misinformative product. The most recommended product present in CBB is a misinformative Kindle book titled \textit{Miller's Review of Critical Vaccine Studies: 400 Important Scientific Papers Summarized for Parents and Researchers} (B07NQW27VD).

\subsubsection{Recommendation type- Customers who viewed this item also viewed (CVV)}

Misinformation bias scores of CVV recommendations are significantly different for debunking, neutral and promoting products (KW H(2, N=4485) = 2673.95, p=0.0) . Post hoc test indicates that CVV recommendations of misinformative products have more misinformation than CVV recommendations of debunking and neutral products.  Notably, CVV recommendations of debunking products contain more misinformation than CVV recommendations  of neutral products. This is troubling since users who are clicking on products that present scientific information are pushed more misinformation in this recommendation type. In the recommend\-ation graph (Figure \ref{view_view} ), we see edges connecting multiple red nodes supporting our finding that CVV recommendations of mis\-informa\-tive products mostly contain other misinformative products. 
The most recommended product occurring in this recommendation type is a misinfor\-mative Kindle book titled \textit{Dissolving Illusions} (B00E7FOA0U).

\subsubsection{Recommendation type- Frequently bought together (FBT)}

Misinformation bias scores of FBT recommendations are significan\-tly different for debunking, neutral and promoting products (KW H(2, N=388) = 277.08, p=6.8e-61). Post hoc tests reveal that amount of misinformation in FBB recommendations of misinformative prod\-ucts is significantly more than the FBB recommendations of neutral and debunking products. The finding is also evident from the graph (Figure \ref{fig:search_results_freq}). There are large sized red nodes attached to other red nodes and several green nodes attached together indicating the presence of a strong filter-bubble effect. 
``Frequently bought together'' can be considered an indicator of buying patterns on the platform. The post hoc tests indicate that people buy multiple misinformative products together. The most recommended product present in this recommendation type is a misinformative Paperback book titled \textit{Dissolving Illusions: Disease, Vaccines, and The Forgotten His\-tory } (1480216895). 

\subsubsection{Recommendation type- Sponsored products related to this item}
Most of the sponsored recommendations are either neutral or promoting (Figure \ref{Sponsored} and Table \ref{RQ1b}). Statistical test reveals that the misinformation bias score of sponsored recommendations are significantly different among debunking, neutral and promoting products (KW H(2, N=6575) = 628.52, p=3.2e-137). Post hoc tests reveal same results as for CVV recommendations. There are two most recommended sponsored books. First is a misinformative paperback book titled \textit{Vaccine Epidemic: How Corporate Greed, Bias\-ed Science, and Coercive Government Threaten Our Human Rights, Our Health, and Our Children} (1620872129). Second is a neutral Kindle book titled \textit{SPANISH FLU 1918: Data and Reflections on the Consequences of the Deadliest Plague, What History Teaches, How Not to Repeat the Same Mistakes} (B08774MCVP).

\subsubsection{Recommendation type- What other items customers buy after viewing this item (CBV)}

Misinfor\-mation bias scores of CBV reco\-mmendations are significantly different for debunking, neutral and promoting products (KW H(2, N=2234) = 1611.34, p=0.0). 
Results of post hoc tests are same as that of CVV recommendations. The presence of an echo chamber is quite evident in the recommendation graph (see  Figure \ref{buy_view}). The graph has two disconnected components, one comprising a mesh of  misinformative products indicating a cluster of misinformative products that keep getting recommended. 
CBV is also indicative of buying patterns of Amazon users. The algorithm has learnt that people viewing misinformative products end up purchasing them. Thus, it pushes more misinformative items to users that click on them, creating a problematic feedback loop. 
The most recommended product  in this recommendation type is a misinformative Kindle book titled \textit{Miller's Review of Critical Vaccine Studies: 400 Important Scientific Papers Summarized for Parents and Researchers} (B07NQW27VD).

\begin{scriptsize}
\begin{table*}[t]
\resizebox{\textwidth}{!}{%
\begin{tabular}{l|l|l|l|l|l|l|l|l|l|l|l|l|c|c|c|ccc|l|c|c|l|l|l}
\hline
                                                                                                                   & \multicolumn{12}{c|}{\textbf{RQ2a}}                                                                                                                                                                                                                                                                                                                                       & \multicolumn{9}{c|}{\textbf{RQ2b}}                                                                                                                                                                                                                                                                                                                                                                                     & \multicolumn{3}{c}{\textbf{RQ2c}}                                                             \\ \cline{2-25} 
                                                                                                                   & \multicolumn{12}{c|}{\textbf{Search results}}                                                                                                                                                                                                                                                                                                                             & \multicolumn{9}{c|}{\textbf{Recommendations}}                                                                                                                                                                                                                                                                                                                                                                          & \multicolumn{3}{l}{\textbf{\begin{tabular}[c]{@{}l@{}}Auto complete\\ suggestions\end{tabular}}} \\ \cline{2-25} 
                                                                                                                   & \multicolumn{3}{l|}{\textbf{Featured}}                                         & \multicolumn{3}{l|}{\textbf{\begin{tabular}[c]{@{}l@{}}Avg. \\ customer \\ reviews\end{tabular}}} & \multicolumn{3}{l|}{\textbf{\begin{tabular}[c]{@{}l@{}}Price low \\ to High\end{tabular}}} & \multicolumn{3}{l|}{\textbf{\begin{tabular}[c]{@{}l@{}}Newest\\ Arrivals\end{tabular}}} & \multicolumn{3}{c|}{\textbf{Homepage}}                                                                                               & \multicolumn{3}{c}{\textbf{Pre-purchase}}                                                                                          & \multicolumn{3}{c|}{\textbf{Product page}}                                                                                                 &                                &                               &                               \\ \cline{2-22}
\multirow{-4}{*}{\textbf{\begin{tabular}[c]{@{}l@{}}Actions performed\\ to build account \\ history\end{tabular}}} & \textbf{D}               & \textbf{N}               & \textbf{M}               & \textbf{D}                      & \textbf{N}                     & \textbf{M}                     & \textbf{D}                   & \textbf{N}                   & \textbf{M}                   & \textbf{D}                  & \textbf{N}                  & \textbf{M}                  & \multicolumn{1}{l|}{\textbf{D}}            & \multicolumn{1}{l|}{\textbf{N}}            & \multicolumn{1}{l|}{\textbf{M}}            & \multicolumn{1}{l|}{\textbf{D}}            & \multicolumn{1}{l|}{\textbf{N}}           & \multicolumn{1}{l|}{\textbf{M}}           & \textbf{D}                              & \multicolumn{1}{l|}{\textbf{N}}                 & \multicolumn{1}{l|}{\textbf{M}}                & \multirow{-2}{*}{\textbf{D}}   & \multirow{-2}{*}{\textbf{N}}  & \multirow{-2}{*}{\textbf{M}}  \\ \hline
Search product                                                                                                     & IR \cellcolor[HTML]{FFD9D7} & IR \cellcolor[HTML]{FFD9D7} & IR \cellcolor[HTML]{FFD9D7} & NP \cellcolor[HTML]{E3DCDC}        & NP \cellcolor[HTML]{E3DCDC}       & NP \cellcolor[HTML]{E3DCDC}       & NP \cellcolor[HTML]{E3DCDC}     & NP \cellcolor[HTML]{E3DCDC}     & NP \cellcolor[HTML]{E3DCDC}     & NP \cellcolor[HTML]{E3DCDC}    & NP \cellcolor[HTML]{E3DCDC}    & NP \cellcolor[HTML]{E3DCDC}    & -                                          & -                                          & -                                          & \multicolumn{1}{c|}{X}                     & \multicolumn{1}{c|}{X}                    & X                                         & \multicolumn{1}{c|}{X}                  & X                                               & X                                              & NP \cellcolor[HTML]{E3DCDC}       & NP \cellcolor[HTML]{E3DCDC}      & NP \cellcolor[HTML]{E3DCDC}      \\ \hline
\begin{tabular}[c]{@{}l@{}}Search \& click\\ product\end{tabular}                                                  & IR \cellcolor[HTML]{FFD9D7} & IR \cellcolor[HTML]{FFD9D7} & IR \cellcolor[HTML]{FFD9D7} & NP \cellcolor[HTML]{E3DCDC}        & NP \cellcolor[HTML]{E3DCDC}       & NP \cellcolor[HTML]{E3DCDC}       & NP \cellcolor[HTML]{E3DCDC}     & NP \cellcolor[HTML]{E3DCDC}     & NP \cellcolor[HTML]{E3DCDC}     & NP \cellcolor[HTML]{E3DCDC}    & NP \cellcolor[HTML]{E3DCDC}    & NP \cellcolor[HTML]{E3DCDC}    & \multicolumn{3}{l|}{\begin{tabular}[c]{@{}l@{}}KW H(2, N=42) = 32.07,\\ p = 1.08e-07\\ M\textgreater{}N\textgreater{}D\end{tabular}} & \multicolumn{1}{c|}{X}                     & \multicolumn{1}{c|}{X}                    & X                                         & \multicolumn{3}{l|}{\begin{tabular}[c]{@{}l@{}}KW H(2, N=42) = 24.89,\\ p = 3.94e-06\\ M\textgreater{}D \& M\textgreater{}N\end{tabular}}  & NP \cellcolor[HTML]{E3DCDC}       & NP \cellcolor[HTML]{E3DCDC}      & NP \cellcolor[HTML]{E3DCDC}      \\ \hline
\begin{tabular}[c]{@{}l@{}}Search + click \&\\ add to cart product\end{tabular}                                      & IR \cellcolor[HTML]{FFD9D7} & IR \cellcolor[HTML]{FFD9D7} & IR \cellcolor[HTML]{FFD9D7} & NP \cellcolor[HTML]{E3DCDC}        & NP \cellcolor[HTML]{E3DCDC}       & NP \cellcolor[HTML]{E3DCDC}       & NP \cellcolor[HTML]{E3DCDC}     & NP \cellcolor[HTML]{E3DCDC}     & NP \cellcolor[HTML]{E3DCDC}     & NP \cellcolor[HTML]{E3DCDC}    & NP \cellcolor[HTML]{E3DCDC}    & NP \cellcolor[HTML]{E3DCDC}    & \multicolumn{3}{l|}{\begin{tabular}[c]{@{}l@{}}KW H(2, N=42) = 33.48,\\ p = 5.38e-08\\ M\textgreater{}N\textgreater{}D\end{tabular}} & \multicolumn{3}{l|}{\begin{tabular}[c]{@{}l@{}}KW H(2, 42) = 32.63,\\ p = 8.19e-08\\ M\textgreater{}N\textgreater{}D\end{tabular}} & \multicolumn{3}{l|}{\begin{tabular}[c]{@{}l@{}}KW H(2, N=42) = 24.05, \\ p = 5.98e-06\\ M\textgreater{}D \& M\textgreater{}N\end{tabular}} & NP \cellcolor[HTML]{E3DCDC}       & NP \cellcolor[HTML]{E3DCDC}      & NP \cellcolor[HTML]{E3DCDC}      \\ \hline
\begin{tabular}[c]{@{}l@{}}Search + click \&\\  mark “Top rated,\\ All positive review” \\ as helpful\end{tabular}   & IR \cellcolor[HTML]{FFD9D7} & IR \cellcolor[HTML]{FFD9D7} & IR \cellcolor[HTML]{FFD9D7} & NP \cellcolor[HTML]{E3DCDC}        & NP \cellcolor[HTML]{E3DCDC}       & NP \cellcolor[HTML]{E3DCDC}       & NP \cellcolor[HTML]{E3DCDC}     & NP \cellcolor[HTML]{E3DCDC}     & NP \cellcolor[HTML]{E3DCDC}     & NP \cellcolor[HTML]{E3DCDC}    & NP \cellcolor[HTML]{E3DCDC}    & NP \cellcolor[HTML]{E3DCDC}    & \multicolumn{3}{l|}{\begin{tabular}[c]{@{}l@{}}KW H(2, N=42) = 32.33,\\ p = 9.52e-08\\ M\textgreater{}N\textgreater{}D\end{tabular}} & \multicolumn{1}{c|}{X}                     & \multicolumn{1}{c|}{X}                    & X                                         & \multicolumn{3}{l|}{\begin{tabular}[c]{@{}l@{}}KW H(2, 42) = 23.36,\\ p = 8.44e-06\\ M\textgreater{}N \& M\textgreater{}D\end{tabular}}    & NP \cellcolor[HTML]{E3DCDC}       & NP \cellcolor[HTML]{E3DCDC}      & NP \cellcolor[HTML]{E3DCDC}      \\ \hline
\begin{tabular}[c]{@{}l@{}}Following \\ contributor\end{tabular}                                                   & IR \cellcolor[HTML]{FFD9D7} & IR \cellcolor[HTML]{FFD9D7} & IR \cellcolor[HTML]{FFD9D7} & NP \cellcolor[HTML]{E3DCDC}        & NP \cellcolor[HTML]{E3DCDC}       & NP \cellcolor[HTML]{E3DCDC}       & NP \cellcolor[HTML]{E3DCDC}     & NP \cellcolor[HTML]{E3DCDC}     & NP \cellcolor[HTML]{E3DCDC}     & NP \cellcolor[HTML]{E3DCDC}    & NP \cellcolor[HTML]{E3DCDC}    & NP \cellcolor[HTML]{E3DCDC}    & -                                          & -                                          & -                                          & \multicolumn{1}{c|}{X}                     & \multicolumn{1}{c|}{X}                    & X                                         & \multicolumn{1}{c|}{X}                  & X                                               & X                                              & NP \cellcolor[HTML]{E3DCDC}       & NP \cellcolor[HTML]{E3DCDC}      & NP \cellcolor[HTML]{E3DCDC}      \\ \hline
\begin{tabular}[c]{@{}l@{}}Search product\\ on Google\end{tabular}                                                 & IR \cellcolor[HTML]{FFD9D7} & IR \cellcolor[HTML]{FFD9D7} & IR \cellcolor[HTML]{FFD9D7} & NP \cellcolor[HTML]{E3DCDC}        & NP \cellcolor[HTML]{E3DCDC}       & NP \cellcolor[HTML]{E3DCDC}       & NP \cellcolor[HTML]{E3DCDC}     & NP \cellcolor[HTML]{E3DCDC}     & NP \cellcolor[HTML]{E3DCDC}     & NP \cellcolor[HTML]{E3DCDC}    & NP \cellcolor[HTML]{E3DCDC}    & NP \cellcolor[HTML]{E3DCDC}    & -                                          & -                                          & -                                          & \multicolumn{1}{c|}{X}                     & \multicolumn{1}{c|}{X}                    & X                                         & \multicolumn{1}{c|}{X}                  & X                                               & X                                              & NP \cellcolor[HTML]{E3DCDC}       & NP \cellcolor[HTML]{E3DCDC}      & NP \cellcolor[HTML]{E3DCDC}      \\ \hline
\end{tabular}%
}
\caption{\textbf{RQ2:} Table summarizing RQ2 results. \fcolorbox{labelpink}{labelpink}{IR} suggests noise and inconclusive results, i.e search results of control and its twin seldom matched. Thus, difference between treatment and control could either be attributed to noise or personalization, making it impossible to study the impact of personalization on misinformation. \fcolorbox{labelgrey}{labelgrey}{NP} denotes little to no personalization. - indicates that the given  activity had no impact on the component. X indicates that component was not collected for the activity. M, N and D indicate average per day bias in the component collected by accounts that built their history by performing actions on misinformative, neutral or debunking products. Higher mean value indicates more misinformation. For example, consider the cell corresponding to action ``search + click \& add to cart product'' and ``Homepage'' recommendation. M>N>D indicates that accounts adding misinformative products to cart ends up with more misinformation in their homepage recommendations in comparison to accounts that add neutral or debunking products to cart. }
\label{RQ2}
\end{table*}
\end{scriptsize}

\begin{figure*}
  \centering
  \begin{subfigure}{0.48\textwidth}
      \centering
      \includegraphics[scale=0.45]{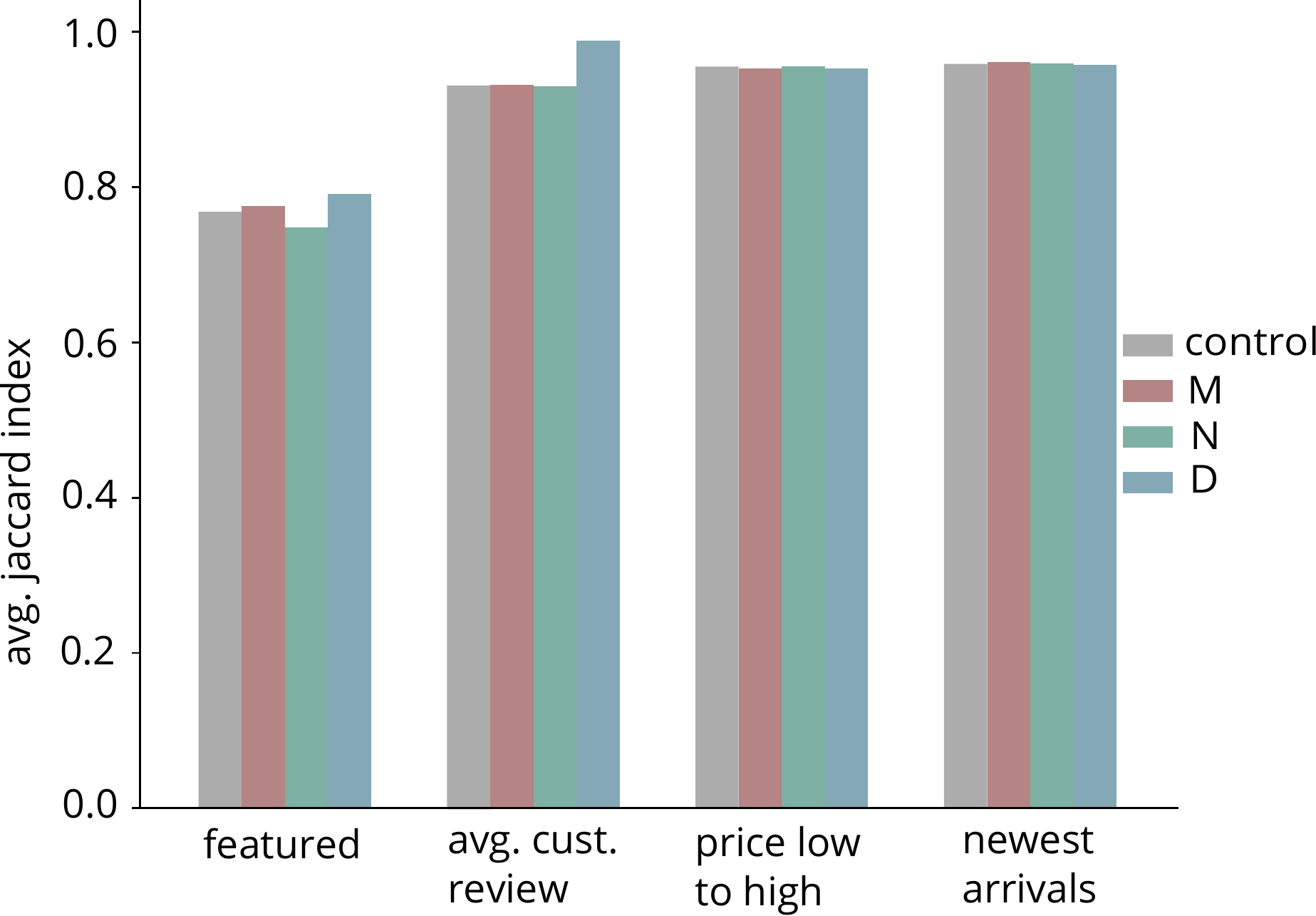}
      \caption{}
      \label{jaccard}
      \Description[Average jaccard index of treatment and control accounts]{Bar chart illustrating jaccard index between search results of control and its twin as well as control and the treatment accounts that performed ``following contributors'' action on misinformative, neutral and debunking products and later searched and sorted results using four Amazon filters. The jaccard index for control and its twin for filter ``featured'' is low (<0.8). For other three search filters, “average customer review”,“price low to high” and “newest arrivals”, we see high (>0.8) jaccard index and between and control and its twin and the metric values for treatment-control comparison are similar to that of control-twin comparison.  }
  \end{subfigure}
\hspace{0.2cm}
  \begin{subfigure}{0.48\textwidth}
      \centering
      \includegraphics[scale=0.45]{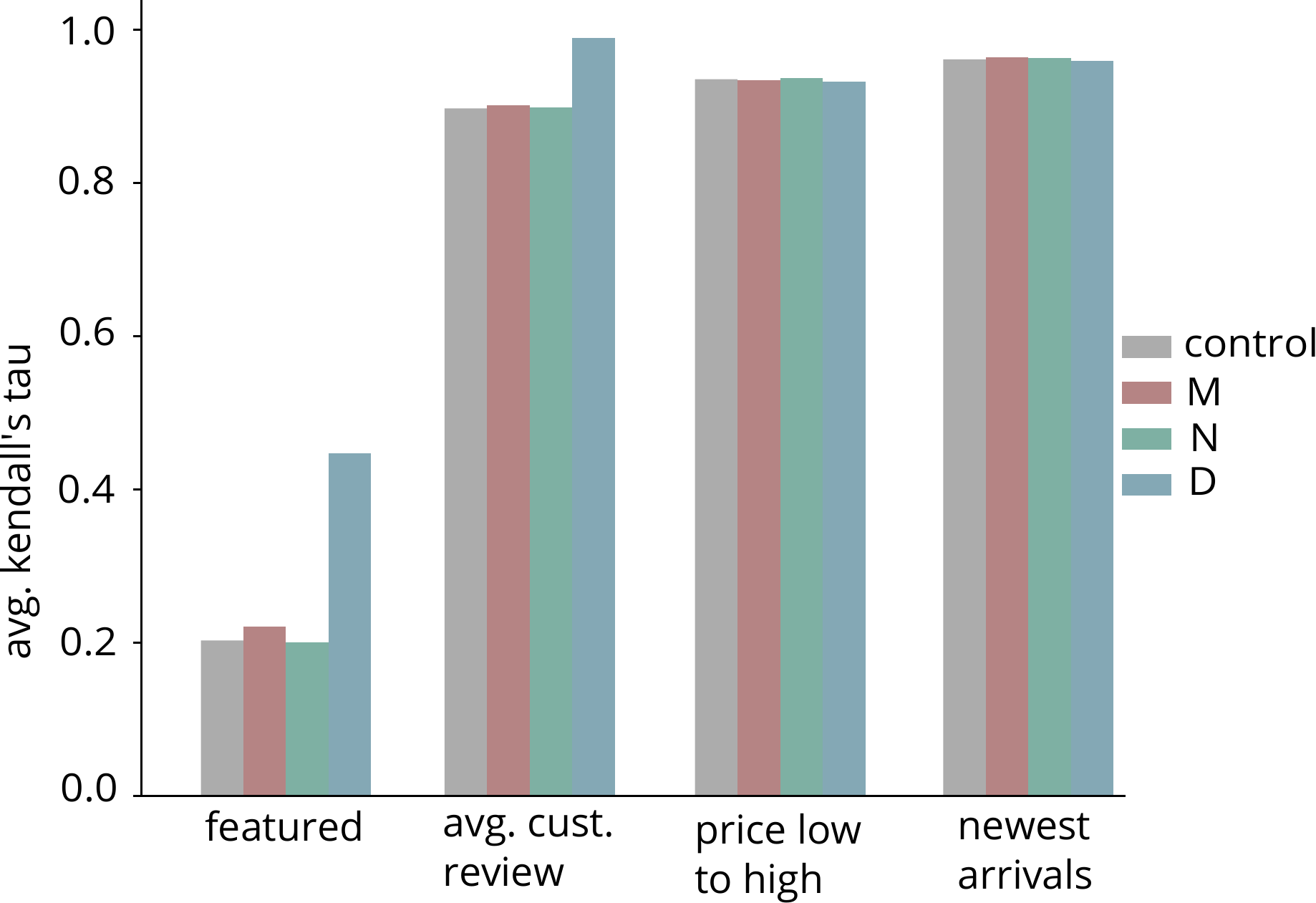}
      \caption{}
      \label{kendall}
      \Description[Average kendall's tau of treatment and control accounts]{Bar chart illustrating kendall's tau index between search results of control and its twin as well as control and the treatment accounts that performed ``following contributors'' action on misinformative, neutral and debunking products and later searched and sorted results using four Amazon filters. The kendall's tau coefficient for control and its twin for filter ``featured'' is low (<0.2). For other three search filters, “average customer review”,“price low to high” and “newest arrivals”, we see high (>0.8) kendall's tau coefficient between and control and its twin and the coefficient values for treatment-control comparison are similar to that of control-twin comparison.}
  \end{subfigure}

  \caption{Investigating the presence and amount of personalization due to ``following contributors'' action by calculating (a) Jaccard index and (b) kendall's tao metric between search results of treatment and control. M, N and D indicate results for accounts that follow contributors of misinformative, neutral and debunking products respectively. }
  \label{ppp}
\end{figure*}

\begin{figure*}
  \centering
  \begin{subfigure}[b]{0.33\textwidth}
      \centering
      \includegraphics[width=1\textwidth,keepaspectratio]{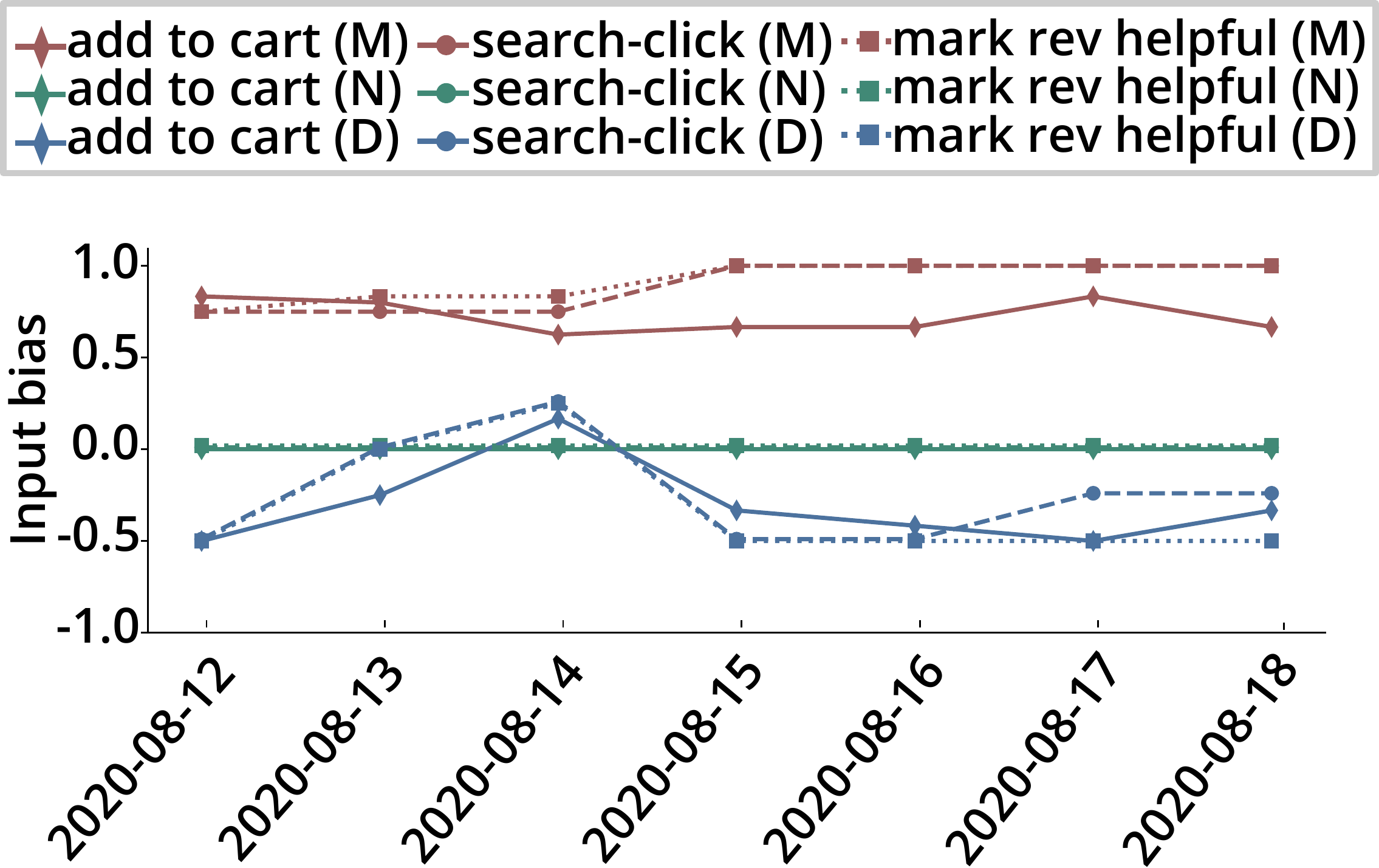}
      \caption{}
      \label{p_home}
      \Description[Input bias in homepages]{Line graph showing input bias on the y axis and dates of experiment run (2020-08-12 to 2020-08-18) on x axis. Input bias for accounts performing actions on neutral products is 0 for all seven days. Input bias for accounts performing actions search+click and mark-review on misinformative products is greater than 0 for all seven days and becomes 1 (max value) from fourth day onwards. Input bias in homepages for accounts adding misinformative products to their cart is also greater than 0 for all seven days but the value is less than the bias in homepages for accounts performing actions search+click and mar review helpful from third day onwards. Input bias in homepages of accounts performing actions on debunking  products becomes positive on the third day and after that drops below 0.}
  \end{subfigure}
  \begin{subfigure}[b]{0.33\textwidth}
      \centering
      
      \includegraphics[width=1\textwidth,keepaspectratio]{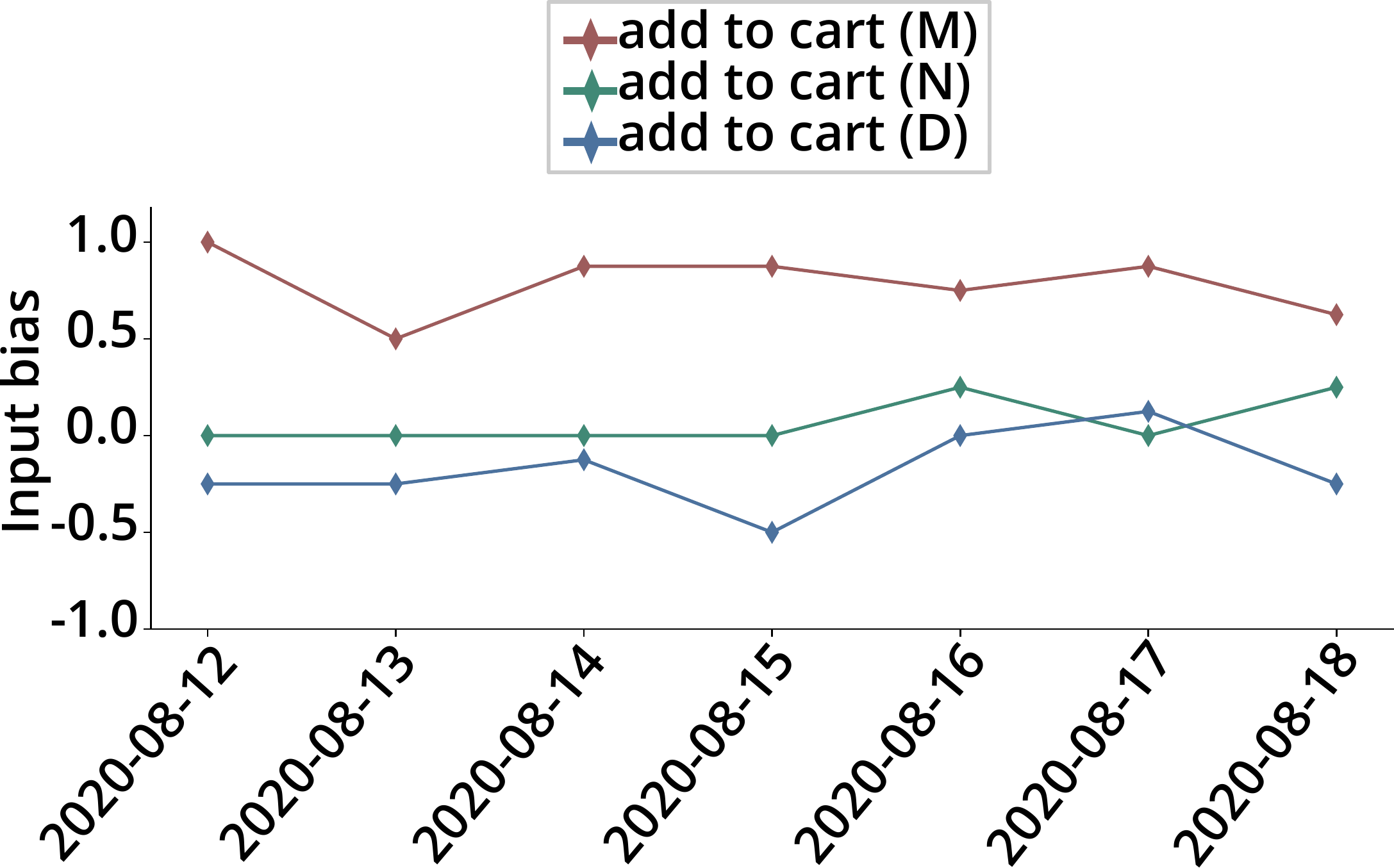}
      \caption{}
      \label{p_prepurchase}
      \Description[Input bias in pre-purchase recommendations]{Line graph showing input bias on the y axis and dates of experiment run (2020-08-12 to 2020-08-18) on x axis. Input bias for accounts that added neutral products to their cart remains 0 for all days except fifth and seventh when its 0.25. For accounts that added misinformative products to their cart, the bias value is greater than 0 for all seven days and for accounts adding debunking products to their cart, the bias value is less than 0 for all days except on fifth day when its 0 and sixth day when its 0.125}
  \end{subfigure}
\begin{subfigure}[b]{0.33\textwidth}
      \centering
      \includegraphics[width=1\textwidth,keepaspectratio]{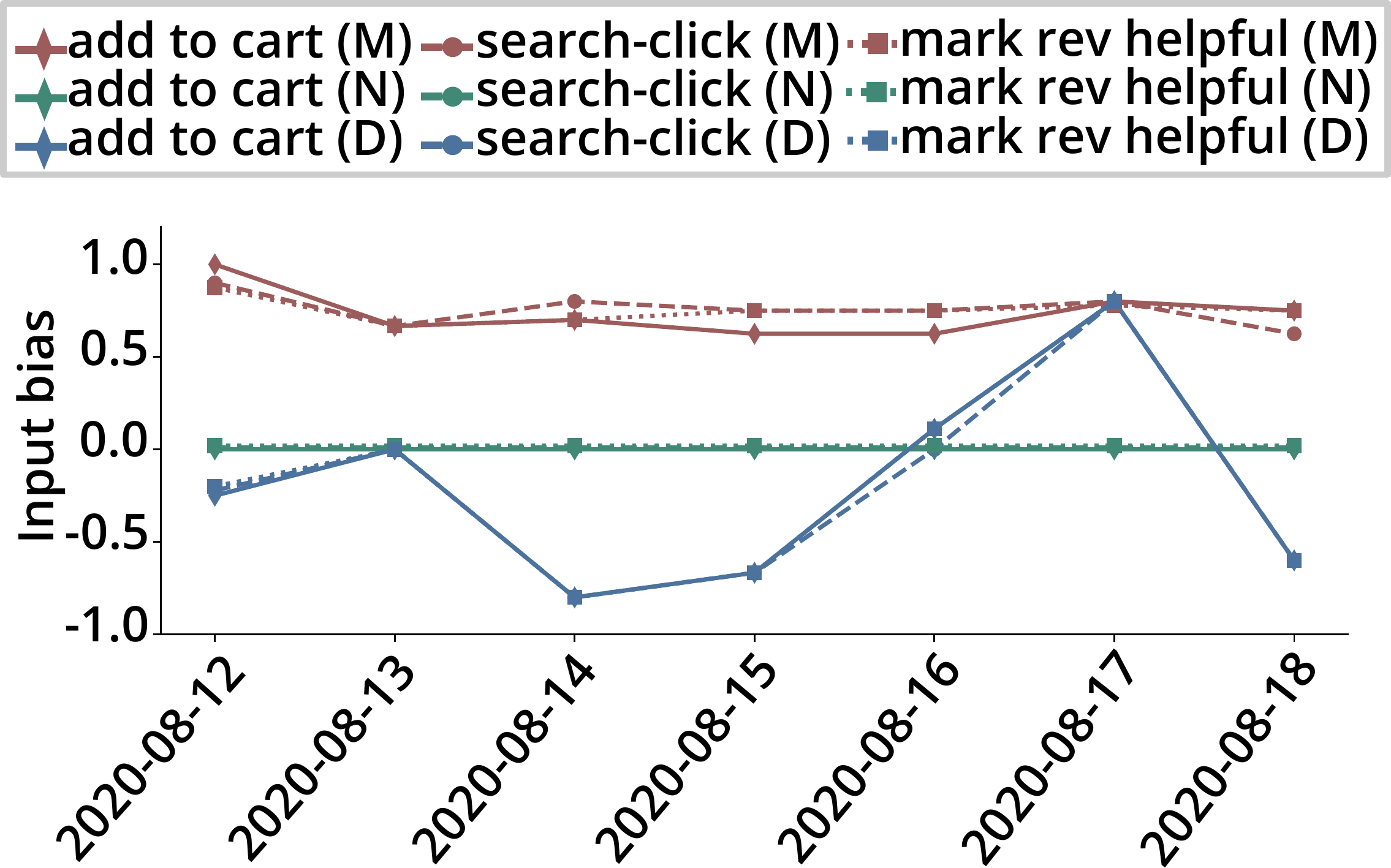}
      \caption{}
      \label{p_productpage}
      \Description[Input bias in product pages ]{Line graph showing input bias on the y axis and dates of experiment run (2020-08-12 to 2020-08-18) on x axis. Input bias for accounts performing actions on neutral products is 0 while its greater than 0 for accounts performing actions on misinformative products for all seven days. Input bias in product pages of accounts performing actions on debunking products is less than zero on first, third, fourth and seventh day and unusually high (>0) on sixth day.}
  \end{subfigure}
  \caption{(a) Input bias in homepages of accounts performing actions `add to cart'', ``search + click'' and ``mark top rated all positive review'' for seven days of experiment run. (b) Input bias in pre-purchase recommendations of accounts for 7 days experiment run. These recommendations are only collected for accounts adding products to their carts.
  (c) Input bias in product pages of accounts performing actions ``add to cart'', ``search + click'' and ``mark top rated all positive review'' for 7 days of experiment run.
  M, N and D indicate that the accounts performed actions on misinformative, neutral and debunking products respectively.}
  \label{ppp_bias}
\end{figure*}

\section{RQ2 Results [Personalized audit]: Effect of personalization} 

The aim of our Personalized audit was to determine the effect of personalization due to account history on the amount of mis\-infor\-mation returned in search results and various recommendations. Table \ref{RQ2} provides a summary. Below, we explain the effect of perso\-nalization on each component.

\subsection{RQ2a: Search Results} 

We measure personalization in search results for each Amazon filter using two metrics: Jaccard index and Kendall $\tau$ coefficient. Jaccard index determines similarity between two lists. 
A Jaccard index of 1 indicates that the two lists have same elements and zero indicates that the lists are completely different. 
On the other hand, Kendall $\tau$ coefficient, also known as Kendall rank correlation coefficient determines the ordinal correlation between two lists. It can take values between [-1,1] with -1 indicating that lists have inverse ordering, 0 signifying no correlation and 1 suggesting that items in the list have same ranks. 

First, we compare search results of control account and its twin.  Recall we created twins for our 2 control accounts in the \emph{Personalized audit} to establish the baseline noise. Ideally, both should have  Jaccard and Kendall rank correlation coefficient closer to 1 since the accounts do not build any history, are set up in a similar manner, perform searches at the same time and are in the same geolocation. Next, we compare search results of control account with treatment accounts that built account histories by performing different actions. If personalization is occurring, 
the difference between search results of treatment and control should be more than the baseline noise (or Jaccard index and Kendall $\tau$ should be less). Whereas, if the baseline noise itself is large, it indicates inconsistencies and randomness in the search results. Interestingly, we found significant noise in search results of control and its twin for ``featured'' filter with jaccard index <0.8 and Kendall's rank correlation coefficient <0.2, that is, control and its twins seldom matched. Presence of noise suggests that Amazon is injecting some randomness in the ``featured'' search results. Unfortunately, this means that we would not be able to study the effect of personalization on the accounts for the ``featured'' search filter setting. 

For the other three search filters, ``average customer review'', ``price low to high'' and ``newest arrivals'', we see high (>0.8)  jaccard index and kendall $\tau$ metric values between and control and its twin. Additionally, we do not see any personalization for these filters since metrics values for treatment-control comparison are similar to that of control-twin comparison. Figure \ref{ppp} shows the metrics calculation for control account and treatments that have built their search histories by following contributor's of misinformative, neutral and debunking products. 
We see two minor inconsistencies for filter ``average customer review'' in accounts building their history on debunking products where  
treatment received more similar results to control than its twin account. In any case, the treatment does not see more inconsistency than the control and its twin indicating no personalization. Other user actions show similar results, hence, we have removed their results for brevity.

\subsection{RQ2b: Recommendations} We investigated the occurrence of personalization and its impact on the amount of misinformation in three different recommendation pages. We discuss them below. \\

\noindent{\textbf{Homepage recommendations:}}
 We find that homepages are per\-sonalized only when a user performs click actions on the search results. Thus, actions ``add to cart'', ``search + click'' and ``mark top rated most positive review helpful'' led to homepage personalization. On the other hand, homepages were not personalized for actions ``follow contributor'', ``search product'' and ``google search'' actions.
After identifying the actions leading to personalized homepages, we investigate the impact of personalization on the amount of misinformation. In other words, we investigate how  misinformation bias in homepages is different for accounts building their  history by performing actions on misinformative, neutral and debunking products.  For each action, we had 6 accounts, two replicates for each action and product type (misinformation, neutral and debunking). For example, for action ``add to cart'' two accounts built their history by adding misinformative products to cart for 7 days, two added neutral products and two accounts added debunking products to their carts. We calculate per day input bias (ib) in homepages by averaging the misinformation bias scores of each recommended product present in the homepage. Therefore, for every account we have seven bias values. We  consider only top two products in each recommendation type. Recall, homepages could contain three different types of recommendations `Inspired by your shopping trends'', ``Recommended items other customers often buy again'' and ``Related to items you've viewed''. All the different types are considered together for analysis.

Statistical tests reveal significant differences in the amount of misinformation present in  homepages of accounts that built their histories by performing actions on misinformative, neutral and debunking products (see Table \ref{RQ2}). This observation holds true for all three activities ``add to cart'', ``search + click'' and ``mark top rated most positive review helpful''.  Post hoc test reveals an echo chamber effect. Amount of misinformation in recommendations of products performing actions on misinformative products is more than the amount of misinformation in homepages of accounts performing actions on neutral products which in turn is more than the misinformation present in homepages of accounts performing actions on debunking products. 

Figure \ref{p_home} shows per day input bias of homepages of different accounts performing different actions. We take an average of the replicates for plotting the graph. Surprisingly, performing actions ``mark top rated most positive review helpful'' and ``search + click''  on a misinformative product leads to highest amount of misinformation in the homepages, even more than the homepages of accounts adding misinformative products to the cart. This means that amount of misinformation present in homepage is comparatively less once a user shows an intention to purchase a misinformative product but high if a user shows interest in the misinformative product but doesn't show an indication to buy it. Figure \ref{p_home}  also shows that amount of misinformation present in homepages of accounts performing actions ``mark top rated most positive review helpful'' and ``search + click'' on misinformative products  gradually increases and becomes 1 on day 4 (2020-08-15). Bias value 1 indicates that all analysed products in homepages  were misinformative. Homepage recommendations of products performing actions on neutral objects show 0 bias constantly indicating all recommendations on all days were neutral. On the other hand, average bias in homepages of accounts building history on debunking accounts rose a little above 0 in the first three days but eventually fells below 0 indicating a debunking lean.\\

\noindent{\textbf{Pre-purchase recommendations:}} These recommendations are only presented to users that add product(s) to their Amazon cart. Therefore, they were collected for 6 accounts, 2 of which added misinformative products to cart, 2 added neutral products and the other 2 added debunking products. These recommendations could be of several types. See Figure \ref{Pre-purchase recommendations} for an example of pre-purchase page. For our analysis, we consider the first product present in each recommendation type. Statistical tests reveal significant difference in the amount of misinformation present in pre-purchase reco\-mmendations of accounts that added misinformative, neutral and debunking products to cart (KW H(2, 42) = 32.63, p = 8.19e-08). Those adding misinformative products to cart contain more misinfo\-rmation than the accounts adding neutral or debunking products to their carts. Figure \ref{p_prepurchase} shows the input bias in the pre-purchase reco\-mmendations for all the accounts. There is no coherent temporal trend, indicating that the input bias in this recommendation type depends on the particular product being added to cart. However, an echo chamber effect is evident. For example, bias in pre-purchase recommendations of accounts adding misinformative products to cart is above 0 for all 7 days.\\

\noindent{\textbf{Product recommendations:}} We collect product recommenda\-tions for accounts performing  ``add to cart'', ``search + click'' and ``mark top rated most positive review helpful'' actions. We find significant difference in the amount of misinformation present in product page recommendations when accounts performed these actions on misinformative, neutral, and debunking products (refer Table \ref{RQ2}). Post hoc analysis reveals that product page recommenda\-tions of misinformative products contain more misinformation than those of neutral and debunking products. Figure \ref{p_productpage} shows the input bias present in product pages across accounts. The bias for neutral products is constantly 0 across the 7 days, but for misinformative products, it is constantly greater than 0 for all actions. We see an unusually high bias value on the 6th day (2020-08-17) of our experiment for accounts performing actions on debunking product titled \textit{Reasons to Vaccinate: Proof That Vaccines Save Lives} (B086B8M\-M71). We checked the product page recommendations of this parti\-cular debunking book and found several misinformative recomm\-endations on its product page. 

\subsection{RQ2c: Auto-complete suggestions} We audited auto-complete suggestions to investigate how personal\-ization affects the change in search query suggestions. Our initial hypothesis was that performing actions on misinformative products could increase the  auto-complete suggestions of anti-vaccine search queries. However, we found little to no personalization in the auto-complete suggestions indicating that account history built by performing actions on vaccine-related misinformative, neutral or debunking products have little to no effect on how auto-complete suggestions of accounts change. In interest of brevity, we do not add the results and graphs for this component.

\section{Discussion} \label{discussion}

There is a growing concern that e-commerce platforms are becom\-ing hubs of dangerous medical misinformation. Unlike search engin\-es where the motivation of the platform is to show relevant search results to sell advertisements, goal of e-commerce platforms is to sell products.
The motivation to increase sales means that relevance in recommendations and search suggestions is driven by what people purchase after conducting a search or viewing an item, irrespective of whether the product serves credible information or not. As a result, due to  lack of regulatory policies, websites like Amazon are providing a platform to people who are making money by selling misinformation---dangerous anti-vaccine ideas, pseudoscience treatments, or unproven dietary alternatives---some of which could have dangerous effects on people's health and well-being. With a US market share of 49\%, Amazon is the leading  product search engine in the United States  \cite{amazonstats_2}. Therefore, any misinfor\-mation present in its search and recommendations could have a far reaching influence where they can negatively shape  users' viewing and purchasing patterns.  
Thus, in this paper we audited Amazon for the most dangerous form of health misinformation---vaccine misinformation. 
Our work resulted in several critical find\-ings with far reaching implications. We discuss them below.

\subsection{Amazon: a marketplace of multifaceted health misinformation}
Our analysis shows that Amazon hosts a variety of health misinfor\-mative products. Maximum number of such products belong to the category Books and Kindle eBooks (Figure \ref{cat}).  Despite the enormous amount of information available online, people still turn to books to gain information. A Pew Research survey revealed that 73\% of Americans read atleast one book in a year \cite{pewsurveybooks}. Books are considered ``intellectual heft'', have more presence than scientific journals and thus, leave ``a wider long lasting wake'' \cite{herr2017writing}. Thus, anti-vaccine books could have a wider reach and can easily influence the audience negatively. Moreover, it does not help that a large number of anti-vaccine books are written by authors with medical degrees \cite{shin2020algorithms}. Not just anti-vaccine books, there are abundant pseudoscience books on the platform, all suggesting unproven methods to cure diseases. We found diet books suggesting recipes with colloidal silver---an unsafe product, as an ingredient. Some of the books proposing cures for incurable diseases, like autism and auto immune diseases, can have a huge appeal for people suffering with such diseases \cite{wired2}. Thus, there is an urgent need to check the quality of health books presented to the users.

The next most prominent category of health misinformative products is Amazon Fashion. Numerous apparels are sold on the platform with innovative anti-vaccine slogans, giving tools to the anti-vaccine propagandists to advocate their anti-vaccine agenda and gain visibility, not just in the online world, but in the offline world. During our annotation process, we also found many dietary supplements claiming to treat and cure diseases---a direct violation of Amazon's policy on dietary supplements. Overall, we find that health misinformation exists on the platform in various forms---books, t-shirts and other merchandise. Additionally, it is very easy to sell problematic content because of lack of appropriate quality-control policies and their enforcement. 


\subsection{Amazon search results: a stockpile of health misinformation }
Analysis of our \emph{Unpersonalized audit} revealed that  10.47\% of search results promote vaccine and other health-related misinformation.  Notably, the higher percentage of products promoting misinforma\-tion compared to debunking suggests that anti-vaccine and problem\-atic health-related content is churned out more and the attempts to debunk the existing misinformation is less. We also found that Amazon's search algorithm puts more health misinformative prod\-ucts in search results than debunking products leading to high input bias for topics like ``vaccination'', ``vaccine controversies'', ``hpv vaccine'', etc. This is specifically true for search filters ``featured'' and ``average customer reviews''. Note, that ``featured'' is the default search filter indicating that by default users will see more misinfor\-mation when they search for the aforementioned topics. On the other hand, if users want to make a purchase decision based on product ratings, again users will be presented with more misinform\-ation since our analysis indicates that sorting by filter "average customer reviews" leads to highest misinformation bias in the search results.
We also found a ranking bias in Amazon's search algorithm with misinformative products getting ranked higher. Past research has shown that people trust higher ranked search results \cite{guan2007eye}. Thus, more number of higher ranked misinformative products can make problematic ideas in these products appear mainstream. The only positive finding of our analysis was the presence of more debunking products in search results sorted by filter ``newest arrivals''. This might indicate that more higher quality products are being sold  on the platform in recent times. However, since there are no studies/surveys indicating which search filters are mostly used by people while making purchase decisions, it is difficult to conclude how beneficial this finding is.

\subsection{Amazon recommendations: problematic echo chambers}
Many search engines and social media platforms employ personali\-zation to enhance users' experience on their platform by recommen\-ding them items that the algorithm think they will like based on their past browsing or purchasing history. But on the downside, if not checked, personalization can also lead users into a rabbit hole of problematic content. Our analysis of \emph{Personalized audit} revealed that an echo chamber exists on Amazon where users performing real-world actions on misinformative books are presented with more misinformation in various recommendations. Just a single click on an anti-vaccine book could fill your homepage with several other similar anti vaccine books. And if you proceed to add that book in your cart, Amazon again presents more anti-vaccine books, nudging you to purchase even more problematic content. The worst discovery is that your homepages get filled with more misinforma\-tion if you just show an interest in a misinformative product (by clicking on it) compared to when you show an intention to buy it by adding product to your cart. Additionally on the product page itself, you are presented 5 different kinds of recommendations each of which contains equally problematic content. In a nutshell, once you start engaging with misinformative products on the platform, you will be presented with more misinformative stuff at every point of your Amazon navigation route and at multiple places. These findings would not have been concerning if buying a milk chocolate would lead to recommendations of other chocolates of different brands. The problem is that Amazon is blindly applying its algorithms on all products including problematic content. Its algorithms do not differentiate or give special significance to vaccine-related topics. 
Amazon has learnt from users' past viewing and purchasing behaviour and has categorized all the anti-vaccine and other problematic health cures together. It presents the problem\-atic content to users performing actions on any of these products, creating a dangerous recommendation loop in the process. There is an urgent need for the platform to treat vaccine and other health related topics differently and ensure high quality searches and recommendations. In the next section, we present a few ways, based on our findings, that could assist the platform in combating health misinformation.

\subsection{Combating health misinformation} \label{interventions}
Tackling online health misinformation is a complex problem and there is no easy  silver-bullet solution to curb its spread. However, the first step towards addressing is accepting that there is a problem. Many tech giants have acknowledged their social responsibility in ensuring high quality in health-related content and are actively taking many steps to ensure the same. For example, Google's policy ``Your Money Or Your Life'' classifies medical and health-related search pages as pages of particular importance, whose content should come from reputable websites \cite{yourmoneyyourlife}. Pinterest completely hobbled the search results of certain queries such as `anti-vax' \cite{pinterest} and limited the search results for other vaccine-related queries to content from officially recognized health institutions \cite{pinterest2}. Even Facebook---a platform known to have questionable content modera\-tion policies---banned anti-vaccine advertisements and demoted the anti-vaccine content in its search results to make its access difficult \cite{facebook}. Therefore, given the massive reach and user base of Amazon---206 million website visits every month \cite{amazonstats}---it is disconcerting to see that Amazon has not yet joined the bandwagon. Till date, it has not taken any concrete steps towards addressing the problem of anti-vaccine content on its platform. Through our findings, we recommend several short-term and long-term strategies that the platform can adopt.

\subsubsection{Short term strategies: design interventions.} The simplest short term solution would be to introduce design interventions. Our \textit{Unpersonalized audit} revealed high misinformation bias in search results. 
The platform can use interventions as an opportunity to communicate to users the quality of data presented to them by signalling mis\-informa\-tion bias. The platform could introduce a bias meter or scale that signals the amount of misinformation present in search results every time it detects a vaccine-related query in its search bar. The bias indicators could be coupled with informational interventions like showing Wikipedia and encyclopedia links, that have already been proven to be effective in reducing traffic to anti-vaccine content \cite{kim2020effects}. The second intervention strategy could  be to recognise and signal source bias. During our massive annotation process, we realized that several health misinformative books have been written by known anti-vaxxers like Andrew Wakefield, Jenny Mccarthy, Robert S. Mendelsohn, etc. We also present a list of authors who have contributed to most misinformative books in Table \ref{tab:contributors}. Imagine a design where  users are presented with a message ``The author is a known anti-vaxxer and is known to write books that might contain health minformation'' every time they click a book written by these authors. An another extreme short term solution could be to either enforce a platform-wide ban prohibiting sale  of any anti-vaccine product or hobble search results for anti-vaccine search queries.

\subsubsection{Long term strategies: algorithmic modifications and policy changes.} Long term interventions would include modification of search, ranking and recommendation algorithm. Our investigations revealed that Amazon's algorithm has learnt problematic patterns through consumer's past viewing and buying patterns. It has cate\-g\-orized all products of similar stance together (see several edges connecting red nodes--- products promoting misinformation in Figure \ref{un_reco_graphs}). In some cases, it has also associated some misinformative products with neutral and debunking products (refer Figure \ref{un_reco_graphs})
Amazon needs to ``unlearn'' this categorization. Additionally, the platform  should incorporate misinformation bias in their search and recommendation algorithms to reduce the exposure to misinfo\-rmative content. 
There is also an urgent need to introduce some policy changes. First and foremost, Amazon should stop promoting health misinformative books by sponsoring them. We found 98 misinformative products in the sponsored recommendations indicat\-ing that today, anti-vaccine outlets can easily promote their products by spending some money. Amazon should also introduce some minimum quality requirements that should be met before a product is allowed to be sponsored or sold  on its platform. 
It can  employ search quality raters to rate the quality of search results for various health-related search queries. Google  has already set an example with its extensive Search Quality Rating process and guidelines \cite{googlelink,searchqualityrating}. In recent times Amazon introduced several policy and algorithmic changes including roll out of a new feature ``verified purchase'' to curb fake reviews problem on its platform \cite{amazonfakereviewcombatsteps}. Similar efforts are required to ensure  product quality as well. Amazon can introduce a similar ``verified quality'' or ``verified claims'' tag with health-related products once they are evaluated by experts. 
Having a product base of millions of products can make any kind of review process tedious and challenging. Amazon can start by targeting specific health and vaccine related topics that are most likely to be searched. Our work itself presents a list of most popular vaccine-related topics that can be used as a starting point. {Can we expect Amazon to make any changes to its current policies and algorithms without sustained pressure? We believe audit studies like ours are the way to reveal biases in the algorithms used by commercial platforms so that there is more awareness about the issues which in turn would create pressure on the organization to act. In the past, such audit studies have led platforms to make positive changes to their algorithms \mbox{\cite{raji2019actionable}}}. We hope our work acts as a call to action for Amazon and also inspires vaccine and health audits on other platforms.

\section{Limitations}

Our study is not without limitations. First, we only considered top products in each recommendation-type present on a page while determining bias of the entire page. Annotating and determining bias of all the recommendations occurring in a page would give a much more accurate logic of recommendation algorithms. However, past studies have shown that the top results receive the highest number of clicks, thus, are more likely to receive attention from users \cite{top3ctr}. Second, search queries themselves have inherent bias. For example query `anti vaccine t-shirt' suggests that user is looking for anti-vax products. Higher bias in search results of neutral queries is much worse than that of biased queries. We did not segregate our analysis based on search query bias. Although, we did notice two neutral search queries namely `vaccine' and `varicella vaccine' appearing in the list of most problematic search-query and filter combinations. Third, while we audited various reco\-mmendations present on the platform, we did not analyse the email recommendations---product recommendations present outside the platform. A journalistic report pointed that email recommendations could be contaminated too if a user shows an interest in a misinfor\-mative product but leaves the platform without buying it \cite{wired1}. We leave investigation of these recommendations to future work. Fourth, in our \emph{Personalized audit}, accounts only built history for a week. Moreover, experiments were only run on Amazon.com.  We plan to continue to run our experiments and explore features such as geolocation for future audits. Fifth, our audit study only targeted results returned in response to vaccine-related queries. Since, Amazon is a vast platform that hosts variety of products and sellers, we cannot claim that our results are generalizable for other misinformative topics or conspiracy theories. However, our methodology is generic enough to be applied to other misinforma\-tive topics. Lastly, another major limitation of the study is that in the \emph{Personalized audit} account histories were built in a very conservative setting. Accounts  performed actions on only one product each day. Additionally, the actions were only performed on products with the same stance. In real-world it will be tough to find users who only add misinformative products in their carts for seven days continuously. But in spite of this limitation, our study still provides a peek into the workings of Amazon's algorithm and has paved way for future audits that could use our audit methodology and extensive qualitative coding scheme to perform experiments considering complex real world settings.

\section{Conclusion}
In this study, we conducted two sets of audit experiments on a popular e-commerce platform, Amazon to empirically determine the amount misinformation returned by its search and recommend\-ation algorithm. We also investigated whether personalization due to user history plays any role in amplifying  misinformation. Our audits resulted in a dataset of 4,997 Amazon products annotated for health misinformation. We found that search results returned for many vaccine-related queries contain large number of mis\-informa\-tive products leading to high misinformation bias. Moreover, misin\-forma\-tive products are also ranked higher than debunking products. Our study also suggests presence of a filter-bubble effect in reco\-mmendations, where users performing actions on misinformative products are presented with more misinformation in their home\-pages, product page recommendations and pre-purchase recomm\-endations. We believe, our proposed methodology to audit vaccine misinformation can be applied to other platforms to investigate health misinforma\-tion bias. Overall, our study brings attention to the need for search engines to ensure high standards and quality of results for health related queries. 


\bibliographystyle{ACM-Reference-Format}
\bibliography{sample-base}

\appendix
 \section{appendix}
 The appendix contains a table (Table \ref{tab:books}) of books annotated as promoting, neutral and debunking that were selected to build history of accounts in the \emph{Personalized audit} as well as illustration of our multi-stage iterative coding process (Figure \ref{q_c}).  Additionally, we give details about our Amazon Mechanical Turk (AMT) task in Appendix, Section \ref{amt_job}.

\begin{table*}
\centering
{ \scriptsize
\begin{tabular}{l|m{3cm}ll|m{3cm}ll|m{3cm}ll}
\hline
\multirow{2}{*}{\textbf{\#}} & \multicolumn{3}{l}{\textbf{Debunking products}}                                                                                                  & \multicolumn{3}{l|}{\textbf{Neutral products}}                                                                                                                                                                                                  & \multicolumn{3}{l}{\textbf{Misinformative products}}                                                                                                                                            \\ \cline{2-10} 
                             & \textbf{title (url code)}                                                                                              & \textbf{S} & \textbf{R} & \textbf{title (url code)}                                                                                                                                                                                             & \textbf{S} & \textbf{R} & \textbf{title (url code)}                                                                                                                                             & \textbf{S} & \textbf{R} \\ \hline
1                            & Vaccinated: One Man's Quest to Defeat the World's Deadliest Diseases (006122796X)                                      & 4.7        & 134        & Baby's Book: The   First Five Years (Woodland Friends) 144131976X                                                                                                                                                     & 4.9        & 614        & Dissolving Illusions:   Disease, Vaccines, and The Forgotten History (1480216895)                                                                                     & 4.9        & 953        \\ \hline
2                            & Epidemiology and Prevention of Vaccine-Preventable Diseases, 13th Edition (990449114)                                  & 4.5        & 11         & My Child's Health Record Keeper (Log   Book) (1441313842)                                                                                                                                                             & 4.8        & 983        & The Vaccine Book: Making the Right   Decision for Your Child (Sears Parenting Library) (0316180521)                                                                   & 4.8        & 1013       \\ \hline
3                            & The Panic Virus: The True Story Behind the Vaccine-Autism Controversy (1439158657)                                     & 4.4        & 175        & Ten Things Every Child with Autism Wishes   You Knew, 3rd Edition: Revised and Updated paperback (1941765882)                                                                                                         & 4.8        & 792        & The Vaccine-Friendly Plan: Dr. Paul's   Safe and Effective Approach to Immunity and Health-from Pregnancy Through   Your Child's Teen Years (1101884231)              & 4.8        & 877        \\ \hline
4                            & Vaccines: Expert Consult - Online and Print (Vaccines (Plotkin)) (1455700908)                                          & 4.4        & 18         & Baby 411: Your Baby, Birth to Age 1!   Everything you wanted to know but were afraid to ask about your newborn:   breastfeeding, weaning, calming a fussy baby, milestones and more! Your baby   bible! (1889392618)) & 4.8        & 580        & How to End the Autism Epidemic (1603588248)                                                                                                                           & 4.8        & 717        \\ \hline
5                            & Bad Science (865479186)                                                                                                & 4.3        & 967        & Uniquely Human: A Different Way of Seeing   Autism (1476776245)                                                                                                                                                       & 4.8        & 504        & How to Raise a Healthy Child in Spite of   Your Doctor: One of America's Leading Pediatricians Puts Parents Back in   Control of Their Children's Health (0345342763) & 4.8        & 598        \\ \hline
6                            & Reasons to Vaccinate: Proof That Vaccines Save Lives (B086B8MM71)                                                      & 4.3        & 232        & The Whole-Brain Child: 12 Revolutionary   Strategies to Nurture Your Child's Developing Mind (0553386697)                                                                                                             & 4.7        & 2347       & Miller's Review of Critical Vaccine   Studies: 400 Important Scientific Papers Summarized for Parents and   Researchers (188121740X)                                  & 4.8        & 473        \\ \hline
7                            & Deadly Choices: How the Anti-Vaccine Movement Threatens Us All (465057969)                                             & 4.2        & 223        & We're Pregnant! The First Time Dad's   Pregnancy Handbook (1939754682)                                                                                                                                                & 4.7        & 862        & Herbal Antibiotics, 2nd Edition: Natural   Alternatives for Treating Drug-resistant Bacteria (1603429875)                                                             & 4.7        & 644  \\    
\hline
       
\end{tabular}}
\caption{Books corresponding to each annotation value shortlisted to build account histories in our \emph{Personalized audit}. S represents the star rating of the product and R denotes the number of ratings received by the book.}
\label{tab:books}
\end{table*}

\begin{figure*}
  \centering
      \includegraphics[scale=0.5]{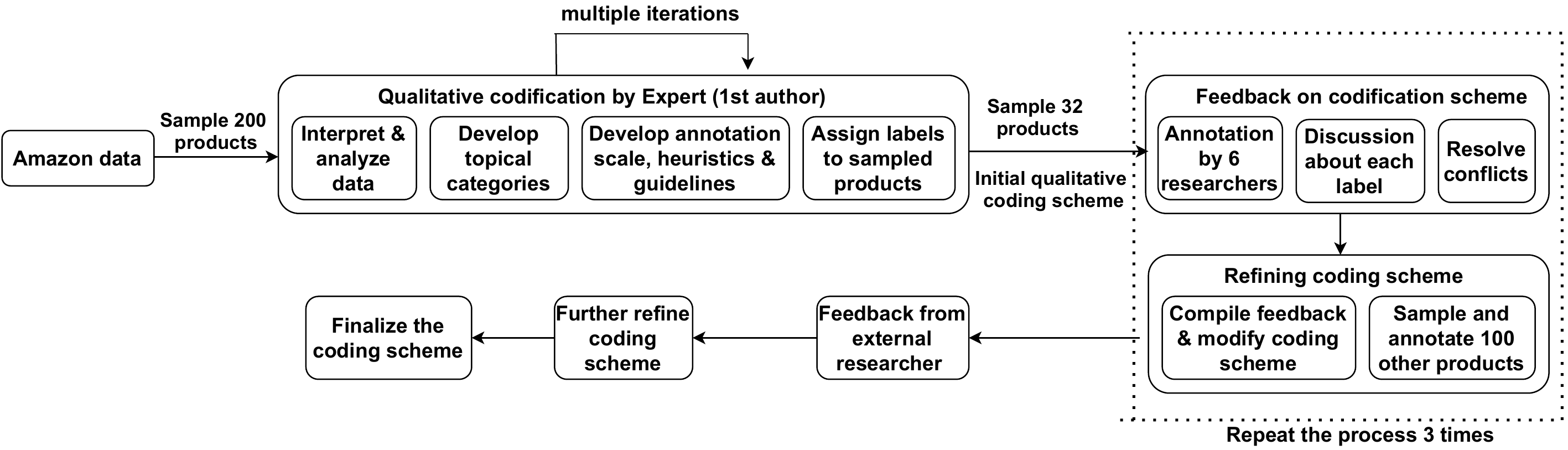}
  \caption{Our multi-stage iterative qualitative coding process to obtain a coding scheme for annotating Amazon products for health misinformation. }
  \label{q_c}
  \Description[Qualitative coding process]{Three stages: (1) multiple iterations of qualitative codification by the first author, (2) multiple iterations of refining the codification scheme based on feedback obtained from 6 researchers followed by (3) feedback from an external researcher.}
\end{figure*}


\subsection{Amazon Mechanical Turk Job} \label{amt_job}
\subsubsection{Turk job description}
In this section, we describe how we obtained annotations for our study from Amazon Mechanical Turk workers (MTurks). 
Past research has shown that it is possible to get good data from crowd-sourcing platforms like Amazon Mechanical Turk (AMT) if the workers are screened and trained for the crowd-sourced task \cite{mitra2015comparing}. Below we describe the screening process and our annotation task briefly.

\subsubsection{Screening} 
\begin{figure*}
  \centering
      \includegraphics[scale=0.35]{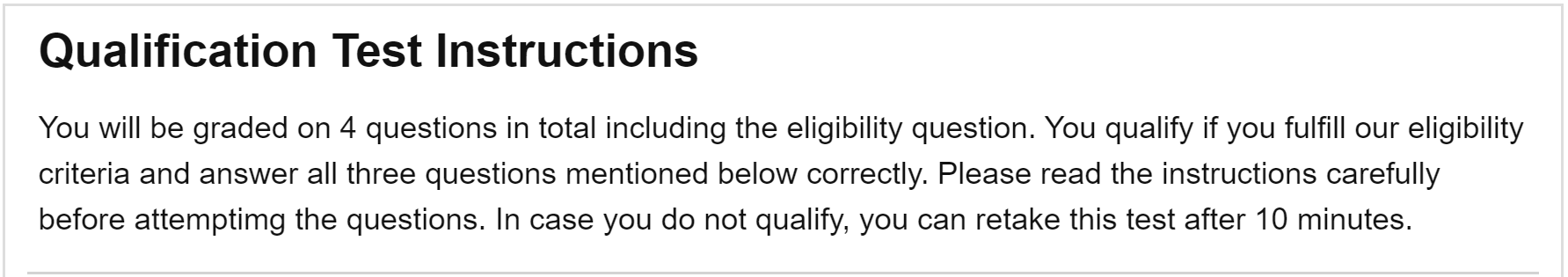}
  \caption{Figure illustrating Qualification Test instructions. Test included 4 questions including one eligibility question required to be added by authors' University. A full score of 100 was required to qualify the test. }
  \label{qual_instructions}
  \Description[Qualitative test instructions:]{The instructions were worded as ``You will be graded on 4 questions in total including the eligibility question. You qualify if you fulfill our eligibility criteria and answer all three questions mentioned below correctly. Please read the instructions carefully before attempting the questions. In case you do not qualify, you can retake this test after 10 minutes.''}
\end{figure*}

\begin{figure*}
  \centering
      \includegraphics[scale=0.35]{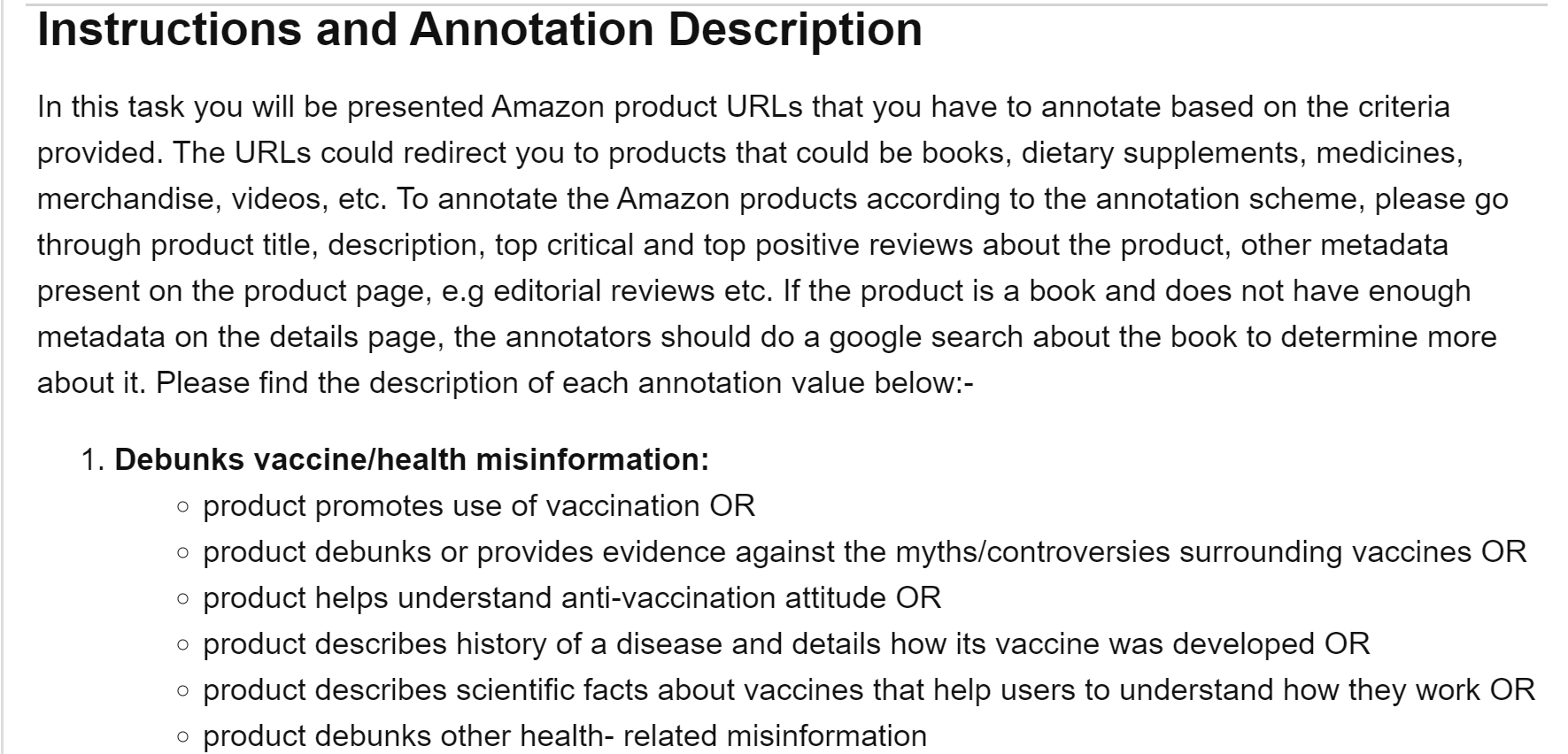}
  \caption{Task description in the Qualification test. Same instructions were provided in the actual task.}
  \label{qual_desc}
  \Description[Instructions and annotation description]{The figure is a snapshot of the annotation instructions provided to MTurks including detailed description of each annotation value. }
\end{figure*}

\begin{figure*}
  \centering
      \includegraphics[scale=0.35]{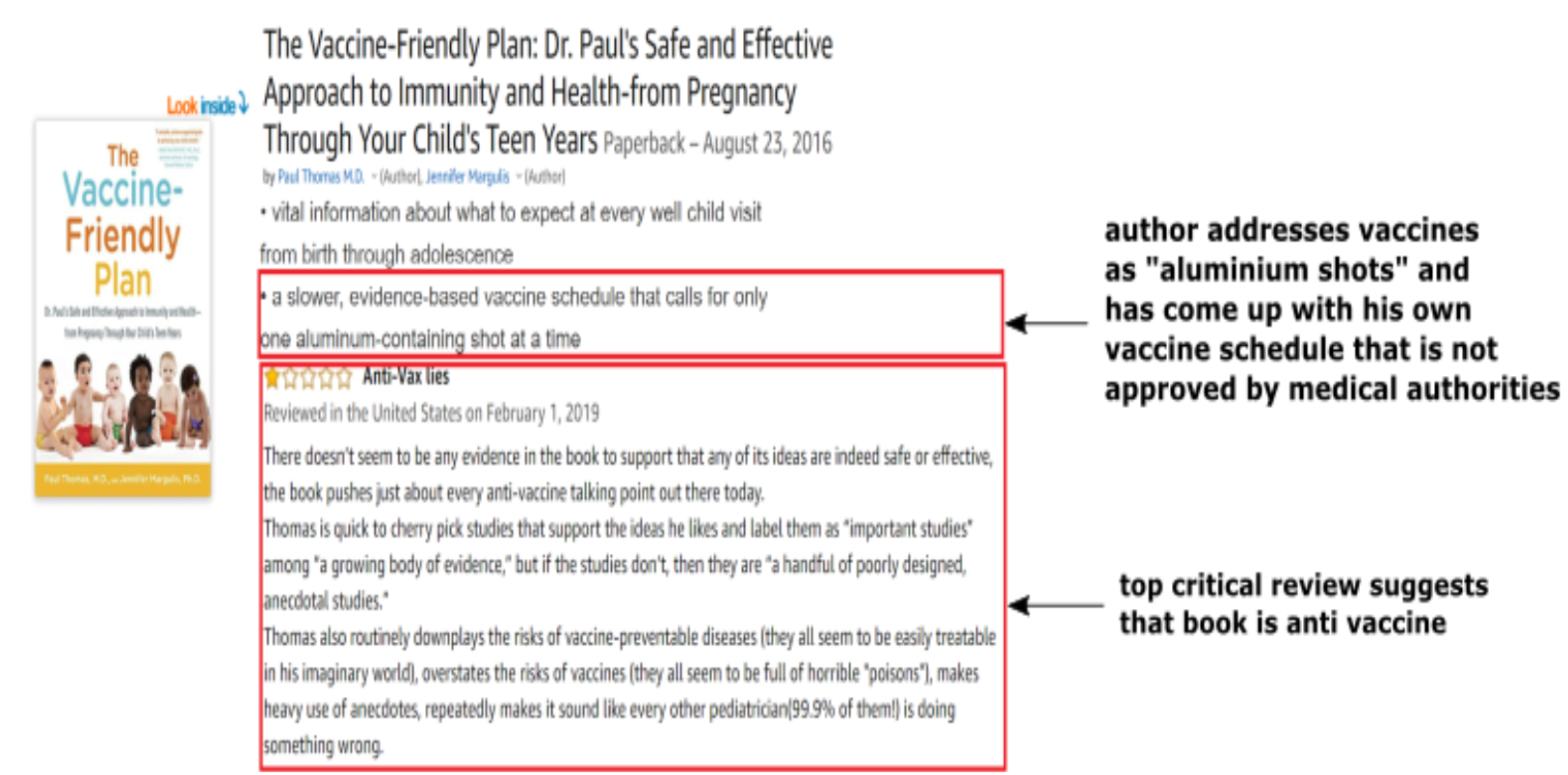}
  \caption{Example explaining Turks how to determine the annotation value.}
  \label{qual_examp}
  \Description[Figure illustrates how user can assign annotation value to a product by looking at its metadata]{In the figure we highlight the description of an Amazon book \textit{The vaccine-friendly plan} where the author addresses vaccines as aluminium shots and has come up with a vaccine schedule that is not approved by medical authorities. We also show the top critical review which suggests that book is anti-vaccine.}
\end{figure*}

\begin{figure*}
  \centering
      \includegraphics[scale=0.35]{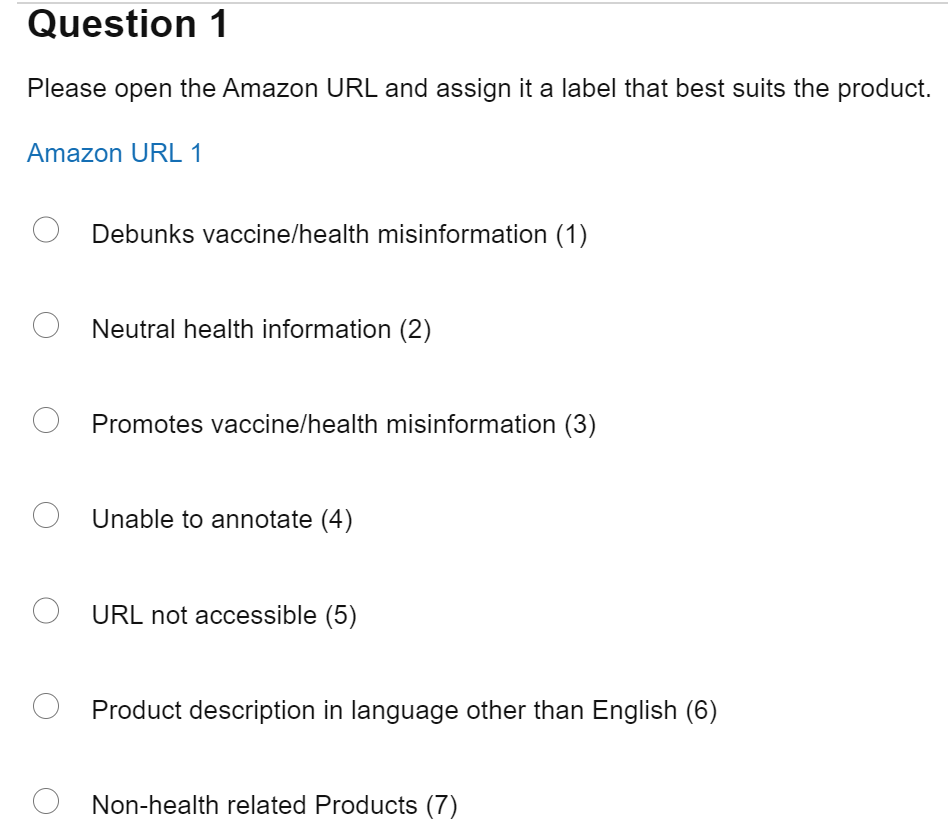}
  \caption{Example of Qualification Test question.}
  \label{qual_quest}
  \Description[Qualification Test question]{Figure illustrates a qualification test question which has URL of the Amazon product along with radio buttons enlisting all the annotation values. MTurks had to select the annotation value that best suits the Amazon product.}
\end{figure*}

To get high quality annotations, we screened MTurks by adding 3 qualification requirements. First, we required MTurks to be Masters. Second, we required them to have atleast 90\% approval rating. And lastly, we required them to get a full score of 100 in a Qualification Test. We introduced a test to ensure that  MTurks attempting our annotation job had a good understanding of the annotation scheme. The test had one eligibility question asking them to confirm whether they are affiliated to authors' University. Other three questions involved Mturks to annotate three Amazon products  (see Figure \ref{qual_quest} for a sample question). First author annotated these products and thus, their annotation values were known. To ensure MTurks understood the task and annotation scheme, we gave detailed instructions  and described each annotation value in detail with various examples of Amazon products in the qualifying test (Figures \ref{qual_instructions},  \ref{qual_desc} and \ref{qual_examp}). Examples were added as visuals. In each example, we marked the meta data used used for the annotation and explained why a particular annotation value was assigned to the product (see Figure \ref{qual_examp}).

We took two steps to ensure that instructions and test questions were easy to understand and attempt. First, we posted the test on subreddit  r/mturk\footnote{https://www.reddit.com/r/mturk/}---a community of MTurks, to obtain feedback. Second, we did a pilot run by posting ten tasks along with the aforementioned screening requirements. After obtaining positive feedback from the community and successful pilot-run, we released   our AMT job titled ``Amazon product categorization task''. We paid the Turks according to the United States federal minimum wage (\$7.25/hr). Additionally, we did not disapprove any worker's responses.

\subsubsection{Amazon product categorization task}
\begin{figure*}
  \centering
      \includegraphics[scale=0.35]{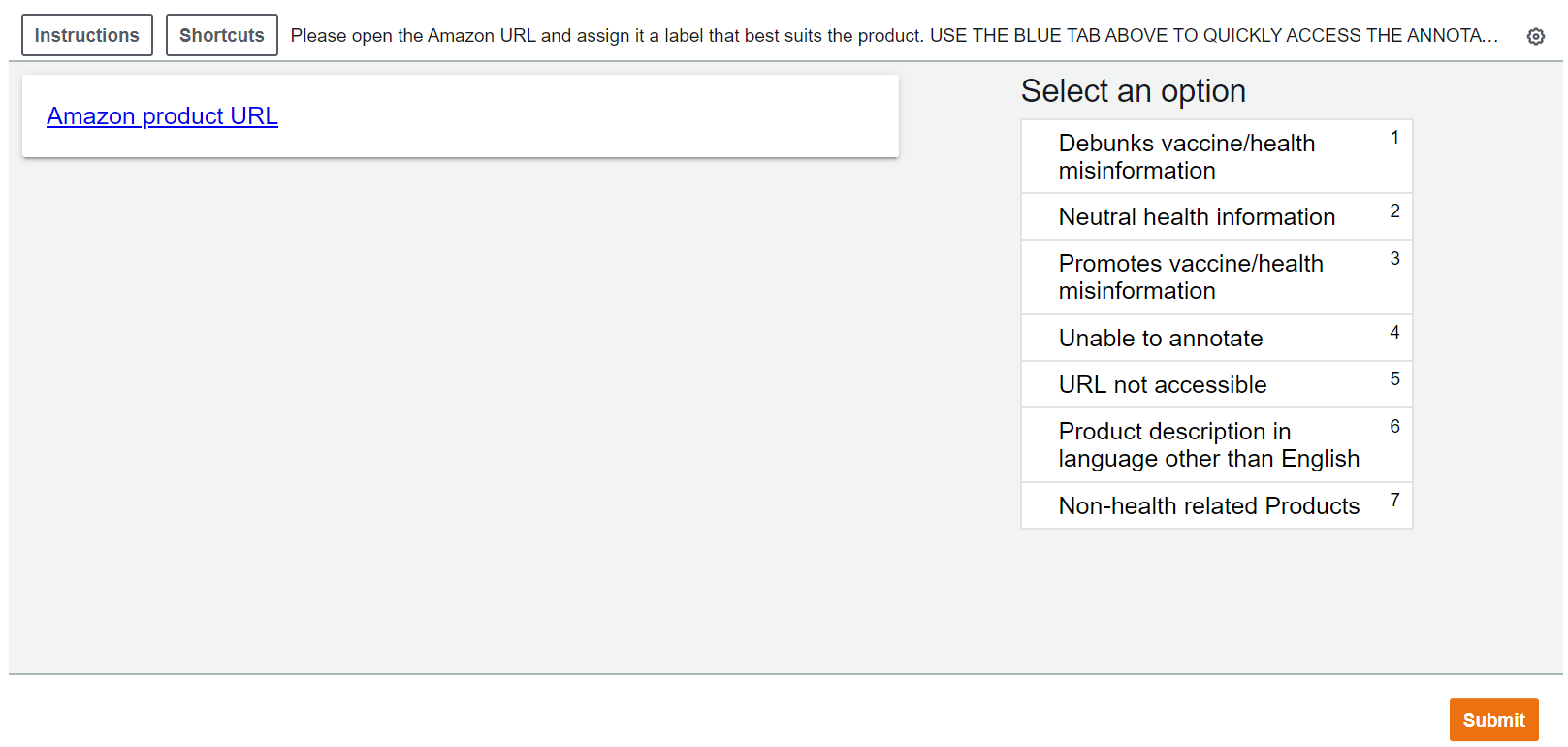}
  \caption{Interface of our Amazon product categorization task.}
  \label{amt_actual_task_interface}
  \Description[Interface of our AMT product categorization task]{Each task had a URL of the Amazon product along with radio buttons enlisting all the annotation values.}
\end{figure*}

We posted 1630 anno\-tations (tasks) in batches of 50 at a time. The job was setup to get three responses for each annotation value. The majority response was selected to label the Amazon product. To avoid any MTurk bias, we did not explicitly reveal that the idea behind the task was to get misinformation annotations. We used the term "Amazon product categorization" to describe our project and task throughout. For 34 products, all three MTurk responses differed. The first author then annotated these products to get annotation values. Figure \ref{amt_actual_task_interface} shows the interface of our AMT job.

\end{document}